\documentclass[english,12pt]{article}
\pdfoutput=1
\usepackage[T1]{fontenc}
\usepackage[latin9]{inputenc}
\usepackage{amstext}
\usepackage{babel}
\usepackage{verbatim}
\usepackage{amsfonts}
\usepackage{mathrsfs}
\usepackage{amsmath}
\usepackage{amssymb}
\usepackage{bm}
\usepackage{extarrows}
\usepackage{slashed}
\usepackage{isodateo}
\usepackage{xcolor}
\usepackage{graphicx}
\usepackage[low-sup]{subdepth}
\usepackage{float}
\usepackage{isodateo}
\usepackage{tensor}
\usepackage{multirow}
\usepackage{enumitem}
\usepackage{ytableau}
\usepackage[linktocpage=true]{hyperref}
\hypersetup{
colorlinks=true,
citecolor=blue,
linkcolor=blue,
urlcolor=blue,
pdfauthor={},
pdftitle={},
pdfsubject={}
}
\makeatletter
\def\simgt{\mathrel{\lower2.5pt\vbox{\lineskip=0pt\baselineskip=0pt
\hbox{$>$}\hbox{$\sim$}}}}
\def\simlt{\mathrel{\lower2.5pt\vbox{\lineskip=0pt\baselineskip=0pt
\hbox{$<$}\hbox{$\sim$}}}}
\numberwithin{equation}{section}

\usepackage{titlesec}
\titleformat*{\section}{\large\bfseries}
\titleformat*{\subsection}{\normalsize\bfseries}
\titleformat*{\subsubsection}{\small\bfseries}
\makeatletter
\newsavebox\myboxA
\newsavebox\myboxB
\newlength\mylenA
\newcommand*\xoverline[2][0.63]{%
\sbox{\myboxA}{$\m@th#2$}%
\setbox\myboxB\null% Phantom box
\ht\myboxB=\ht\myboxA%
\dp\myboxB=\dp\myboxA%
\wd\myboxB=#1\wd\myboxA% Scale phantom
\sbox\myboxB{$\m@th\overline{\copy\myboxB}$}%  Overlined phantom
\setlength\mylenA{\the\wd\myboxA}%   calc width diff
\addtolength\mylenA{-\the\wd\myboxB}%
\ifdim\wd\myboxB<\wd\myboxA%
\rlap{\hskip 1.2\mylenA\usebox\myboxB}{\usebox\myboxA}%
\else
\hskip -0.5\mylenA\rlap{\usebox\myboxA}{\hskip 0.5\mylenA\usebox\myboxB}%
\fi}
\definecolor{Pp}{RGB}{255,0,0}

\definecolor{Green}{RGB}{199,238,206}
\newcommand{\nod}[1]{:\!{#1}\!:}
\newcommand{\ehmd}[2]{\hat{k}_{#1} \cdot\hspace*{0.5mm}\hat{k}_{#2}}
\renewcommand{\thefootnote}{\arabic{footnote}} 
\newcommand{\eqrefe}{Eq.\eqref}
\newcommand{\beq}{\begin{equation}}
\newcommand{\eeq}{\end{equation}}
\newcommand{\ba}{\begin{array}}
\newcommand{\ea}{\end{array}}
\newcommand{\beqa}{\begin{eqnarray}}
\newcommand{\eeqa}{\end{eqnarray}}
\newcommand{\beqs}{\begin{subequations}}
\newcommand{\eeqs}{\end{subequations}}
\def\dis{\displaystyle}
\newcommand{\la}{\langle}
\newcommand{\ra}{\rangle}
\newcommand{\bla}{\big\langle}  
\newcommand{\bra}{\big\rangle}
\newcommand{\fr}[2]{\mbox{$\frac{\,{#1}\,}{#2}$}}
\renewcommand{\rm}{\mathrm}
\def\geqq{\geqslant}
\def\({\left(}
\def\){\right)}
\def\[{\left[\,}
\def\]{\,\right]}
\def\LB{\left\{}
\def\RB{\right\}}
\def\nn{\nonumber}
\def\pd{\partial}
\def\pp{\prime}
\def\to{\rightarrow}
\def\ito{\!\rightarrow\!}

\def\over{\overline}
\def\Tr{\text{Tr}}

\def\diag{\text{diag}}
\def\MP{M_{\text{Pl}}^{}}
\def\MPl{M_{\text{Pl}}}

\def\ba{\bar{a}}

\def\A{\mathcal{A}}

\def\CC{\mathcal{C}}

\def\td{\text{d}}

\def\ii{\text{i}}

\def\KK{\mathcal{K}}
\def\KKt{\widetilde{\mathcal{K}}}
\def\dKK{\delta\mathcal{K}}
\def\dKKt{\delta\widetilde{\mathcal{K}}}
\def\hk{\hat{k}}

\def\M{\mathcal{M}}
\def\MT{\widetilde{\mathcal{M}}}
\def\dM{\delta\mathcal{M}}
\def\dMT{\delta\widetilde{\mathcal{M}}}
\def\NN{\mathcal{N}}
\def\NNt{\widetilde{\mathcal{N}}}
\def\dNN{\delta\mathcal{N}}
\def\dNNt{\delta\widetilde{\mathcal{N}}}

\def\tN{\widetilde{N}}
\def\mO{\mathcal{O}}

\def\Qt{\widetilde{Q}}

\def\bR{\mathbb{R}}
\def\bs{\bar{s}}

\def\mS{\mathcal{S}}

\def\SS{\mathcal{S}}

\def\dT{\delta\mathcal{T}}
\def\TT{\mathcal{T}}
\def\tT{\widetilde{\mathcal{T}}}

\def\mX{\mathcal{X}}
\def\mY{\mathcal{Y}}

\def\Z{\mathbb{Z}}
\def\ZZ{\mathbb{Z}_2^{}}
\def\SS{\mathbb{S}}
\def\bz{\bar{z}}
\def\NH{\widehat{N}}
\def\vrhohat{\widehat{\varrho}}
\def\al{\alpha}
\def\alp{\alpha^\prime}
\def\be{\beta}

\def\ka{\kappa}

\def\mn{\mu\nu}

\def\da{\delta}

\def\lam{\lambda}

\def\si{\sigma}

\def\ct{c_\theta}
\def\st{s_\theta}

\def\cct{c_\theta^2}
\def\sst{s_\theta^2}

\def\ctt{c_{2\theta}}
\def\cttt{c_{3\theta}}
\def\ctf{c_{4\theta}}

\def\cts{c_{6\theta}}

\def\sz{s_0^{}}
\def\tz{t_0^{}}
\def\uz{u_0^{}}
\def\bsz{\bar{s}_0}

\def\rrp{r_{\!+}^2}

\def\qb{\bar{q}}
\def\qqbp{\bar{q}^{\prime 2}}

\def\op{\text{op}}
\def\cl{\text{cl}}
\def\Mn{M_n}

\def\sgn{\text{sgn}}
\def\Mnn{M_n^2}
\def\hs{\hspace*{0.3mm}}
\def\hsx{\hspace*{0.5mm}}
\def\hsm{\hspace*{-0.3mm}}
\def\hsmx{\hspace*{-0.5mm}}

\allowdisplaybreaks[4]
\begin{document}

\interfootnotelinepenalty=10000
\baselineskip=17pt

\vspace*{-2cm}
\thispagestyle{empty}

\begin{center}
{\Large\bf %\boldmath
Scattering Amplitudes of Kaluza-Klein Strings
\\[1mm] 
and Extended Massive Double-Copy}

%\end{center}
%\end{document}

\vspace*{10mm}

{\bf Yao Li}\,$^{a,}$\footnote{Email: {neolee@sjtu.edu.cn}},
~
{\bf Yan-Feng Hang}\,$^{a,}$\footnote{Email: {yfhang@sjtu.edu.cn}},
~
{\bf Hong-Jian He}\,$^{a,b,}$\footnote{Email: {hjhe@sjtu.edu.cn}},
~
{\bf Song He}\,$^{c,d,}$\footnote{Email: {songhe@itp.ac.cn}}

\vspace*{3mm}
$^a$\,Tsung-Dao~Lee Institute \& 
School of Physics and Astronomy, 
\\[-1mm]
Key Laboratory for Particle Astrophysics and Cosmology (MOE),
\\[-1mm]
Shanghai Key Laboratory for Particle Physics and Cosmology,
\\[-1mm]
Shanghai Jiao Tong University, Shanghai 200240, China
\\[0.5mm]
$^b$\,Institute of Modern Physics and Department of Physics,
\\[-1mm]
Tsinghua University, Beijing 100084, China;
\\[-1mm]
Center for High Energy Physics, Peking University, 
Beijing 100871, China
\\[0.5mm]
$^c$\,CAS Key Laboratory of Theoretical Physics, 
Institute of Theoretical Physics,
\\[-1mm] 
Chinese Academy of Sciences, Beijing 100190, China
\\
$^d$\,School of Fundamental Physics and Mathematical Sciences,
\\[-1mm]
Hangzhou Institute for Advanced Study, UCAS, Hangzhou 310024, China
%School of Physical Sciences, University of Chinese Academy of Sciences, %Beijing 100049, China

\vspace{11mm}
%\today

\end{center}

%\vspace{8mm}
%\noindent
%{\bf Abstract.}
\begin{abstract}
\baselineskip 16pt
\noindent
We study the scattering amplitudes of massive Kaluza-Klein (KK) states
of open and closed bosonic strings under toroidal compactification. 
We analyze the structure of vertex operators for the KK strings and  
derive an extended massive KLT-like relation which connects the 
$N$-point KK closed-string amplitude to the products of 
two KK open-string amplitudes at tree level. 
Taking the low energy field-theory limit of vanishing Regge slope, 
we derive double-copy construction formula of the $N$-point
massive KK graviton amplitude from the sum of proper products
of the corresponding KK gauge boson amplitudes.\ 
Then, using the string-based massive double-copy formula, 
we derive the exact tree-level four-point KK gauge boson amplitudes and KK graviton amplitudes, which fully agree with those given by   
the KK field-theory calculations.\ With these, we give an explicit prescription on constructing the exact four-point KK graviton 
amplitudes from the sum of proper products of the corresponding color-ordered KK gauge boson amplitudes.\ We further analyze the 
string-based double-copy construction of five-point and six-point 
scattering amplitudes of massive KK gauge bosons and KK gravitons.
\\[4mm] 
\\[2mm]
JHEP 02 (2022) 120 [arXiv:2111.12042\,[hep-th]].
\end{abstract}

\newpage
\baselineskip 17.5pt

\linespread{.9}
\tableofcontents

\vspace{8mm}
%\newpage

\setcounter{footnote}{0}
\renewcommand{\thefootnote}{\arabic{footnote}}

\section{\hspace*{-3mm}Introduction}
\label{sec:1}

Early attempts of unifying the electromagnetic and 
gravitational forces pointed to a truly 
fundamental possibility of a higher dimensional spacetime structure 
of 5d with a single extra spatial dimension compactified on 
a circle \`{a} la Kaluza-Klein (KK)\,\cite{KK}.\ 
This intriguing avenue was seriously pursued and
widely explored in various contexts, including the
string/M theories\,\cite{string} and extra dimensional field theories 
with large or small extra dimensions\,\cite{Exd}.\
In fact, the unification among the conventional gauge forces
was first realized through the electroweak theory\,\cite{SM} 
of the standard model (SM) and subsequently extended to the 
grand unification (GUT) of the electroweak and strong forces\,\cite{GUT}.

\vspace*{1mm}

The big obstacle to further unification between the gauge forces and 
gravity force lies in the apparently distinctive natures of  
Einstein's generality relativity (GR) including its intricate 
nonlinearity and perturbative nonrenormalizability.
However, the conjectured double-copy relation of  
$\,\text{GR}\hsm =\hsmx (\text{Gauge~Theory})^2\,$
points to fundamental clues to the deep gauge-gravity connection.
The Kawai-Lewellen-Tye (KLT) relation\,\cite{KLT} 
was constructed
to connect the scattering amplitudes of closed strings to  
the products of scattering amplitudes of open strings at tree level.
In the low energy field-theory limit, the KLT relation leads to
the connection of scattering amplitudes of massless gravitons
to the products of scattering amplitudes of massless gauge bosons
with proper kinematic factors.
This was then extended to the field theory framework through
the double-copy method of color-kinematics duality of
Bern-Carrasco-Johansson (BCJ)\,\cite{BCJ}\cite{BCJ-Rev}
which links the scattering amplitudes of
massless gauge theories to that of the massless gravity.   
Analyzing the properties of the heterotic string and open string 
amplitudes can prove and refine parts of the 
BCJ conjecture\,\cite{Tye-2010}.\
The Cachazo-He-Yuan (CHY) formalism\,\cite{CHY}\cite{CHY14} 
shows that the KLT kernel can be interpreted as the inverse amplitudes 
of bi-adjoint scalars, and this can be generalized to 
double-copy relations for other field 
theories\,\cite{CHY14}.\
So far substantial efforts have been made to formulate and test
the double-copy constructions between the massless gauge theories and
massless GR\,\cite{BCJ-Rev}, 
and some recent works attempted to extend the double-copy method to
the 4d massive Yang-Mills (YM) theory versus Fierz-Pauli-like
massive gravity\,\cite{dRGT}, 
to the KK-inspired effective gauge theory with extra
global U(1)\,\cite{DC-5dx}, 
to compactified 5d KK gauge/gravity theories\,\cite{Hang:2021fmp},
and to the 3d Chern-Simons theories with
or without supersymmetry\,\cite{3dCS-susy}\cite{3dCS1}\cite{3dHHS}.

\vspace*{1mm}

But the extensions of conventional double-copy method to 
massive gauge/gravity theories are generally difficult,
because many such theories (including the massive YM 
theory and massive Fierz-Pauli gravity) violate gauge symmetry and
diffeomorphism invariance (which are the key for successful double-copy
construction). 
The two important candidates with promise include the compactified KK 
gauge/gravity theories and the topologically massive Chern-Simons (CS) 
gauge/gravity theories.\
The massive KK gauge bosons and KK gravitons acquire their masses 
from geometric 
``Higgs'' mechanisms\,\cite{5DYM2002}\cite{GHiggs}\cite{Hang:2021fmp} 
of the KK compactifications which spontaneously break the 
higher dimensional gauge symmetry and diffeomorphism invariance 
to that of 4d by boundary conditions.\ 
Such geometric ``Higgs'' mechanisms
can be quantitatively formulated at the scattering $S$-matrix level
by the KK gauge boson equivalence theorem (KK\,GAET)\,\cite{5DYM2002}\cite{5DYM2002-2}\cite{KK-ET-He2004} 
and KK gravitational ET (KK\,GRET)\,\cite{Hang:2021fmp}, 
which generally
ensure much better high energy behaviors of the KK scattering
amplitudes than that of other ill-defined massive theories (with
explicitly broken gauge/gravity symmetries) and thus hold real promise 
for successful double-copy construction.\ 
The 3d CS gauge/gravity theories\,\cite{3dCS0} naturally realize
topological mass-generation for the gauge bosons (gravitons) 
in a gauge-invariant (diffeomorphism-invariant) way, which can also
ensure good high energy behaviors of 
the scattering amplitudes\,\cite{3dHHS} and 
realize successful double-copy constructions\,\cite{3dCS-susy}\cite{3dCS1}\cite{3dHHS}.

\vspace*{1mm}

A recent work\,\cite{Hang:2021fmp} systematically studied
the extended BCJ-type double-copy construction between the
scattering amplitudes of the massive KK gauge bosons and of the
massive KK gravitons in the KK YM gauge theory and
KK gravity theory under the 5d compactification of 
$\SS^1\hsm /\ZZ\hs$.\
The double-copy construction for the scattering amplitudes of
massive KK gauge bosons and KK gravitons is highly nontrivial 
even for the four-point elastic KK amplitudes 
due to the presence of double-pole-like structure with exchanges
of both zero-modes and KK-modes.\ 
Ref.\,\cite{Hang:2021fmp} first proposed an improved double-copy method
for massive KK amplitudes by using the 
{\it high energy expansion order by order.}
The leading-order (LO) KK gauge boson amplitudes 
were shown\,\cite{Hang:2021fmp} to be mass-independent
and their numerators obey the kinematic Jacobi identity, so
the extended BCJ-type double-copy construction can be universally
realized to reconstruct the correct LO KK graviton amplitudes.\ 
But the next-to-leading-order (NLO) KK gauge boson amplitudes
were found\,\cite{Hang:2021fmp} 
to be mass-dependent and the corresponding double-copied KK
graviton amplitudes do not always match the exact KK 
graviton amplitudes at the NLO. 
So the naive extension of the BCJ double-copy method 
could not fully work and a modified
massive double-copy construction was proposed\,\cite{Hang:2021fmp} 
for the NLO KK amplitudes$\hs$\footnote{%
As we will show in Appendix\,\ref{app:B} of the present paper,
we can construct a new type of numerators by making generalized gauge
transformations and properly choosing the energy expansion parameter
such that the kinematic Jacobi identity is obeyed.
But further improvements for realizing precise BCJ-type double-copy are still needed.}$\!$,$\hs$ 
but this is yet to be established for 
all KK scattering processes and for going beyond the NLO. 
Hence, it is truly attractive and important 
to establish the double-copy construction, 
from the first principle of KK string theory formulation,  
for the exact tree-level massive KK gauge-boson/graviton amplitudes 
and in a universal way.

\vspace*{1mm}

In this work, we take the simplest KK compactification of
the 26d bosonic string theory\,\cite{string} as a tool
to derive the extended massive KLT-like relations for 
KK closed/open-string amplitudes and then achieve the doubel-copy
construction of the realistic 5d KK gauge boson/graviton amplitudes 
in the low energy field-theory limit.\  
The essential advantage of the compactified KK string theory 
is that {\it the connection between the KK closed-string amplitudes 
and the proper products of KK open-string amplitudes 
can be intrinsically built in from the start.} 
A recent literature\,\cite{wKLT} 
studied the general KLT factorization
of winding string amplitudes in the 
bosonic string theory and computed explicitly  
the four-point tachyon amplitudes, 
but did not consider the amplitudes in the 
low energy field-theory limit.
We will study the scattering amplitudes of massive KK states
of open and closed bosonic strings, and derive the corresponding
scattering amplitudes of KK gauge-bosons and of KK gravitons 
in the field-theory limit. 
With these, we derive the extended KLT-like relations 
which connect the $N$-point KK closed-string amplitude 
to the product of the 
two corresponding open string amplitudes at tree level.\
Taking the field-theory limit of vanishing Regge slope, 
we derive double-copy construction which formulates the general $N$-point
massive KK graviton amplitude as the sum of proper products
of the corresponding KK gauge boson amplitudes.\ 
Then, using the string-based KLT-like massive KK double-copy formula, 
we derive the exact four-point elastic and inelastic 
scattering amplitudes of KK gravitons from the sum of the proper
products of the relevant color-ordered amplitudes 
of KK gauge bosons at tree level. 
We verify that the reconstructed four-point elastic KK graviton 
amplitude fully agrees with that given by the available Feynman-diagram 
calculations\,\cite{Chivukula:2020S}\cite{Chivukula:2020L} 
in the 5d KK field theory of GR.  
Based on our KK-string formulation, 
we give an explicit prescription 
on constructing the exact four-point KK graviton scattering 
amplitudes from the sum of relevant products of the corresponding color-ordered KK gauge boson scattering amplitudes.\ 
We further analyze the 
string-based double-copy construction of five-point and six-point 
scattering amplitudes of massive KK gauge bosons and KK gravitons.

\vspace*{0.5mm}
This paper is organized as follows. 
In section\,\ref{sec:2}, we analyze the structure of vertex operators 
for the KK strings and derive an extended massive KLT-like relation 
which connects the $N$-point KK closed-string amplitude to the product 
of the two corresponding open-string amplitudes at tree level. 
Then, we take the low energy field-theory limit of vanishing Regge slope
and present the double-copy construction of the $N$-point
massive KK graviton amplitude by the sum of relevant products
of the color-ordered KK gauge boson amplitudes.
In section\,\ref{sec:3}, we systematically derive the four-point 
elastic and inelastic color-ordered scattering amplitudes of 
KK gauge bosons from the field-theory limit of the corresponding 
scattering amplitudes of KK open strings. 
We study the structure of these color-ordered massive KK gauge boson 
amplitudes and demonstrate that they can be obtained from
the scattering amplitudes of massless zero-mode gauge bosons
under proper shifts of the Mandelstam variables. This gives
an elegant and efficient way to compute any color-ordered massive 
KK gauge boson amplitudes.
In section\,\ref{sec:4}, applying the string-based massive 
double-copy formula, we derive the exact tree-level four-point 
elastic and inelastic KK graviton amplitudes. 
Then, we give an explicit prescription 
on constructing the exact four-point KK graviton 
amplitudes from the sum of relevant products of the corresponding color-ordered KK gauge boson amplitudes.\
We also use our general string-based double-copy construction to
obtain the five-point and six-point scattering amplitudes of massive 
KK gauge bosons and KK gravitons.
Finally, we conclude in section\,\ref{sec:5}.
Appendix\,\ref{app:A} provides the notational setup and
kinematics formulas for the elastic and inelastic scattering processes 
of four KK states. In Appendix\,\ref{app:B} we present the 
exact four-point amplitudes of the elastic and inelastic scattering
of KK gauge bosons at tree level, which are shown to fully agree with
those obtained from the corresponding scattering amplitudes 
of KK open strings under low energy field-theory limit as given by
section\,\ref{sec:3}. This also serves as a systematic  
consistency check of our open-string calculations in the main text.

\section{\hspace*{-3mm}KK String Amplitudes and Extended Massive KLT-Like Relation}
\label{sec:2}

In this section, we consider the compactifications of the 26d bosonic string theory with one relatively large extra spatial dimension compactified to a circle and with all other extra spatial dimensions
decoupled due to their extremely small radii of $\mO(\MPl^{-1})$. 
We study the mass spectra of both KK open and closed strings. 
Then, we explicitly compute the $N$-point KK open-string amplitudes under compactification by using the relevant compact photon vertex operators 
and construct the amplitudes of KK closed-strings by products of 
two KK open-string amplitudes. 
The scattering amplitudes of KK closed-strings take 
the KLT-like form.
Finally, taking the low energy field-theory limit 
$\hs \alp \!\ito 0\,$, 
we derive the formulas of general $N$-point amplitudes in the 
compactified KK field theories, so the KLT-like relation
of KK string amplitudes will result in the double-copy formula
of the corresponding KK graviton amplitudes.

\subsection{\hspace*{-3mm}Compactification of Bosonic Strings}
\label{sec:2.1}

For the sake of the present study, 
we consider the bosonic strings propagating in a 26-dimensional 
spacetime background 
$\,\mathbb{R}^{1,24} \hsmx\times \SS^1\,$.\
We can first compactify the extra spatial dimensions of coordinates $\{X^{4},\cdots\hsm ,X^{24}\}$ with their radii 
$\hs r_j^{}\!=\hsm\mO(\MPl^{-1})\!\ll\!R\,$ 
($j\!=\!4,\cdots\hsm ,24$), so they are fully decoupled 
at energy scales much below the reduced Planck scale 
$\MP\hsmx =\hsmx (8\pi G)^{-1/2}$.
Thus, we only need to study the toroidal compactification of the single
extra spatial dimension of coordinate $X^{25}$ on a circle $\hs\SS^1$
with radius $R$\,, 
which is much larger than the Planck length $\MPl^{-1}$ 
and does not decouple in our low energy effective theory.
(Here by low energy, we mean the
energy scale scales which are lower than the reduced Planck scale 
$\MPl^{}$ by about two orders of magnitude or more.)
For the present study, we take bosonic string as a
computational tool for establishing the massive KLT-like relations 
of KK string states and for deriving the low energy KK graviton 
scattering amplitudes.

\vspace*{1mm} 

Then, we can identify the coordinate $X^{25}\hs (\hs\equiv\! X\hs )$  
as a scalar field on the string worldsheet  
and it obeys the periodic boundary condition on the circle $\SS^1$\,:
%%%
\begin{equation} 
\label{eq:Compact}
X   \,\cong\, X   + 2\pi R  \,.
\end{equation}
%%%
For the closed strings,
without loss of generality, the periodic boundary condition further 
takes the following form:
%%%
\begin{equation}
\label{eq:Compact-close}
X(\tau, \si \!+\! 2\pi) \,=\, X(\tau,\si) + 2\pi w R  \,,
\end{equation}
%%%
where $\,w \!\in\! \Z\,$ is the winding number describing 
how many times a closed string winds around the circle $\hs\SS^1$. 
Thus, the eigenvalues of generators of Virasoro algebra
$(L_0^{},\hs \over{L}_0^{})$ for 
the oscillation modes of closed string are derived as follows:
%%%
\beqs
\begin{align}
L_{0} \,&=\,
\frac{\alp}{4} p^{\mu} p_{\mu}^{}+\frac{\,\alp}{4}
\!\(\frac{\,\hat{n}}{R}
+\frac{\hs wR\hs}{\,\alp}\)^{\!\hsm\!2} \!+(N \!-\! 1) \,,
\\
\bar{L}_0 \,&=\,
\frac{\,\alp}{4} p^{\mu} p_{\mu}+\frac{\,\alp}{4}
\!\(\frac{\,\hat{n}\hs}{R}
-\frac{\hs wR\hs}{\alp}\)^{\!\!2} \!+(\tN \!-\! 1) \,,
\end{align}
\eeqs
%%%
where $\alp$ is the Regge slope,
$\,\hat{n} \in \Z\,$ denotes the Kaluza-Klein (KK) level,
and $(N,\hs\tN)$ represent the string level.
Hence, the mass spectrum of the KK state of a closed string can be obtained by imposing the physical conditions 
$\hs L_0 = \bar{L}_0 =0\,$:
%%%
\begin{align}
\label{eq:KK-mass-close} 
M^2_{\cl} \,= \(\!p_{\hat{n}}^{} +\! \frac{w R}{\alp}\)^{\!\!\!2} 
\!+\! \frac{4}{\alp}(N \!-\! 1)
\,=\, \(\!p_{\hat{n}}^{} -\! \frac{w R}{\alp}\)^{\!\!\!2} 
\!+\! \frac{4}{\alp} (\tN \!-\! 1) \,,
\end{align}
%%%
where 
\begin{equation}
p_{\hat{n}}^{} = \frac{\hat{n}}{\,R\,} =\, \sgn\hsm\times\!\Mn \hs ,
\end{equation}
with $\hs \hat{n}\!\in\!\Z\,$ and
$\,\sgn \!\equiv\text{sign}(\hat{n})\hsmx =\hsmx \pm 1\hs$.
In the above, $\,\Mn \!=\hsm |\hat{n}|/\hsm R\,$
is the KK mass-parameter.
Note that in \eqrefe{eq:KK-mass-close}, 
the second equality is realized 
by imposing the level matching condition 
$\,nw \hsmx =\hsm \tN\hsm -\hsm N\hs$.  
We can further decompose the mass spectrum \eqref{eq:KK-mass-close} 
as follows:
%%%
\begin{equation}
M^2_{\cl} \,=\, \Mn^2 + 
\frac{\,w^2\hsm R^2\,}{\al^{\prime\hs 2}} + 
\frac{2}{\,\alp\,}(N\hsm +\hsm \tN\hsm -\hsm 2) \,,
\end{equation}
%%%
where the right-hand-side (RHS) contains the contributions from 
the squared KK-mass $M_n^2 \hsmx =\hsm n^2\hsm /\hsm R^2\hs$ 
and a squared topological mass 
$(wR/\alpha')^2$ (related to the string winding number).
For studying the low energy limit of string theory 
as a field theory, we will set $\hs w\!=\!0\hs$ and thus 
$\hs N\hsm\!=\!\tN \hsm\!=\!1\hs$ for the (KK) gravitons.

\vspace*{1mm}

The open string satisfies the Dirichlet boundary condition 
for the compactified dimensions and the Neumann boundary condition for non-compactified dimensions.\ Each ending point of the 
open string is attached to a D-brane\,\cite{string}. 
The boundary condition for the open string takes the following form:
%%%
\begin{equation}
X(\tau,\hs\si\hsm +\hsm \pi) \,=\,
X(\tau,\hs\si) + 2\pi wR + q\hs L \,,
\end{equation}
%%%%
where $q \!\in\! \Z$ labels the D-branes transverse to the 
compactified circle $\SS^1$ with equal distance $L\hs$. 
We make the following mode expansion for open string: 
%%%
\begin{equation}
X(\tau,\si)\,=\, x_q^{} + \frac{\si}{\,\pi\,} 
(q' L \hsm +\hsm 2\pi w R) - \sqrt{2\alp\,} \sum_{m \neq 0} 
\frac{\,\al_{m}^{}\,}{m} e^{- \ii\hs m \tau} \sin (m\si) \,,
\end{equation}
%%%
where we set the two ends of the open string at the $q$-th brane
and $(q\hsm +\hsm q')$-th brane: 
%%%
\begin{align}
& X(\tau,0)=x_q^{} \,, \quad~~ 
X(\tau,\pi)= x_q^{} \hsm + q' L \hs .
\end{align}
%%%
Thus, the mass spectrum of open strings is derived as follows:
%%%%
\begin{eqnarray} 
\label{eq:KK-mass-open}
M_{\op}^{2}\,=\, 
\(\!\!\frac{~\hat{n}L \!+\! 2\pi wR~}{2\pi \alp}\!\)^{\hsmx\!\!2} +\frac{1}{\,\alp\,}(N\hsm -1) \,.
\end{eqnarray}
%%%%
For studying the low energy limit of string theory 
as a field theory, we only need to consider the 
$\,w\hsm =\hsm 0\,$ case and thus 
$\hs N\hsm\!=\!1\hs$ for the (KK) gauge bosons.
The mass spectra \eqref{eq:KK-mass-close} and
\eqref{eq:KK-mass-open} are not necessarily identical in general. 
But, for the consistent realization of double-copy construction, 
we can impose the following matching condition on 
\eqrefe{eq:KK-mass-open}:
%%%
\begin{equation}
R\hs L\,=\,2\pi \alp \,,
\end{equation}
%%%%
and make the rescaling 
$\,(\alp, R,\hs L) \ito \fr{1}{4}\!\(\alp, R, \hs L\)\hs$,\, 
such that the mass spectrum \eqref{eq:KK-mass-open} 
of open strings coincides with the mass spectrum 
\eqref{eq:KK-mass-close} of closed strings\,\cite{wKLT}.

\subsection{\hspace*{-3mm}Vertex Operators of KK String States}
\label{sec:2.2}

In this subsection, we present the vertex operators for the KK 
string states. 
For the closed and open strings under compactification, 
we can write down the integrated vertex operators for their KK states:
\\[-9mm]
%%%%
\beqs
\begin{align}
V_{\op}(\zeta,k,\hat{n}) \,&=~
\ii\hs g_{\op}^{}\!
\int \!\! \td y \, \hs \zeta_\mu \nod{\pd X^\mu \hs e^{\ii k\cdot X} 
e^{ \ii p_{\hat{n}}^{}\mY}} \,,
\\
V_{\cl}^{}(\zeta ,k,\hat{n}) \,&=~  
\ii\hs g_{\cl}^{}\!
\int \!\! \td^2 z \,  \zeta_{\mn}
\nod{\pd X^\mu \bar{\pd} X^\nu  \hs e^{\ii k \cdot X} 
e^{\ii p_{\hat{n}}^{}\mX}}\,,
\end{align}
\eeqs
%%%
where $k$ denotes the momentum in the noncompactified spacetime and
$\,p_{\hat{n}}^{} \! =\hsm {\hat{n}}/\hsm{R}\,$ 
($\hat{n}\! \in\hsm \Z$) is the quantized momentum in 26d.  
The compactified 26d string coordinates $(\mX,\mY)$ are defined as:
%%%
\begin{align}
\mY(y) \hsm= X_L(y) \!-\! X_R(y) \,, \quad 
\mX(z,\bz) \hsm= X_L(z) \!+\! X_R(\bz) \,, 
\end{align}
%%%
where $X_L(z)$ and $X_R(\bz)$ denote the left-moving and right-moving
string coordinates in the 26d, respectively.

\vspace*{1mm}

Then, for a noncompactified spatial dimension,
we can write down the Green functions for the open strings
under the Neumann boundary condition:
%%%
\beqs
\begin{align}
\label{open-OPE}
\bla X^\mu_L(y_1) X^\nu_R(y_2) \bra \, &= \, \bla X^\mu_R(y_1) X^\nu_L(y_2)\bra  \,=\, 0 \,,
\\[1mm]
\bla X^\mu_L(y_1) X^\nu_L(y_2)\bra\, &= \, \bla X^\mu_R(y_1) X^\nu_R(y_2)\bra  \,=\, -\alp  \eta^{\mu \nu}\ln\hsm |y_1 \hsm -\hsm y_2|  \,,
\end{align}
\eeqs
%%%
where the Lorentz indices $\mu,\nu=0,1,\ldots,24$\,.
While for a compactified spatial dimension, we have the Green functions
of open strings under the Dirichlet boundary condition:
%%%
\beqs
\begin{align}
\bla X_L(y_1) X_R(y_2)\bra\, &= \, \la X_R(y_1) X_L(y_2)\bra
\,=\, \alp \ln\hsm |y_1 \hsm -\hsm y_2|  \,,
\\[1mm]
\bla X_L(y_1) X_L(y_2)\ra\, &= \, \bla X_R(y_1) X_R(y_2)\bra
\,=\, -\alp \ln\hsm |y_1^{} \hsm -\hsm y_2^{}|  \,.
\end{align}
\eeqs
%%%%%
And the Green functions for closed string is  given by
%%%
\begin{equation}
\label{close-OPE}
\bla X^M (z_1,\bar{z}_1) X^N (z_2,\bar{z}_2) \bra \, =\, -\frac{\alp}{2} \eta^{MN} \ln(|z_1-z_2|^2) \,,
\end{equation}
%%%
where the 26d Lorentz indices $M,N=(0,1, \ldots,25) $\,.

\vspace*{1mm}

In string theory, imposing the orbifold compactification $\SS^1/\Z_2$ will generate an anti-periodic boundary condition, 
which lifts the vacuum energy on the worldsheet and modifies  
the mass spectrum\,\cite{string}. 
In consequence, the mass of each KK open-string state gets a shift
$\,\Delta M_{\op}^2\hsm =\hsm\frac{1}{\,16\alpha'\,}\,$, and 
the mass of each KK closed-string state receives an increase 
$\Delta M_{\cl}^2\hsm =\hsm\frac{1}{\,4\alpha'\,}$.  
Thus they will be decoupled in the field-theory limit of
$\,\alp\hsm\ito 0\,$.
Hence, we will first make our analysis 
by using the periodic boundary condition
on $\SS^1$ without imposing the $\ZZ\hs$.\  
Then, we can construct the vertex operators for KK open strings
and KK closed strings with specified $\ZZ$ parity:
%%%
\beqs
\label{eq:Vpm}
\begin{align}
\label{eq:Voppm}
V_{\op}^{\pm} (\zeta,k,n) \,&=\, \frac{1}{\sqrt{2\,}\,}\! 
\[\hsm V_{\op}^{}(\zeta,k,n) \pm V_{\op}^{}(\zeta,k,-n)\hsm\] ,
\\[1.5mm]
V_{\cl}^{\pm} (\zeta , k , n) \,&=\, \frac{1}{\sqrt{2\,}\,} \! 
\[\hsm V_{\cl}^{}(\zeta, k,n) \pm V_{\cl}^{}(\zeta ,k,-n)\hsm\] ,
\label{eq:Vclpm}
\end{align}
\eeqs
%%%
where $\,n \!\in\! \Z^+$ and the superscript ``$+\hs (-)$'' 
stands for the $\Z_2$-even ($\Z_2$-odd) state. 
We note that for the zero-modes ($n\!=\!0$), vertex operators for
open and closed strings,
$V_{\op}^{}(\zeta,k,0)$ and $V_{\cl}^{}(\zeta,k,0)$, 
are always $\Z_2$-even.
In particular, we will be interested in the above vertex operators 
having $\ZZ$-even parity because their scattering amplitudes will give, 
in the low energy field-theory limit, 
the corresponding scattering amplitudes of the KK gauge bosons and 
of the KK gravitons in the KK gauge/gravty theories under the 
5d compactification of $\hs\SS^1/\ZZ\hs$. This will also be valuable for
comparison with the literature\,\cite{Hang:2021fmp}\cite{5DYM2002}\cite{Chivukula:2020L} 
which computed some of these KK amplitudes by Feynman diagram approach
in the KK gauge/gravity field theories under 5d compactification of
$\hs\SS^1/\ZZ\hs$.

\subsection{\hspace*{-3mm}Open and Closed String Amplitudes for Massive KK States}
\label{sec:2.3}

For a $N$-point KK string amplitude, the conservation of 
the compactified momentum (KK-number) is achieved by the neutral condition\,\cite{CFT} of the vertex operators on the string worldsheet,
i.e., the sum of  KK numbers ($\hat{n}_j^{}$) of the external states 
should vanish:
%%%%%%
\begin{equation}
\label{KK-num-sum}
\sum_{j=1}^N \hat{n}_j^{} \,=\, 
\sum_{j=1}^N \sgn_j^{} \!\times n_j^{} \,=\, 0\,, 
\end{equation}
%%%
where $\hs\hat{n}_j^{}\!\in\! \Z\,$ is the KK number of the $j$-th 
external state,  
$\hs n_j^{} \!=\!\hsm |\hat{n}_j^{}|\!\in\!\Z^+\hs$, 
and the sign of $\hs\hat{n}_j^{}\hs$ is denoted as 
$\,\sgn_j^{}\!=\text{sign}(\hat{n}_j^{})\hs$. 
Thus we can write 
$\,p_{\hat{n}_j^{}}^{}\!\!=\sgn_j^{}\!\hsmx\times\! M_{n_j}^{}$.
Hence, the tree-level $N$-point open-string amplitude 
can be written as follows:
%%%%
\begin{align}
\hspace*{-4mm}
&\A^{(N)}_{\op}(\zeta, k,n) =~
\frac{e^{-\lam}}
{\,\rm{Vol}[\hs\rm{SL}(2,\bR)]\,}
\prod_{j=1}^{N}\int \! \td y_j
\, \bla V_{\op}^\pm (\zeta _1 ,k_1, n_1)
\cdots
V_{\op}^\pm (\zeta_N^{},k_N^{}, n_N^{}) \bra
\nn\\[0.5mm]
\hspace*{-4mm}
&=~
\frac{\,(-)^{\omega}\hs 2^{-\NH /2}\hs e^{-\lam}\,}
{\,\rm{Vol}[\hs\rm{SL}(2,\bR)]\,}
\prod_{j=1}^{N} \sum_{\{\sgn_j^{}\}}\!
\int \!\! \td y_j^{}  \,
\bla V_{\op}(\zeta_1^{} ,k_1^{}, \sgn_1^{}\hs n_1^{})
\cdots
V_{\op} (\zeta_N^{} ,k_N^{}, \sgn_N^{}\hs n_N^{} ) \bra \hs,~~
\label{eq:open-amp-1}
\end{align}
%%%%
where $\,\lam\,$ denotes the vacuum expectation value of the dilaton 
and in the second line we have further expressed the amplitude 
in terms of  open-string vertex operators 
from the RHS of \eqrefe{eq:Voppm}.
Here we choose each external state to be $\ZZ$ even (odd),\ 
corresponding to its vertex operator 
$V_{\op}^\pm (\zeta_j ,k_j^{}, n_j^{})$ being $\ZZ$ even (odd)
as indicated by its superscript $+\,(-)\hs$. 
[An external state of the amplitude
$\A^{(N)}_{\op}(\zeta, k,n)$ 
can also be chosen as non-eigenstate of $\ZZ$ parity and
thus its corresponding vertex operator is 
$V_{\op}^{} (\zeta_j ,k_j^{}, \pm n_j^{})$.
Such cases can be studied by our formulation as well,
although our present study will focus on 
the $N$-point amplitude like Eq.\eqref{eq:open-amp-1}.]
In the second line of Eq.\eqref{eq:open-amp-1}, the overall coefficient
contains the sign factor $(-)^{\omega}$, 
where $\hs\omega \!=\!0\hs$
for all vertex operators being $\ZZ$-even and $\hs\omega \neq 0\hs$
for $N_-$ number of vertex operators being $\ZZ$-odd.
These $N_-$ number of $\ZZ$-odd vertex operators will contribute a nontrivial sign factor $(-)^{\omega}$, where  
$\hs\omega \!=\!\sum_{j=1}^{N_-}[1\hsmx +\hsmx H(\sgn_j^{})]\hs$
and the Heaviside step function $H(+)\!=\!1$ and  $H(-)\!=\!0\hs$.
In Eq.\eqref{eq:open-amp-1}, we use $\NH$ to denote the number of 
the external KK excitation states 
(with $n_j^{}\!\!>\!0$) and the factor 
$\,2^{-\NH /2}$ 
arises from the overall coefficient $1/\!\sqrt{2}$
of \eqrefe{eq:Voppm}.\ 
So the number of possible external zero-mode states
equals $\hs (N\!-\!\NH)\hs$. 
In the second line of Eq.\eqref{eq:open-amp-1}, 
the summation over $\hs\{\sgn_j^{}\}\hs$ means to sum up all allowed
sign-combinations of KK numbers of  external states
which obey the condition \eqref{KK-num-sum}.   
Integrating over $\hs y_j^{}\hs$,\, we can reexpress the $N$-point 
KK open string amplitude as follows:
%%%%
\begin{equation}
\label{eq:open-amp-2}
\hspace*{-2mm} 
\A_{\op}^{(N)}(\zeta , k) \,=\, 
\(\hsm\frac{1}{2}\hsm\)^{\!\!\!\hsm\frac{\NH}{2}}\!\!
\sum_{\{ \sgn_j^{}\!\}} \sum_{\al \in S_{\!N\hsm-\hsm1}} \!\!
\A^{(N)}_{\op} \big[ \zeta_j^{} ,\hs \hk_j^{} \hs
\big|\{1,\al(2 \cdots N)\} \big]\!\times\! 
T \!\[\!1,\al(2 \cdots N)\!\] \!,~
\end{equation}
%%%
where 
$\,\hk_j^{}\!=\!(k_j^\mu,\,\sgn_j^{}\hsm\times\hsm M_{n_j}^{}\hsm )\hs$
is the 26d momentum and the notation 
$\,T[1,\al(2\cdots N)]\!=\!\Tr(T^1T^{\al(2)}$ $\cdots T^{\al(N)})\,$
denotes the Chan-Paton factor.\ 
The partial amplitudes on the RHS of \eqrefe{eq:open-amp-2} 
are not fully independent, among which only 
$(N\hsm\!-\!3)!$ partial amplitudes are independent\,\cite{BCJ}\cite{BCJ-Rev}. 
We further express the color-ordered partial amplitude
on the RHS of \eqrefe{eq:open-amp-2} as follows:
%\\[-5mm]
%%%%%%
\begin{equation}
\A_{\op}^{(N)}\big[\zeta_j^{} , \, \hk_j \hs
\big| \{1,\al(2 \cdots N)\} \big] 
=\ii\hs\hs g_{\op}^N \hs C_{D_2} (2\pi)^{26} \hs 
\delta^{(26)}\!\Big(\!\sum\nolimits_j\!\hk_j^{}\hsm\Big)\hsx
\xoverline{\A}^{\hs (N)}_{\op}\big[\zeta_j^{} ,\hs \hk_j^{} \hs
\big|\{1,\al(2 \cdots N)\} \big]  ,~
\end{equation}
%%%%
where the coefficient $C_{\!D_2}^{}$ is given by 
$\,C_{D_2}\hsm\!=\! e^{-\lam}(C_{D_2}^g C_{D_2}^X)\hs$.
The constants $C_{D_2}^g$ and $C_{D_2}^X$ are given 
by the path integral of the zero mode of the $bc$ ghost 
and $X$ scalar field on the string worldsheet, respectively.
We will further compute the reduced amplitude
$\xoverline{\A}^{\hs (N)}_{\op}$ explicitly for the 4-point
scattering in section\,\ref{sec:3}.

\vspace*{1mm}

The closed-string amplitude can be derived from the product of
two open-string amplitudes\,\cite{KLT}.
Using the open-string amplitude \eqref{eq:open-amp-2}, 
we construct the $N$-point massive KK closed-string amplitude 
at tree level:
%%%%
\beqs 
\label{eq:close-amp}
\begin{align}
\label{eq:close-amp-1}
{\A}_{\cl}^{(N)}(\zeta, k) \,=&~ \ii\hs (2\pi)^{26} \delta^{(26)}\!\Big(\!\sum\nolimits_j\!\hk_j^{}\hsm\Big) \xoverline{\A}_{\cl}^{(N)}(\zeta, k) \,,
\\[1mm]
\xoverline{\A}_{\cl}^{(N)}(\zeta, k)
\,=&~ g_{\cl}^N \hs C_{S_2}\!
\(\!\frac{1}{\,2\,}\hsmx\)^{\!\hsm\!\NH\hsmx/2} \!\!\(\hsm\!-\frac{\pi\alp}{2}\hsm\)^{\!\!\hsmx N\hsm-3}
\!\!\sum_{\{a_{\hsm j},\hs b_{\hsm j}\}}  
\sum_{\{\sgn_j^{}\hsm\}} \sum_{\{\al,\be\} \in S_{\!N\hsm-\hsm3}} 
\hspace*{-2mm}\big\{ \vrhohat_{ab}^{}\, 
{\mS_{\rm{ST}}^{\alp}[\al | \be]_{\hk_1}^{}}
\nn\\[0mm] 
& \hspace*{8mm}
\times \xoverline{\A}^{\hs (N)}_{\op} 
\big[\zeta_j^{a_j} , \hk_j^{} \big| 
\{1,\al(2 \cdots N\!-\!2),N\!-\! 1,N \} \big]
\nn\\[1mm]
& \hspace*{8mm}
\times \xoverline{\A}^{\hs (N)}_{\op} \big[\zeta_j^{b_j} , \hk_j^{} \big| 
\{N\!-\! 1, N, \be(2 \cdots N\!-\! 2)  ,1 \} \big] \big\} 
\big|_{\alp \to \hs \alp\hsm/4} \,,
\label{eq:close-amp-2}
\end{align}
\eeqs 
%%%  
where 
$\,C_{S_2} \!\!=\! e^{-2\lam}(C_{S_2}^g C_{S_2}^X)\! 
   \!=\!{32\pi}/(\alpha^{\prime\hs 3} g_{\cl}^2)\hs$.
In the above Eq.\eqref{eq:close-amp-2},
the polarization tensor $\,\zeta_{\mn}^{}\,$  
of closed strings is expressed as a sum of the products of  
polarization vectors of two open strings:
\begin{equation}
\zeta_{\mn}^{} \,=\,  \varrho_{ab}^{}\, 
\zeta_{\mu}^a \hs \zeta_{\nu}^b \,, 
\end{equation}
%%%
where the coefficient 
$\,\varrho_{ab}^{} \hsm\in\hsm \mathbb{R}\,$. 
In \eqrefe{eq:close-amp-2}, the coefficient $\hs\vrhohat_{a_j^{}b_j^{}}\hs$ is defined as the product of 
$\,\varrho_{ab}^{}\,$ for all external graviton states:
%\\[-8mm]
%
\begin{equation}
\vrhohat_{ab}^{} =\, \prod_{j=1}^N
\varrho_{a_j^{}b_j^{}}^{} \,.
\end{equation}
%
%\\[-5mm]
The string momentum kernel 
$\,\mS_{\rm{ST}}^{\alp}[\al | \be]_{\hk_1}^{}$ 
connects the two open-string amplitudes and takes the following
explicit form\,\cite{Sondergaard:2011iv}:
%%%
\begin{equation}
\label{eq:Kernel-ST}
\hspace*{-2mm}
\mS^{\alp}_{\rm{ST}}[\al_1\cdots \al_j | \be_1\cdots \be_j]_{\hk_1}^{}
= \hs
\(\!\frac{\,\pi\alp\,}{2}\)^{\hsm\!\!-j}\!\prod_{i=1}^j
\sin \! \bigg[\pi\alp\bigg(\! 
\hk_1^{} \!\hsm\cdot\hsm\hk_{\al_i^{}}^{}  
\!+\! \sum_{\ell>i}^j \Theta(\al_i^{} , \al_\ell^{}) \hs
\hk_{\al_i^{}}^{} \!\!\hsm\cdot\hsm 
\hk_{\al_\ell^{}}^{} \hsm\bigg)\bigg] ,~~
\end{equation}
%%%%
where the step function 
$\,\Theta(\al_i^{},\al_\ell^{})\!=\!1\,$ 
when the order of $\,(\al_i^{},\al_\ell^{})\,$ 
is opposite in $\(\{\al_i^{}\},\{\be_i^{}\}\)$, while $\,\Theta(\al_i^{},\al_\ell^{})\!=0\,$ when the order of $(\al_i^{},\hs\al_\ell^{})$\, 
is the same as in $\(\{\al_i^{}\},\{\be_i\}\)$\,.

\subsection{\hspace*{-3mm}Low Energy Scattering Amplitudes of 
KK Gauge Bosons and Gravitons}
\label{sec:2.4}

In this subsection, we will derive the extended massive KLT-like 
relations of KK states for the low energy field theory.
For this purpose,  we take the limit of zero Regge slop 
$\,\alp \!\hsm\ito\hsm 0\,$ for the closed-string amplitude in 
Eqs.\eqref{eq:close-amp} and \eqref{eq:Kernel-ST}. 
Then, the open/closed-string amplitudes and the string momentum kernel 
will reduce to their corresponding field-theory expressions:
%%%%
\begin{equation}
\A_{\op} \!\to \TT \,,\quad
\A_{\cl} \!\to \M \,, \quad
\mS_{\rm{ST}}^{\alp}  \!\to  \mS_{\rm{FT}} \,.
\end{equation}
%%%%%
With these, we can derive the following low energy $N$-point 
graviton scattering amplitude:
%%%
\begin{align}
\label{eq:Amp-N-Gr}
\M^{(N)} (\zeta, k)
\,=\, &  \hspace{1.5mm}
\(\!\frac{\,\ka\,}{\,4\,}\!\)^{\!\!N-2}\!
2^{-{\NH}/{2}}\hs(-1)^{N+1}\!
\sum_{\{a_{\hsm j},\hs b_{\hsm j}\}} 
\sum_{\{\sgn_j^{}\}}\! \sum_{\{\al,\be\}\in S_{\!N\hsm-\hsm3}}\!\! 
\hspace*{-2mm}\big\{ \vrhohat_{ab}^{} \hsx
\mS_{\rm{FT}}[\al |\be]_{\hk_1}^{}
\nn\\
& \hspace*{6mm}
\times\TT^{(N)} \!\big[\zeta^{a_j^{}}_j, \, \hk_j^{} \big|
\{1,\al(2 \cdots N\!-\!2),N\!-\! 1,N \}  \big]
\nn\\[1mm]
&\hspace*{6mm}
\times\TT^{(N)} \!\big[\zeta^{b_j^{}}_j, \, \hk_j^{} \big|
\{N\!-\! 1, N, \be(2 \cdots N\!-\! 2),1 \} \big]\hs \big\}  \hs ,
\end{align}
%%%
where the gravitational coupling $\,\ka\,$ and the closed-string 
coupling $\,g_{\cl}^{}\,$ are connected by the relation 
$\,\ka\hsm =\hsm 2\pi\alp g_{\cl}^{}\hs.\hs$ 
In the above, $\hs\mS_{\rm{FT}}[\al |\be]_{\hk_1}^{}$ is the
momentum kernel in the field theory limit and 
takes the following form\,\cite{Bern:1998sv}\cite{Sondergaard:2011iv}:
%%%
\begin{equation}
\mS_{\rm{FT}}[\al_1^{}\cdots\al_j^{}|\be_1^{}\cdots\be_j^{}]_{\hk_1}^{}
=~ \prod_{i=1}^j\!
\bigg(\! 2 \hs \hk_1^{}\!\hsm\cdot\! \hk_{\al_i^{}}^{}
\!+\hs 2
\sum_{\ell>i}^j \Theta(\al_i^{} , \al_\ell^{})\hs \hk_{\al_i^{}}\!\!\cdot\hsm \hk_{\al_\ell^{}}^{}\!\bigg) .
\end{equation}
%%%

\section{\hspace*{-3mm}Massive KK Open String Amplitudes and Field Theory Limit}
\label{sec:3}

In this section, we compute explicitly the four-point 
color-ordered elastic and inelastic scattering amplitudes 
of KK open strings and derive the corresponding KK gauge boson
scattering amplitudes in the low energy field-theory limit.

\vspace*{1mm}

Thus, we compute the four-point color-ordered partial amplitudes of 
KK open-string scattering with three fixed points 
$(y_1^{},\hs y_2^{},\hs y_3^{}) = (0,1,\infty)\hs$:
%%%
\beqs
\label{eq:Ab1234}
\begin{align}
g_{\op}^4 \hs C_{D_2}\hs\xoverline{\A}_{\op}^{}\hs [1243]  \,&=\, \frac{g_{\op}^2}{\,2\al^{\prime\hs 2}\,} 
\int_{1}^\infty \!\! F(y_4^{}) \, \td y_4^{} \,,
\\[1mm]
g_{\op}^4 \hs C_{D_2}\hs
\xoverline{\A}_{\op}^{}\hs [1423] \,&=\, 
\frac{g_{\op}^2}{\,2\al^{\prime\hs 2}\,} 
\int_{0}^1 \!F(y_4^{}) \,\td y_4^{} \,,
\\[1mm]
g_{\op}^4 \hs C_{D_2}\hs
\xoverline{\A}_{\op}^{}[4123]  \,&=\,  
\frac{g_{\op}^2}{\,2\al^{\prime\hs 2}\,}  
\int_{-\infty}^0 \!\!F(y_4^{}) \,\td y_4^{} \,,
\end{align}
\eeqs
%%%
where the superscript ``$(4)$'' in each amplitude 
$\,\xoverline{\A}_{\op}^{}\,$ is not displayed for simplicity and the 
relation $\,g_{\op}^2 C_{D_2} \!=\! 1/(2\al'^{\hs2})\,$  is imposed.
In the above, the function $F(y_4^{})$  is defined as:
\vspace*{.5mm}
%%%
\beqs 
\begin{align}
F(y_4^{}) \,&=\,
f(y_4^{}) \hs |y_4^{}|^{-2\alp (\ehmd{1}{4}) }\hs 
|1\!-\!y_4^{}|^{-2\alp (\ehmd{1}{3})} \,,
\\[2mm]
f(y_4)  \,&= \,
\lim_{y_1^{}\to 0}\lim_{y_2^{}\to 1}\lim_{y_3^{}\to\infty}
y_3^2 \hs \prod_{i\neq j}
\exp\!\hsm\[\!\! \frac{\,2\alp (\zeta_i^{} \!\cdot\zeta_j^{})\,}
{(y_i^{}\!- y_j)^2} - 
\frac{\,2\alp (\zeta_i^{}\!\cdot\hk_j^{})\,}
{y_i^{}\!- y_j^{}}\!\!\] \! ,
\end{align}
\eeqs 
%%%
where we only need to expand up to the linear term of each external
polarization vector. 
The other three color-ordered partial amplitudes 
([1342], [1432], [4132])
can be obtained by exchanging $(2,3)$.\ 
In this section, we focus on the two partial amplitudes with the 
color-ordering [1234] and [1243] for the sake of our later 
double-copy construction closed-string amplitudes.

\subsection{\hspace*{-3mm}Elastic Amplitudes of KK Gauge Bosons from KK Open Strings}
\label{sec:3.1}

\begin{figure}[t]
\centering
\hspace*{-6mm} 
\includegraphics[width=14.7cm]{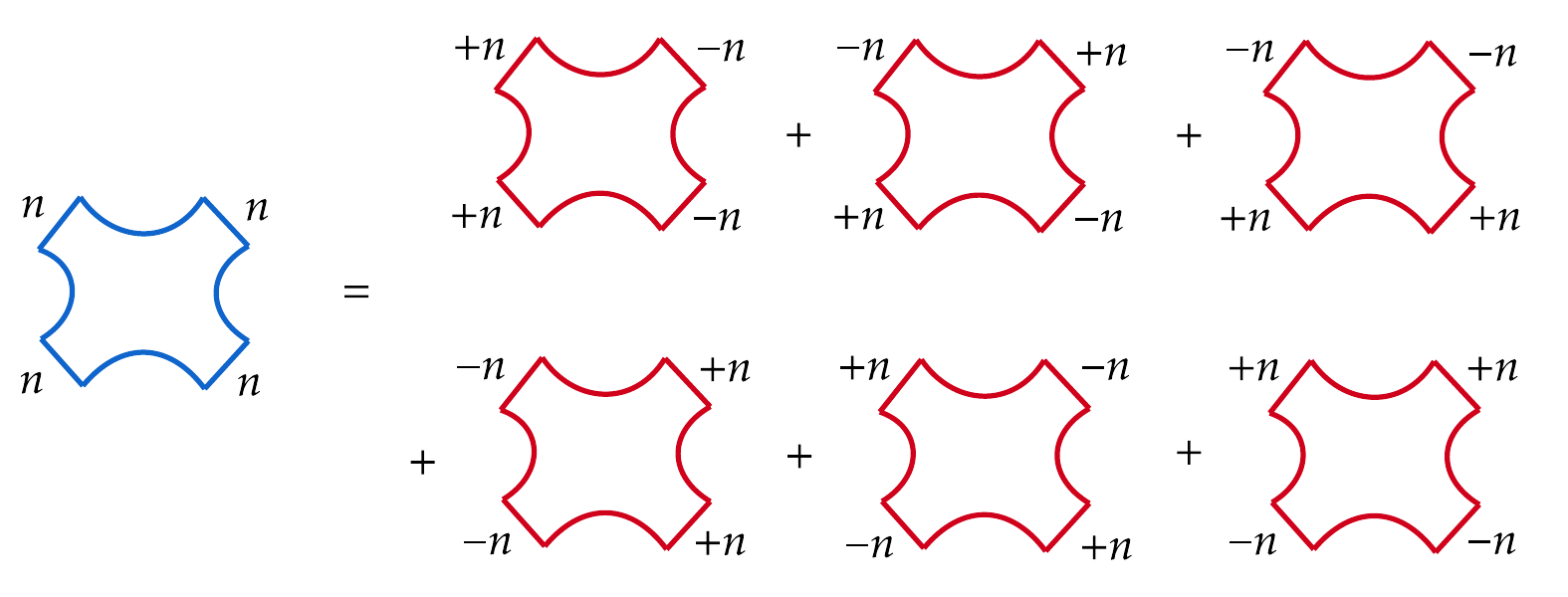}
\vspace*{-1.5mm}
\caption{\small{Elastic scattering amplitude of $(n,n)\ito (n,n)$ 
for massive KK open strings with $\ZZ$-even parity (blue color) can be decomposed to a sum of six sub-amplitudes of massive KK open strings 
(red color) under the $\SS^1$ compactification of 26d 
for bosonic strings.}}
\label{fig:1}
\end{figure}

In this subsection, we study the four-point elastic KK scattering
process $(n,n)\hsm\ito (n,n)$
with all external states being $\ZZ$-even.\ 
We observe that \eqrefe{KK-num-sum} allows six different 
combinations of the KK numbers for the external states of 
the four-point sub-amplitudes
(which are originally defined under the $\SS^1$ compactification):
%%%
\begin{equation}
\begin{aligned}
\label{eq:KK-num-4-point}
&\{+n,\hs+n,\hs-n,\hs-n\} \hs , \hspace{3mm}
\{+n,\hs-n,\hs+n,\hs-n\} \hs , \hspace{3mm}
\{+n,\hs-n,\hs-n,\hs+n\} \hs ,
\\
&\{-n,\hs-n,\hs+n,\hs+n\} \hs , \hspace{3mm}
\{-n,\hs+n,\hs-n,\hs+n\} \hs , \hspace{3mm}
\{-n,\hs+n,\hs+n,\hs-n\} \hs ,
\end{aligned}
\end{equation}
%%%
with $n \in \Z^+$.\
This means that the four-point elastic KK open-string amplitude
$\A_{\op}^{(4)}$ can be decomposed into 
a sum of the six sub-amplitudes, as presented in Fig.\,\ref{fig:1}. 
We note that the compactification under $\SS^1$ respects the
$\ZZ$ parity, so among the above six combinations of KK numbers
only three are independent, where the three combinations in the
first row of \eqrefe{eq:KK-num-4-point} are connected to the other 
three corresponding combinations in the second row
by $\ZZ$ parity transformation. From the above, our key insight
is that even though the external states on the LHS of 
\eqrefe{eq:KK-num-4-point} are all $\ZZ$-even, the external states
on the RHS contain two $\ZZ$-even states and two $\ZZ$-odd states
such that the condition \eqref{KK-num-sum} is obeyed. 
This is because our string compactification of 26d is 
under $\SS^1$ with the periodic boundary condition \eqref{eq:Compact}
(without having $\ZZ$ orbifold). Hence, even for a scattering amplitude
with $\ZZ$-even external states \eqref{eq:Voppm}, it contains
the combination of individual amplitudes whose external states
include both positive and negative KK numbers, as shown in
\eqrefe{eq:KK-num-4-point} for the case of four-point amplitudes.

%%%%%
%%%%%%

\vspace*{1mm}

Then, we compute the color-ordered partial amplitudes 
from \eqrefe{eq:Ab1234} under the field theory limit 
$\hs \alp\ito 0\hs$. We present their explicit expressions 
in Appendix\,\ref{app:B}, where the open-string coupling 
$\hs g_{\op}^{}\hs$ is replaced by the gauge coupling $\hs g\hs$ 
of the YM theory.
Thus, substituting the momenta in \eqrefe{eq:Momenta}
and the longitudinal polarization vectors 
$\hs\zeta^\mu_{j,L}\hs$ in \eqrefe{eq:Pol}
into \eqrefe{eq:T1234-1243-FT}, 
we derive the sub-amplitudes with color ordering $[1234]$\hs:
%%%%%
\beqs
\label{eq:T1234-KK}
\begin{align}
\hspace*{-2mm}
\TT[1^{\pm n}_L \hs 2^{\pm n}_L \hs 3^{\mp n}_L \hs4^{\mp n}_L] 
\,&= \, g^2 \hsm \frac{7\!+\hsm\ctt}{\,(1\!+\!\ct)\,}  \,,
\\[1mm]
\hspace*{-2mm}
\TT[1^{\pm n}_L \hs 2^{\mp n}_L \hs 3^{\pm n}_L \hs4^{\mp n}_L] 
\,&= \,   g^2 \hsm
\frac{~[\hs 7 \bs^2\!-\!24\bs\!+\!48 \!-\! 16 (\bs\!-\!4) \ct \!+\! (\bs\!+\!4)^2 \ctt \hs]
\sec^2 \!\frac{\theta}{2}~}
{2\bs(\bs\!-\! 4)} \,,
\\[1mm]
\hspace*{-2mm}
\TT[1^{\pm n}_L \hs 2^{\mp n}_L \hs 3^{\mp n}_L \hs4^{\pm n}_L] 
\,&= \,   g^2 \hsm
\frac{~7 \bs^2\!-\!24\bs\!+\!48 \!+\! 16 (\bs\!-\!4) \ct \!+\! (\bs\!+\!4)^2 \ctt ~}
{\bs \hs[\hs \bs \!+\! 4 \!+\! (\bs\!-\!4) \ct \hs]} \,,
\end{align}
\eeqs
%%%
and the sub-amplitudes with color ordering $[1243]$\hs:
%%%%
\beqs
\label{eq:T1243-KK}
\begin{align}
\hspace*{-2mm}
\TT[1^{\pm n}_L \hs 2^{\pm n}_L \hs 4^{\mp n}_L \hs3^{\mp n}_L] 
\,&=\, g^2 \hsm
\frac{\,(7 \!+\! \ctt)\hsm \csc ^2\! \frac{\theta}{2}~}{2} \,,
\\[1mm]
\TT[1^{\pm n}_L \hs 2^{\mp n}_L \hs 4^{\pm n}_L \hs 3^{\mp n}_L] 
\,&=\, g^2 \hsm
\frac{~7 \bs^2 \!-\! 24 \bs \!+\! 48 \!-\! 16 (\bs\!-\!4) \ct \!+\!  (\bs\!+\!4)^2 \ctt~}
{\bs \hs[\hs \bs \!+\! 4 \!-\! ( \bs\!-\! 4) \ct \hs]} \,,
\\[1mm]
\TT[1^{\pm n}_L \hs 2^{\mp n}_L \hs 4^{\mp n}_L \hs 3^{\pm n}_L] 
\,&=\, g^2 \hsm 
\frac{~[\hs 7 \bs^2 \!-\! 24 \bs \!+\! 48 \!+\! 16 (\bs\!-\!4) \ct \!+\!  (\bs\!+\!4)^2 \ctt \hs]\csc^2\hsm \frac{\theta}{2}~}
{2\bs (\bs\!-\! 4)} \,,
\end{align}
\eeqs
%%%%
where we have defined $\,\bs =s/M_n^2$\,.
With the above, we sum up the four-point amplitudes 
in \eqrefe{eq:T1234-KK} and  \eqrefe{eq:T1243-KK}, and 
derive the following color-ordered full elastic amplitudes 
with all external states being $\ZZ\hsm$ even: 
%%%
\beqs
\label{eq:T1234-1243nnnn}
\begin{align}
\label{eq:T1234}
\TT[1^n_L \hs 2^n_L \hs 3^n_L \hs 4^n_L ] 
&\,=\,  \fr{1}{2}
(\TT[1^{+n}_L \hs 2^{+n}_L \hs 3^{-n}_L \hs 4^{-n}_L]
+ \TT[1^{+n}_L \hs 2^{-n}_L \hs 3^{+n}_L \hs 4^{-n}_L]
+\TT[1^{+n}_L \hs 2^{-n}_L \hs 3^{-n}_L \hs 4^{+n}_L])
\nn\\[1.5mm]
&\,=\, g^2
\frac{~(P_0 \!+\! P_1 \ct \!+\! P_2 \ctt \!+\! P_3 \cttt )
\sec^2\!\fr{\theta}{2}~}
{16 \bs (\bs\!-\! 4)  \hs[\hs \bs\!+\! 4 \!+\! (\bs\!-\! 4) \ct \hs]}  \,,
\\[2mm]
\label{eq:T1243}
\TT[1^n_L \hs 2^n_L \hs 4^n_L \hs 3^n_L ] 
&\,=\,  \fr{1}{2}
(\TT[1^{+n}_L \hs 2^{+n}_L \hs 4^{-n}_L \hs 3^{-n}_L]
+ \TT[1^{+n}_L \hs 2^{-n}_L \hs 4^{+n}_L \hs 3^{-n}_L]
+\TT[1^{+n}_L \hs 2^{-n}_L \hs 4^{-n}_L \hs 3^{+n}_L])
\nn\\[1.5mm]
&\,=\, g^2
\frac{~(P_0 \!-\! P_1 \ct \!+\! P_2 \ctt \!-\! P_3 \cttt )
\csc^2\!\fr{\theta}{2}~}
{16 \bs (\bs\!-\! 4)  \hs[\hs \bs\!+\! 4 \!-\! (\bs\!-\! 4) \ct \hs]}   \,,
\end{align}
\eeqs
%%%
where the polynomials $\{P_j\}$ are given by
%%%
\begin{equation}
\begin{alignedat}{3}
\label{eq:Tnnnn-Pn}
P_0^{} &=42 \bs^3\!-\hsm 96\hs\bs^2\!-\hsm 32\hs\bs\hs, \quad~~
&& 
P_1^{} =45\hs\bs^3\!-\hsm 320\hs\bs^2
\!+\hsm 528\hs\bs\hsm +\hsm 128 \hs,
\\
P_2^{} & =2\hs\bs\hs (3\hs\bs^2\!+\hsm 16\hs\bs\hsm +\hsm 16) \hs, \quad~~
&&
P_3^{} =3\hs\bs^3\!-\hsm 16\hs\bs\!-\hsm 128 \hs.
\end{alignedat}
\end{equation}
%%%%%
The above elastic KK gauge boson amplitudes 
\eqref{eq:T1234}-\eqref{eq:T1243} 
are derived from the KK open-string amplitudes \eqref{eq:Ab1234}.
We inspect these color-ordered KK gauge boson amplitudes
\eqref{eq:T1234-1243nnnn}
based on the KK open-string calculation and 
find that they can be expressed in the following forms:
%%%
\begin{align}
\label{eq:T1234-1243-Kj}
\TT[1^n_L \hs 2^n_L \hs 3^n_L \hs 4^n_L ]  \,=\, g^2 (-\KK_s^{\rm{el}} + \KK_t^{\rm{el}})  \,, \qquad
\TT[1^n_L \hs 2^n_L \hs 4^n_L \hs 3^n_L ]  \,=\, g^2 (\KK_s^{\rm{el}} - \KK_u^{\rm{el}}) \,,
\end{align}
%%%
where the kinematic factors 
$\{\KK_j^{\rm{el}}\}$ are summarized in \eqrefe{eq:K-KT-exact}
of Appendix\,\ref{app:B}.
Impressively, we find that the expressions in Eq.\eqref{eq:T1234-1243-Kj} 
fully agree with the corresponding KK gauge boson amplitudes 
of Ref.\,\cite{Hang:2021fmp} 
which were computed independently within the compactified 
5d KK YM gauge field theory.

\subsection{\hspace*{-3mm}Inelastic Amplitudes of KK Gauge Bosons from KK Open Strings}
\label{sec:3.2}

%%%%%
\begin{figure}[t]
\centering
\includegraphics[width=16cm]{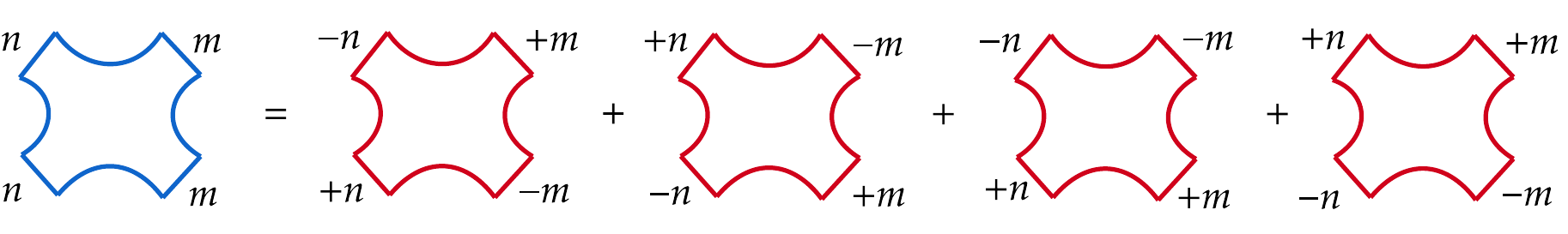}
\vspace*{-1.5mm}
\caption{\small{Inelastic scattering amplitude of 
$\hs (n,n)\ito (m,m)\hs$  
for massive KK open strings with $\ZZ$-even parity (blue color) can be decomposed to a sum of four sub-amplitudes of massive KK open strings 
(red color) under the $\SS^1$ compactification of 26d for 
bosonic strings.}}
\label{fig:2}
\end{figure}
%%%%%%
	
Next, we analyze the four-point color-ordered partial amplitude
for the inelastic channel $(n,\hs n)\ito (m,\hs m)$
with all external states being $\ZZ$-even. 
We inspect \eqrefe{KK-num-sum} and find that 
different from the elastic channel, there are only four allowed
combinations of the external KK states 
(as originally defined under the $\SS^1$ compactification):
%%%
\begin{equation}
\begin{aligned}
\label{eq:KK-num-4-point-2}
& \hspace{3mm}
\{+n,\hs-n,\hs+m,\hs-m\} \hs , \hspace{2mm}
\{+n,\hs-n,\hs-m,\hs+m\} \hs ,
\\
& \hspace{3mm}
\{-n,\hs+n,\hs-m,\hs+m\} \hs , \hspace{2mm}
\{-n,\hs+n,\hs+m,\hs-m\} \hs .
\end{aligned}
\end{equation}
%%%
Thus, the four-point inelastic KK open-string amplitude
is given by a sum of the four sub-amplitudes,
as illustrated in Fig.\,\ref{fig:2}. 

\vspace*{1mm}  

Then, we derive the four-point 
inelastic scattering amplitudes of longitudinal gauge bosons with 
color-ordering $[1234]$\hs:
%%%%%
\beqs
\label{eq:Tnnmm-L1234} 
\begin{align}
\TT[1^{\pm n}_L \hs 2^{\mp n}_L \hs 3^{\pm m}_L \hs4^{\mp m}_L] 
&=  g^2 \hs
\frac{~(7\bs^2\!-\!12\hs\rrp\bs \hsm +\hsm 48\hsx r^2)\hsm
-64\hs r\hs\qb\qb'\ct \!+\!16\hs\qb^2\qqbp\ctt~}   
{2\hs\bs\left[(\bs\hsmx -\hsmx 4r)\hsmx +\hsmx 
4\qb\qb'\ct \right]},
\\[2mm]
\TT[1^{\pm n}_L \hs 2^{\mp n}_L \hs 3^{\mp m}_L \hs4^{\pm m}_L] 
&=  g^2 \hs
\frac{~(7\bs^2\!-\!12\hs\rrp\bs \hsm +\hsm 48\hsx r^2)\hsm
+64\hs r\hs\qb\qb'\ct \!+\!16\hs\qb^2\qqbp\ctt~}   
{2\hs\bs\left[(\bs\hsmx +\hsmx 4r)\hsmx +\hsmx 
4\qb\qb'\ct \right]},
\end{align}
\eeqs
%%%
where we have used the notations,
\begin{align}
& q = (E^2\hsm\!-\!M_n^2)^{\hsmx\frac{1}{2}},~~~
q'\!= (E^2\hsm\!-\!M_m^2)^{\hsmx\frac{1}{2}},~~~
r = {M_m}/\hsmx{M_n}\hs , ~~~		
r_{\!+}^2 = 1 \!+\hsm r^2\hs ,
\nn\\
\label{eq:qsr}
& s = 4E^2,~~~ \sz \!= 4q^2,~~~
\bs = s/M_n^2 = \bsz \hsm +\hsm 4\,,~~~
\bsz = \sz/\hsmx M_n^2 \hs ,
\\
& \bar{q}^2\!=\hsm q^2\hsmx /\hsmx M_n^2=\fr{1}{4}\bs -\!1 \hs,~~~
\bar{q}^{\prime 2} \!=\hsm q^{\prime 2}\!/\hsmx M_n^2
\!=\! \fr{1}{4}\bs \hsm -\hsm r^2 ,~~~
\qb^2\qqbp \hsm =\hsm
\fr{1}{\,16\,}(\bs \!-\!4) (\bs\hsmx -\hsm 4\hs r^2)\hs .
\nn
\end{align}
For the color-ordering $[1243]$\hs, we derive the four-point 
inelastic scattering amplitudes of longitudinal gauge bosons
as follows:
%%%%
\beqs
\label{eq:Tnnmm-L1243}
\begin{align}
\TT[1^{\pm n}_L \hs 2^{\mp n}_L \hs 4^{\pm m}_L \hs3^{\mp m}_L] 
&=  g^2 \hs
\frac{~(7\bs^2\!-\!12\hs\rrp\bs \hsm +\hsm 48\hsx r^2)\hsm
-64\hs r\hs\qb\qb'\ct \!+\!16\hs\qb^2\qqbp\ctt~}   
{2\hs\bs\left[(\bs\hsmx +\hsmx 4r)\hsmx -\hsmx 
4\qb\qb'\ct \right]},
\\[1mm]
\TT[1^{\pm n}_L \hs 2^{\mp n}_L \hs 4^{\mp m}_L \hs3^{\pm m}_L] 
&=  g^2 \hs
\frac{~(7\bs^2\!-\!12\hs\rrp\bs \hsm +\hsm 48\hsx r^2)\hsm
+64\hs r\hs\qb\qb'\ct \!+\!16\hs\qb^2\qqbp\ctt~}  
{2\hs\bs\left[(\bs\hsmx -\hsmx 4r)\hsmx -\hsmx 
4\qb\qb'\ct \right]}.
\end{align}
\eeqs
%%%%

%%%%%
\begin{figure}[t]
\centering
\includegraphics[width=9.5cm]{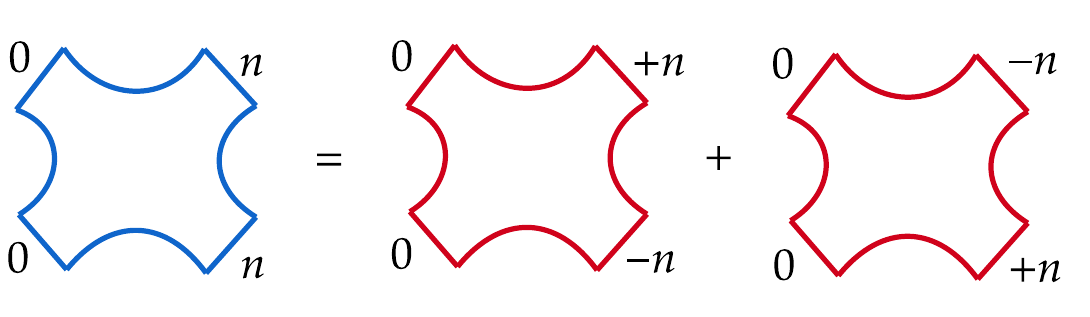}
\vspace*{-3mm} 
\caption{\small{Inelastic scattering amplitudes of 
$\hs (0,0)\ito (n,n)\hs$   
for massive KK open string with $\ZZ$-even parity (blue color) 
can be decomposed into a sum of two sub-amplitudes of 
massive KK open strings (red color) under the 
$\SS^1$ compactification of 26d for bosonic strings.}}
\label{fig:3}
\end{figure}
%%%%%%

Summing up the four-point inelastic amplitudes \eqref{eq:Tnnmm-L1234} 
and \eqref{eq:Tnnmm-L1243}, we derive the following color-ordered 
full amplitudes with all external states being $\ZZ\hsm$ even: 
%%%
\beqs
\label{eq:T1234-1243nnmm}
\begin{align}
\label{eq:T1234nnmm}
\TT[1^n_L \hs 2^n_L \hs 3^m_L \hs 4^m_L ] 
&\,=\,  \fr{1}{2}
(\TT[1^{+n}_L \hs 2^{-n}_L \hs 3^{+m}_L \hs 4^{-m}_L]
+\TT[1^{+n}_L \hs 2^{-n}_L \hs 3^{-m}_L \hs 4^{+m}_L])
\nn\\[1.5mm]
&\,=\, g^2
\frac{~P_0^{}\!+\! P_1^{}\ct \!+\! P_2^{}\ctt \!+\! P_3^{}\cttt ~}
{~\bs\hs(\hs 3\hs\bs^2 \!-\! 4\hs\bs\hs r_+^2 \!-\! 16\hs r^2
\!+\! 16\hs\bs\hs\qb\hs\qb'\ct \hsm +\! 
16\hs\qb^2\hs\qb'^2 \ctt \hs)~}  \,,
\\[2mm]
\label{eq:T1243nnmm}
\TT[1^n_L \hs 2^n_L \hs 4^m_L \hs 3^m_L ] 
&\,=\,  \fr{1}{2}
(\TT[1^{+n}_L \hs 2^{-n}_L \hs 4^{+m}_L \hs 3^{-m}_L]
+\TT[1^{+n}_L \hs 2^{-n}_L \hs 4^{-m}_L \hs 3^{+m}_L])
\nn\\[1.5mm]
&\,=\, g^2
\frac{~P_0^{} \!-\! P_1^{}\ct \!+\! P_2^{}\ctt \!-\! P_3^{}\cttt~}
{~\bs\hs(\hs 3\hs\bs^2 \!-\! 4\hs\bs\hs r_+^2 \!-\! 16\hs r^2\!-\! 
16\hs\bs\hs\qb\hs\qb'\ct\hsm +\! 16\hs\qb^2 \qb'^2 \ctt \hs)}  \,,
\end{align}
\eeqs
%%%
where the polynomials $\{P_j\}$ take the forms:
%%%
\begin{equation}
\begin{alignedat}{3}
P_0^{} &=\bs\hs (7\bs^2\!-\hsm 12\hs\bs\hs r_+^2\hsm +\hsm 48\hs r^2 )\hs, \quad~~
&& 
P_1^{} =
2\hs\qb\hs\qb'(15\hs\bs^2\!-\hsm 20\hs\bs\hs r_+^2 
\hsm -\hsm 16\hs r^2) \hs,
\\
P_2^{} & = \bs\hs (\bs^2 \!+\hsm 4\hs\bs\hs r_+^2\!+\hsm 16r^2) \hs, 
\quad~~
&&
P_3^{} =2\hs\qb\hs\qb' 
(\bs^2\!+\hsm 4\hs\bs\hs r_+^2\hsm +\hsm 16\hs r^2) \hs.
\end{alignedat}
\end{equation}
%%%%%
Inspecting the string-based KK gauge boson amplitudes
\eqref{eq:T1234-1243nnmm}, we can re-express them 
in the following forms:
%%%
\begin{align}
%\label{eq:T1234-1243-Kj}
\TT[1^n_L \hs 2^n_L \hs 3^m_L \hs 4^m_L ]\hs =\hs 
g^2 (-\KK_s^{\rm{in}} \hsm + \KK_t^{\rm{in}})  \hs, \qquad
%\\
\TT[1^n_L \hs 2^n_L \hs 4^m_L \hs 3^m_L ]\hs =\hs 
g^2 (\KK_s^{\rm{in}}\hsm - \KK_u^{\rm{in}}) \hs,
\end{align}
%%%
where the kinematic functions  
$\{\KK_j^{\rm{in}}\}$ are summarized in \eqrefe{eq:K-KT-nnmm}
of Appendix\,\ref{app:B}. We find that these $\hs\{\KK_j^{\rm{in}}\}\hs$
functions fully agree with what we derived    
independently from the compactified 5d KK YM gauge field theory.

\vspace*{1mm}

Next, we study the mixed inelastic channel of gauge boson scattering  
$\hs (0,\hs 0)\ito (n,\hs n)\hs$.
We find that the condition \eqref{KK-num-sum} allows only two
combinations of the external KK states 
(as originally defined under the $\SS^1$ compactification of 
26d bosonic strings), 
%%%
\begin{equation}
\{n_1^{},\hs n_2^{},\hs n_3^{},\hs n_4^{}\} \,=~
\{0,\hs 0, \hs +n, \hs -n\} \hs , \hspace{3mm}
\{0,\hs 0, \hs -n, \hs +n\} \,.
\end{equation}
%%%
Hence, this inelastic KK open-string amplitude
equals a sum of six sub-amplitudes,
as shown in Fig.\,\ref{fig:3}. 
Then, we compute the color-ordered gauge boson amplitudes as follows:
%%%%
\beqs
\label{eq:T00nn-LL}
\begin{align}
\TT[1^0_{+1} \hs 2^0_{-1} \hs 3^{\pm n}_L \hs 4^{\mp n}_L] &\,=\,
-\frac{g^2\left(\bs \hsm +\hsm 4\right) s^2_\theta}
{~\bs \hsm +\hsm 2\hs\bs^{{1}/{2}}\hs\qb\hs\ct~} \,,
\\[1mm]
\TT[1^0_{+1} \hs 2^0_{-1} \hs 4^{\pm n}_L \hs 3^{\mp n}_L] &\,=\,
-\frac{g^2\hs (\bs\hsm +\hsm 4)\hs s^2_\theta}
{~\bs \hsm -\hsm 2\hs\bs^{{1}/{2}}\hs\qb\hs\ct~} \,,
\end{align}
\eeqs
where $\hs\bs\hs$ and $\hs\qb\hs$ are defined in \eqrefe{eq:qsr}.

\vspace*{1mm}

Then, from the four-point inelastic amplitudes in \eqrefe{eq:T00nn-LL}, 
we can obtain the following color-ordered amplitudes 
with all external states being $\ZZ\hsm$ even:
%%%%
\beqs
\label{eq:T00nn-LL-2}
\begin{align}
\TT[1^0_{+1} \hs 2^0_{-1} \hs 3^{n}_L \hs 4^{n}_L] &\,=\,
\TT[1^0_{+1} \hs 2^0_{-1} \hs 3^{+n}_L \hs 4^{-n}_L] \,=\,
-\frac{g^2\hs (\bs \hsm +\hsm 4)\hs s^2_\theta}
{~\bs \hsm +\hsm 2\hs\bs^{{1}/{2}}\hs\qb\hs\ct~} \,,
\\[1mm]
\TT[1^0_{+1} \hs 2^0_{-1} \hs 4^{n}_L \hs 3^{n}_L] &\,=\,
\TT[1^0_{+1} \hs 2^0_{-1} \hs 4^{+ n}_L \hs 3^{- n}_L] \,=\,
-\frac{g^2\hs (\bs\hsm +\hsm 4)\hs s^2_\theta}
{~\bs \hsm -\hsm 2\hs\bs^{{1}/{2}}\hs\qb\hs\ct~} \,.
\end{align}
\eeqs
We can re-express the above string-based inelastic amplitudes 
in the following forms:
%%%
\begin{align}
\TT[1^0_{+1} \hs 2^0_{-1} \hs 3^n_L \hs 4^n_L ]  \,=\, g^2 (-\KK_s^{\rm{in}} \hsm + \KK_t^{\rm{in}})\hs, 
\qquad
\TT[1^0_{+1} \hs 2^0_{-1} \hs 4^n_L \hs 3^n_L ]  \,=\, g^2 (\KK_s^{\rm{in}} \hsm - \KK_u^{\rm{in}}) \hs ,
\end{align}
%%%
where the kinematic functions
$\{\KK_j^{\rm{in}}\}$ are summarized in \eqrefe{eq:K-KT-00nn}
of Appendix\,\ref{app:B} and fully agree with what we derived    
independently from the compactified 5d KK YM gauge field theory.

\subsection{\hspace*{-3mm}Structure of Color-Ordered Massive KK Amplitudes}
\label{sec:3.3}

In this subsection, we study the structure of the color-ordered 
scattering amplitudes of massive KK gauge bosons.
We demonstrate that the tree-level massive KK gauge boson amplitudes 
can be obtained from the corresponding 
color-ordered amplitudes of the massless zero-mode gauge bosons
by making proper shifts of the Mandelstam variables.

\vspace*{1mm} 

From the formulation of the open-string amplitudes 
in section\,\ref{sec:2}, we observe that color-ordered 
massive KK sub-amplitudes
in $d$-dimensions, such as 
{$\hs\TT[1^{+n} \hs 2^{+n} \hs 3^{-n}\hs 4^{-n}]\hs$}, 
can be viewed as the massless amplitudes in $(d+\!1)$-dimensions
with the $(d+\!1)$-th component of each momentum being 
discretized since the $(d+\!1)$-th spatial dimension 
is compactified on $\SS^1$. Namely, we can express the 
$(d\!+\!1)$-dimensional momentum $\hat{k}^{\hat\mu}$ in terms of
the $d$-dimensional momentum $k^{\mu}$ plus an extra discretized
$(d\!+\!1)$-th component:
$\hat{k}^{\hat\mu}\!=\!(k^\mu,\,\hat{n}/\hsm R)\hs$. 
For a given polarization vector $\zeta^{\mu}$ of the on-shell 
gauge boson in $d$-dimensions, we can symbolically express it as a  
$(d+\!1)$-dimensional polarization vector 
$\hs\hat{\zeta}^{\hat\mu}\!=(\zeta^\mu,\hs 0)\hs$.
Thus, we have 
$\hs\hat\zeta_i^{} \hsm\cdot \hat\zeta_j^{}
 =\zeta_i^{} \hsm\cdot \zeta_j^{}\hs$
and
$\hs\hat\zeta_i^{}\hsm\cdot \hk_j^{}=\zeta_i^{}\hsm\cdot k_j^{}\hs$,
where the subscripts $(i,j)$ denote the particle numbers of the 
external states.\footnote{%
Our later explicit calculations of the KK scattering amplitudes will
be always performed in the effective (3+1)-dimensional spacetime 
with $d\!=\hsmx 4\hs$,
and with a single compactified extra spatial dimension of 
coordinate $X^{25}$. The other extra spatial dimensions of coordinates $\{X^{4},\cdots\hsm ,X^{24}\}$ have much smaller radii 
$\hs r_j^{}\!=\hsm\mO(\MPl^{-1})\,$ 
($j\!=\!4,\cdots\hsm ,24$), so they are fully decoupled 
at energy scales much below the reduced Planck scale $\MP\hs$,
as we discussed at the beginning of Sec.\,\ref{sec:2.1}.
Thus, the bosonic strings effectively propagate in (4+1)d spacetime
with the single extra spatial dimension of $X^{25}$ compactified on
$\SS^1$.}
Keeping these in mind, we can first compute a
$(d\!+\!1)$-dimensional massless scattering amplitude
and then we deduce the corresponding $d$-dimensional 
massive KK amplitude by using relations 
\begin{equation}
\hat\zeta_i^{} \hsm\cdot \hat\zeta_j^{}
 =\zeta_i^{} \hsm\cdot \zeta_j^{}
\hs, \hspace*{10mm}
\hs\hat\zeta_i^{}\hsm\cdot \hk_j^{}=\zeta_i^{}\hsm\cdot k_j^{}
\hs,
\end{equation}
and the relation between the two sets of Mandelstam variables
%%%
\begin{equation}
\label{eq:sijhat-sij}
\hat{s}_{ij}^{} =\, -2\hsx\hk_i^{} \hsm\cdot \hk_j^{}  
=\, 
s_{ij}^{} \hsm - \hsm
\(\!\frac{~\hat{n}_i^{} \!+\hsm\hat{n}_j^{}~}{R}\!\)^{\hsm\!\!2} ,
\end{equation}
%%%%
where the $(d\!+\!1)$-dimensional momenta obey the on-shell conditions
$\hs \hk^2_i\!=\hsm\hk^2_j\!=\hsm 0\,$.
Optionally, we can first write a $d$-dimensional massless 
(zero-mode) scattering
amplitude $\,\TT_{(0)}^{}(s_{ij}^{})\,$ with all polarization vectors
and momenta of the external states in symbolic format, 
and then we deduce the corresponding KK sub-amplitude $\,\TT_{\text{KK}}^{\hs\text{sub}}({s}_{ij}^{})\,$ as follows:
\begin{equation}
\TT_{\text{KK}}^{\hs\text{sub}}(s_{ij}^{}) \,=\,
\TT_{(0)}^{}(\hat{s}_{ij}^{}) \hs,
\end{equation}
with each $d$-dimensional 
external state having its momentum obey the on-shell condition
of the massive KK gauge boson ($k_j^2\hsm =\!-M_j^2$) 
and its polarization vector replaced by the polarization vector
$\zeta_j^{}$ of the KK gauge boson. 
In this way, we can derive all the $d$-dimensional massive KK gauge boson 
amplitudes according to the structure of the corresponding $d$-dimensional 
massless (zero-mode) gauge boson amplitudes.
	
\vspace*{1mm}

Next, we compute a four-point massless (zero-mode) gauge boson 
scattering amplitude with color ordering [1234].
This can be done either in the massless YM field theory, or, we
can deduce it by taking the field theory limit 
$\alp\!\ito 0\hs$ of the open-string scattering amplitude \eqref{eq:Ab1234}:
%%%
\begin{align}
\hspace*{-11mm}
\TT[1^0 \hs 2^0 \hs 3^0 \hs 4^0]
=& \hspace*{1.5mm}
g^2 \Big\{
2\big[ ( \zeta_1 \!\cdot\! \zeta_3) (\zeta_2 \!\cdot\! \zeta_4)
\!-\! (\zeta_1 \!\cdot\! \zeta_4) (\zeta_2  \!\cdot\! \zeta_3)
\!-\! (\zeta_1 \!\cdot\! \zeta_2) (\zeta_3 \!\cdot\! \zeta_4)\big]
\nn\\
&\hspace*{5mm} +\! \frac{\,4\,}{\,t\,} \big[
(\zeta_2\!\cdot\!\zeta_3) (\zeta_1\!\cdot\! k_2) (\zeta_4\!\cdot\! k_1)
\!-\! (\zeta_2\!\cdot\!\zeta_3) (\zeta_1\!\cdot\! k_4) 
(\zeta_4\!\cdot\! k_2)
\!-\!(\zeta_1\!\cdot\!\zeta_4) (\zeta_2\!\cdot\! k_4) (\zeta_3\!\cdot\! k_2)
\nn\\
&\hspace*{5mm}
+\!(\zeta_2\!\cdot\!\zeta_4) (\zeta_1\!\cdot\! k_4) (\zeta_3\!\cdot\! k_2)
-(\zeta_1\!\cdot\!\zeta_4) (\zeta_2\!\cdot\! k_1) (\zeta_3\!\cdot\! k_4)
\!-\! (\zeta_1\!\cdot\!\zeta_4) (\zeta_2\!\cdot\! k_4) (\zeta_3\!\cdot\! k_4)
\nn\\
&\hspace*{5mm}
+\! (\zeta_3\!\cdot\!\zeta_4) (\zeta_1\!\cdot\! k_4) (\zeta_2\!\cdot\! k_1)
\!+\! (\zeta_3\!\cdot\!\zeta_4) (\zeta_1\!\cdot\! k_4) (\zeta_2\!\cdot\! k_4)
-(\zeta_1\!\cdot\!\zeta_3) (\zeta_2\!\cdot\!k_1) (\zeta_4\!\cdot\! k_1)
\nn\\
&\hspace*{5mm}
-\! (\zeta_1\!\cdot\!\zeta_3) (\zeta_2\!\cdot\! k_4) (\zeta_4\!\cdot\! k_1)
\!-\!(\zeta_1\!\cdot\!\zeta_2) (\zeta_3\!\cdot\! k_2) (\zeta_4\!\cdot\! k_1)
\!-\! \frac{\,s\,}{\,2\,}
(\zeta_1 \!\cdot\!\zeta_4) (\zeta_2 \!\cdot\!\zeta_3) ]
\label{eq:T1234-0000}
\\[-0.5mm]
&\hspace*{5mm}
+\! \frac{\,4\,}{\,s\,} [\hs 
(\zeta_2\!\cdot\!\zeta_4) (\zeta_1\!\cdot\!k_2)(\zeta_3\!\cdot\!k_4)
\!-\!(\zeta_1\!\cdot\!\zeta_4)(\zeta_2\!\cdot\!k_1)(\zeta_3\!\cdot\!k_4)
\!-\!(\zeta_1\!\cdot\!\zeta_2)(\zeta_4\!\cdot\!k_2)(\zeta_3\!\cdot\!k_4)
\nn\\
&\hspace*{5mm}
+\!(\zeta_3\!\cdot\!\zeta_4)(\zeta_1\!\cdot\!k_4)(\zeta_2\!\cdot\! k_1)
\!-\!(\zeta_3\!\cdot\!\zeta_4)(\zeta_1\!\cdot\!k_2)(\zeta_2\!\cdot\! k_4)
\!-\!(\zeta_1\!\cdot\!\zeta_3)(\zeta_2\!\cdot\!k_1)(\zeta_4\!\cdot\! k_1)
\nn\\
&\hspace*{5mm}
+\!(\zeta_2\!\cdot\!\zeta_3)(\zeta_1\!\cdot\!k_2)(\zeta_4\!\cdot\!k_1)
\!-\!(\zeta_1\!\cdot\!\zeta_2)(\zeta_3\!\cdot\!k_2)(\zeta_4\!\cdot\!k_1)
\!-\!(\zeta_1\!\cdot\!\zeta_3)(\zeta_2\!\cdot\!k_1)(\zeta_4\!\cdot\!k_2)
\nn\\
&\hspace*{5mm} 
+\!(\zeta_2\!\cdot\!\zeta_3)(\zeta_1\!\cdot\!k_2)(\zeta_4\!\cdot\!k_2)
\!-\!(\zeta_1\!\cdot\!\zeta_2)(\zeta_3\!\cdot\!k_2)(\zeta_4\!\cdot\!k_2)
\!-\! \frac{\,t\,}{\,2\,}
(\zeta_1\!\cdot\!\zeta_2)(\zeta_3\!\cdot\!\zeta_4) \hs ] 
\Big\} \,.
\nn 
\end{align}
%%%
For the other massless color-ordered amplitudes, such as the one 
with color ordering [1243],  
they can be obtained through the relation: 
%%%%
\begin{equation}
\label{eq:T1234-1243} 
u \hs \TT[1^0\hs2^0\hs4^0\hs3^0] \,=\, 
t\hs \TT[1^0 \hs 2^0\hs3^0\hs4^0]\,.
\end{equation} 
%%%%
For instance, we consider the elastic KK scattering
$(n,\hs n) \ito (n,\hs n)$ as discussed in section\,\ref{sec:3.1}. 
We can obtain the color-ordered sub-amplitude 
$\hs\TT[1^{+n}\hs 2^{+n} \hs 3^{-n} \hs 4^{-n}]\hs$ 
by the replacement $\,s\ito (s-4M_n^2)\,$ 
in $\hs\TT[1^02^03^04^0]\hs$. 
In general, the color-ordered sub-amplitude  $\TT[1^{\hat{n}_1}2^{\hat{n}_2}3^{\hat{n}_3}4^{\hat{n}_4}]$ can be obtained by the following replacements:
%%%
\begin{equation}
\label{eq:sij-sij}
{s}_{ij}^{} \,\longrightarrow~ 
s_{ij}^{} \hsm - \hsm
\(\!\hsm\frac{~\hat{n}_i^{} \!+\hsm\hat{n}_j^{}~}{R}\hsm\!\)^{\!\!2} .
\end{equation}
%%%%
This procedure can be applied to deriving the general $N$-point
scattering amplitudes of massive KK gauge bosons
and be extended to the case of KK gravitons 
which we will present elsewhere.

\vspace*{1mm}

For the color-ordered combinations
[1234] and [1243], we derive the
following massive KK gauge boson amplitudes
according to Eq.\eqref{eq:KK-num-4-point}:
%%%%%
\beqs
\label{eq:T1234-1243-FT}
\begin{align}
\TT[1^n 2^n 3^n 4^n] &\,=\, \frac{1}{\,2\,}\hsm 
\( \TT_1^{\rm{el}}+\TT_2^{\rm{el}}+\TT_3^{\rm{el}} \)\!,
\\
\TT[1^n2^n4^n3^n] &\,=\, \frac{1}{\,2\,} \!\(\!\frac{\,t\,}{\,u\,}\hs\TT_1^{\rm{el}}
+\frac{t}{\,u\hsm -\hsm 4M_n^2\,}\hs\TT_2^{\rm{el}}
+\frac{\,t\hsm -\hsm 4M_n^2\,}{u}\hs\TT_3^{\rm{el}} \) \!,
\end{align}
\eeqs
%%%
where the basis amplitudes 
$\{\TT_1^{\rm{el}},\hs \TT_2^{\rm{el}},\hs \TT_3^{\rm{el}}\}$ 
are given by the following sub-amplitudes:
%%%%
\beqs
\label{eq:T1234-nnnn}
\begin{align}
\label{eq:TT1-nnnn}
\TT_1^{\rm{el}} &= 
\TT[1^{\pm n} \hs 2^{\pm n} \hs 3^{\mp n} \hs 4^{\mp n}] 
= \TT[1^0 \hs  2^0 \hs 3^0 \hs 4^0]
\Big|_{s\hsx\to\hs (s\hs -\hs 4\Mn^2)} \,,
\\[1mm]
\label{eq:TT2-nnnn}
\TT_2^{\rm{el}} &= 
\TT[1^{\pm n} \hs 2^{\mp n} \hs 3^{\pm n} \hs 4^{\mp n} ]
= \TT[1^0 \hs  2^0 \hs 3^0 \hs 4^0] 
\Big|_{u\hsx\to\hs (u\hs -\hs 4\Mn^2)}\,,
\\[1mm]
\label{eq:TT3-nnnn}
\TT_3^{\rm{el}} &= 
\TT[1^{\pm n} \hs 2^{\mp n} \hs 3^{\mp n}\hs 4^{\pm n}] 
= \TT[1^0 \hs  2^0 \hs 3^0 \hs 4^0]
\Big|_{t\hsx\to\hs (t\hs -\hs 4\Mn^2)} \,.
\end{align}
\eeqs
%%%%
We note that the KK numbers of the sub-amplitude
$\TT[1^{\pm n} \hs 2^{\mp n} \hs 3^{\pm n} \hs 4^{\mp n}]$ 
in Eq.\eqref{eq:TT2-nnnn} makes the shifted mass-term in
Eq.\eqref{eq:sij-sij} vanish, so there is practically 
no replacement needed. 

\vspace*{1mm} 

Following the same procedure, we derive the following color-ordered
amplitudes for the inelastic scattering channel 
$(n,n)\ito(m,m)\hs$: 
%%%%%
\beqs
\begin{align}
\TT[1^n2^n3^m4^m] \,&=\, \frac{1}{2}\!
\( \TT_1^{\rm{in}}+\TT_2^{\rm{in}} \),
\\[1mm]
\TT[1^n2^n4^m3^m] \,&=\, \frac{1}{2} \!\(\!
\frac{t\!-\!M_{n-m}^2}{~u\!-\!M_{n+m}^2~}\hs\TT_1^{\rm{in}} 
+\frac{\,t\!-\!M_{n+m}^2\,}{~u\!-\!M_{n-m}^2~}\hs\TT_2^{\rm{in}}\) ,
\end{align}
\eeqs
%%%
where the basis amplitudes 
$\{\TT_1^{\rm{in}},\hs \TT_2^{\rm{in}}\}$ 
are obtained by the relations:
%%%%%
\beqs
\begin{align}
\TT_1^{\rm{in}} \,&=\, \TT[1^{\pm n}  \hs 2^{\mp n} \hs 3^{\pm m} \hs4^{\mp m}] 
\,=\, \TT[1^0 \hs 2^0 \hs 3^0 \hs 4^0]
\Big|_{t\hs\to\hs (t\hs -\hs M_{n-m}^2)}^{} \hs,
\\[1mm]
\TT_2^{\rm{in}} \,&=\, 
\TT[1^{\pm n} \hs 2^{\mp n} \hs 3^{\mp m} \hs 4^{\pm m}] 
\,=\, \TT[1^0 \hs 2^0 \hs 3^0 \hs 4^0]
\Big|_{t\hs\to\hs (t\hs -\hs M_{n+m}^2)}^{} \hs.
\end{align}
\eeqs
%%%
For the inelastic scattering channel $(0,0)\to(n,n)$,
we can deduce its amplitude from that of $(n,n)\to(m,m)$
by the following replacements: 
%%%%%
\beqs
\begin{align}
\TT[1^0 \hs 2^0 \hs 3^n \hs 4^n] \,&=\,  
\TT[1^{n}  \hs 2^{n} \hs 3^{m} \hs4^{m}]
\Big|_{\{n\to 0 ,\hs m\to n \}} 
\hs,
\\[1mm]
\TT[1^0 \hs 2^0 \hs 4^n \hs 3^n] \,&=\,  
\TT[1^{n}  \hs 2^{n} \hs 4^{m} \hs3^{m}]
\Big|_{\{n\to 0 ,\hs m\to n \}} 
\hs.
\end{align}
\eeqs
% 

%\vspace*{3mm}
\section{\hspace*{-3mm}KK Graviton Amplitudes
from Extended Massive Double-Copy}
\label{sec:4}

\begin{figure}[t]
\vspace*{-2mm} 
\centering
\includegraphics[width=12cm]{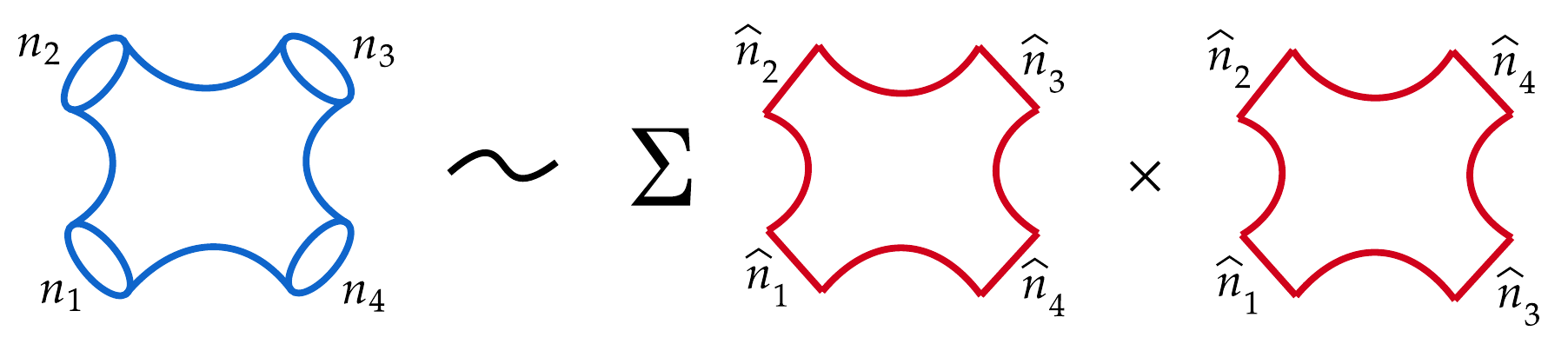}
\vspace*{-3mm} 
\caption{\small{%
Four-point scattering amplitude of massive KK closed-strings
with $\ZZ$-even parity (marked in blue) can be decomposed into a sum
of the products of two color-ordered amplitudes of massive KK open-strings
under the $\SS^1$ compactification of 26d for bosonic strings.}}
\label{fig:4}
%\vspace*{3mm}
\end{figure}

According to the extended massive KLT-like relation \eqref{eq:close-amp-1},
we can construct explicitly the four-point massive KK closed-string
amplitude from the product of the corresponding massive KK open-string
amplitudes as follows:
%%%%
\begin{equation}
\label{eq:closed-N4}
\A_{\cl}(\zeta, k) =
\frac{~8\ka^2\,}{~\pi \al^{\pp\hs 5}~} \!\!
\sum_{\{a_{\hsm j}^{},\hsx b_{\hsm j}^{}\}}\! 
\sum_{\{\sgn_j^{}\}}\!
\vrhohat_{ab}^{} 
\sin(\hsm\pi \alp \hk_1^{} \hsm\!\cdot\hsm \hk_2 )
\Big\{\hsm \xoverline{\A}^{}_{\op} 
[\zeta_j^{a_{\hsm j}^{}}\!,\hs \hk_j^{} \big|1234]  
\!\times\!
\xoverline{\A}^{}_{\op}
\big[\zeta_j^{b_{\hsm j}^{}}\!,\hs \hk_j^{} \big|1243\big] 
\hsm\Big\},
\end{equation}
%%%
where we have replaced the closed-string coupling by the relation
$\hs g_{\cl}^{}\!=\!\hsm\kappa /(2\pi\alpha')\hs$.\ 
For the two massive KK open-string amplitudes inside $\{\cdots\}$,
the Regge slope should be rescaled as
$\,\alp\hsmx\ito\alp\hsmx/4\,$.
This is equivalent to considering the $N$-point low energy
field theory formula \eqref{eq:Amp-N-Gr} and derive
the four-longitudinal KK graviton amplitude for the case of
$\hs N\!=4\hsx$.  We illustrate in Fig.\,\ref{fig:4} 
the extended massive KLT-like relation \eqref{eq:closed-N4} 
between the four-point scattering amplitude of KK closed-strings and 
the products of two color-ordered scattering amplitudes of KK 
open-strings.  

%%%%%
%%%%%%

\subsection{\hspace*{-3mm}Constructing Elastic Scattering Amplitudes of Four KK Gravitons}
\label{sec:4.1}

In this subsection, we construct the four-point
elastic scattering amplitudes of KK gravitons
by using the extended massive KLT-like relation \eqref{eq:closed-N4}
under the low energy field theory limit.

\vspace*{1mm}

Taking the zero-slope limit
$\,\alpha'\ito 0\,$ for the closed-string amplitude
\eqref{eq:closed-N4}, we derive the four-longitudinal
KK graviton scattering amplitude for the elastic channel
$(n,\hs n)\ito (n,\hs n)\hs$.\
Thus, we can express the elastic amplitude
of longitudinal KK gravitons as the sum of products of two
color-ordered massive KK gauge boson amplitudes:
%%%
\begin{align}
\label{eq:KLT-nnnn}
\hspace*{-6mm}
\M[1^n_L\hs 2^n_L \hs 3^n_L \hs 4^n_L]
&\,=\,
\frac{\,\ka^2\,}{\,32\,}
\sum_{\{a_{\hsm j}^{},\hsx b_{\hsm j}^{}\}}
\sum_{\{\sgn_j^{}\}}\!
\vrhohat_{ab}^{}\hs
( -\hk_1^{}\!\hsm\cdot\hsm \hk_2^{}) \hs
{\TT}\big[\hsm\zeta^{a_{\hsm j}^{}}_j \hsm , \hk_j^{}\big| 1234\hs\big]
\!\hsmx\times\!
{\TT}\big[\hsm\zeta^{b_{\hsm j}^{}}_j\hsm , \hk_j^{}\big| 1243\hs\big]
\nn\\
&\,=\,
\frac{\,\ka^2\,}{32}
\sum_{\{a_{\hsm j}^{},\hsx b_{\hsm j}^{}\}} \!
\vrhohat_{ab}^{}\! \LB
\(s \!-\! 4\Mn^2\) \hsm
\TT_{a_{\hsm j}^{}}[1^{+n} \hs 2^{+n} \hs 3^{-n} \hs4^{-n}]\,
\TT_{b_{\hsm j}^{}}[1^{+n} \hs 2^{+n} \hs 4^{-n} \hs3^{-n}]
\right.
\nn\\[-3mm]
& \hspace*{42.1mm}
+ \hsmx s \,
\TT_{a_{\hsm j}^{}}[1^{+n} \hs 2^{-n} \hs 3^{+n} \hs4^{-n}] \,
\TT_{b_{\hsm j}^{}}[1^{+n} \hs 2^{-n} \hs 4^{-n} \hs3^{+n}]
\nn\\
& \hspace{42.5mm} \left.
+\hs  s \,
\TT_{a_{\hsm j}^{}}[1^{+n} \hs 2^{-n} \hs 3^{-n} \hs4^{+n}]  \,
\TT_{b_{\hsm j}^{}}[1^{+n} \hs 2^{-n} \hs 4^{+n} \hs3^{-n}] \RB
\hsm,
\end{align}
%%%%
where in the above second equality we have defined
the short-hand notation for the partial amplitudes as
$\hs\TT_{a_j}[1^{+n} \hs 2^{-n} \hs 3^{+n} \hs 4^{-n}] \!=\!
\TT[1^{+n}_{a_1} \hs 2^{-n}_{a_2} \hs 3^{+n}_{a_3} \hs4^{-n}_{a_4}]
\hs$,$\hs$
and so on.\
Note that a massive KK graviton in 4d has 5 physical helicity states
and their polarization tensors are given by
%%%
\beqs
\label{eq:Gpol}
\begin{align}
\label{eq:Gpol-21}
\zeta^{\mn}_{\pm 2} &=
\zeta_{\pm 1}^{\mu} \zeta_{\pm 1}^{\nu}\,,\qquad
%%%%%%
\zeta_{\pm 1}^{\mn} =
\fr{1}{\sqrt{2\,}\,}\!\(\zeta_{\pm 1}^{\mu}\zeta_{L}^{\nu}
\!+\zeta_{L}^{\mu}\zeta_{\pm 1}^{\nu}\) \!,
\\[1mm]
\label{eq:Gpol-L}
\zeta^{\mn}_L & = \fr{1}{\sqrt{6\,}\,}\!
\(\zeta^\mu_{+1}\zeta^\nu_{-1} + \zeta^\mu_{-1}\zeta^\nu_{+1}
  +2\hs\zeta^\mu_L\zeta^\nu_L\)\!.
\end{align}
\eeqs
%%%
Thus, in \eqrefe{eq:KLT-nnnn} the helicity indices of each
partial amplitude are
$\{a_j^{},b_j^{}\} = \{\pm 1,L\}$.
For the graviton polarization tensor $\zeta^{\mn}_{\pm 2}$,\,
its coefficient
$\varrho_{a_j^{} b_j^{}}^{}\!\!=\!\varrho_{\pm1, \pm1}^{} \!=\!1$;
for the polarization tensor $\zeta^{\mn}_{\pm 1}$,
its coefficient
$\varrho_{a_j^{} b_j^{}}^{}\!\!=\!\varrho_{\pm 1, L}^{}
\!\!=\!\varrho_{L,\pm 1}^{} \!\!=\!\fr{1}{\sqrt{2\,}\,}$;
and for the polarization tensor $\zeta^{\mn}_L$,
its coefficient $\varrho_{a_j^{} b_j^{}}^{}\!$
takes the following values:
\\[-4mm]
%%%
\begin{equation}
\varrho_{a_j^{} b_j^{}}^{} =
\left\{ \begin{array}{rcl}
\sqrt{\frac{1}{6}}\,,~~ & \rm{ for }
& \{a_j, b_j\} = \{\pm 1,\mp 1\} \hs ,
\\[1.5mm]
\sqrt{\frac{2}{3}}\,,~~ &\rm{ for }& \{a_j^{}, b_j^{}\} = L \hs ,
\\[2mm]
0\,,~~ & \rm{ for } & \rm{others} \hs .
\end{array}\right.
\end{equation}
%%%
We note that the longitudinal graviton's
polarization tensor \eqref{eq:Gpol-L} includes 3 products
of two massive spin-1 polarization vectors, among which the
two products $\zeta _\mu^{+1}\zeta _\nu^{-1}$ and
$\zeta _\mu^{-1}\zeta _\nu^{+1}$ arise from the transverse polarization
vectors and are important for the successful construction of
full massive graviton amplitudes.

\vspace*{1mm}

Then, we can compute all the color-ordered four-point
gauge boson amplitudes
from Eqs.\eqref{eq:T1234-1243} and \eqref{eq:T1234-nnnn}
with the relevant external
polarization vectors \eqref{eq:Pol} and momenta \eqref{eq:Momenta}.
Using these amplitudes,
we derive the four-point elastic longitudinal KK graviton
scattering amplitude from the extended massive
double-copy formula \eqref{eq:KLT-nnnn}:
%%%%
\begin{align}
\label{eq:MLnnnn}
\M[1^n_L\hs 2^n_L \hs 3^n_L \hs 4^n_L] \,=\,
- \frac{~\ka^2 M_n^2\hs (X_0 \hsm +\! X_2\ctt^{}
\hsm +\!X_4\ctf^{} \hsm +\! X_6\cts^{} )\hsm\csc^2\!\theta~}
{~512\hs\bs\hs (\bs\hsmx -\hsmx 4)
[\hs\bs^2 \!-\hsm (\bs\hsm -\hsm 4)^2\ctt^{} 
\!+\hsm 24\hs\bs\hsm +\!16]~}  \,,
\end{align}
%%%
where the coefficients $\{X_j^{}\}$ in the numerator are given by
%%%%%%
\begin{equation}
\label{eq:4hL-Pn}
\begin{aligned}
X_0 &= -2\hs (255\hs\bs^5\! +\hsm 2824\hs\bs^4\!-\! 19936\hs\bs^3
\!+\hsm 39936\hs\bs^2 \!-\hsm 256\hs\bs \hsm +\!14336)\hs ,
\\[1mm]
%%%%
X_2 &= 429\hs\bs^5\!-\hsm 10152\hs\bs^4\!+\hsm 30816\hs\bs^3
\!-\hsm 27136\hs\bs^2\!-\hsm 49920\hsm\bs\hsm +\hsm 34816 \hs ,
\\[1mm]
%%%%
X_4 &= 2\hs (39\hs\bs^5\!-\hsm 312\hs\bs^4\!-\hsm 2784\hs\bs^3\!
-\hsm 11264\hs\bs^2\!+\hsm 26368\hsm\bs \hsm-\! 2048) \hs ,
\\[1mm]
%%%%
X_6 &= 3\bs^5\!+\hsm 40\hs\bs^4\!+\hsm 416\hs\bs^3
\!-\hsm 1536\hs\bs^2 \!-\hsm 3328\hs\bs \hsm-\hsm 2048 \hs .
\end{aligned}
\end{equation}
%%%%%%%%
The above formulas take exactly the same form as Eq.(F.3a)
and Eqs.(F.4a)-(F.4d) of Ref.\,\cite{Hang:2021fmp}.
We can compare our above elastic longitudinal KK graviton amplitude
with that obtained by the previous explicit lengthy Feynman diagram
calculations\,\cite{Chivukula:2020L}.
It is truly impressive that we find full agreement
between our Eqs.\eqref{eq:MLnnnn}-\eqref{eq:4hL-Pn}
and the Eq.(71) of Ref.\,\cite{Chivukula:2020L} after taking into
account the notational difference.\
Hence, our string-based massive double-copy construction {\it does
successfully predict} the exact four-point elastic scattering amplitude
of longitudinal KK gravitons at tree level.
In the next subsection, we will further present our
string-based massive double-copy constructions of the
inelastic scattering amplitudes of massive KK gravitons.

\vspace*{1mm}

We can further derive the following LO and NLO scattering amplitudes
of \eqrefe{eq:MLnnnn} under high energy expansion:
\\[-7mm]
%%%
\beqs
\begin{align}
\label{eq:MLnnnn-LO}
\M_0[1^n_L\hsx 2^n_L \hsx 3^n_L \hsx 4^n_L] \,&=\,
\frac{~3\hs\ka^2\,}{\,128\,}\hs s^{}\hs
( 7\hsm +\hsm \ctt^{} )^2 \hsm\csc^2\!\theta  \,,
\\[1mm]
\label{eq:MLnnnn-NLO}
\dM[1^n_L\hsx 2^n_L \hsx 3^n_L \hsx 4^n_L]  \,&=\,
-\frac{\,\kappa^2 M_n^2~}{256} 
(1810+93\hs \ctt^{}\hsm +\hsm 126\hs c_{4\theta}^{}
\hsm +\hsm 19\hs c_{6\theta}^{})\hsm \csc^4\!\theta   \,.
\end{align}
\eeqs
%%%%

\vspace*{1mm}

We stress that {\it the string-based double-copy formula
\eqref{eq:KLT-nnnn} for 4-point amplitudes or
\eqref{eq:Amp-N-Gr} for general $N$-point amplitudes
gives an explicit prescription on how to practically construct the exact tree-level
KK graviton scattering amplitudes from the sum of the products of the corresponding KK gauge boson amplitudes in
the compactified KK field theories.}
Hence, given our string-based double-copy formula
\eqref{eq:Amp-N-Gr} or \eqref{eq:KLT-nnnn},
one can practically follow this explicit prescription
to derive the full KK graviton amplitudes without relying on
computing the original KK string amplitudes.

\vspace*{1mm}

Some further remarks are in order.\
It is instructive to compare our above string-based massive double-copy construction (\`{a} la extended KLT-like relations) with the extended
BCJ-type double-copy construction under high energy expansion as given
by Sec.\,5 of Ref.\,\cite{Hang:2021fmp}.
At the leading order (LO) of the high energy expansion, both the
4-point longitudinal KK gauge boson amplitude and
KK graviton amplitude are {\it mass-independent,}
which are of $\mO(E^0)$ and $\mO(E^2)$, respectively.
We found\,\cite{Hang:2021fmp}
that the BCJ-type numerators in the LO gauge boson amplitude
satisfy the kinematic Jacobi identity at the $O(E^2M_n^0)$,
so the extension from the conventional massless BCJ method 
can be realized directly.\ 
At the next-to-leading order (NLO) of the high energy expansion,
the longitudinal KK gauge boson amplitude
and KK graviton amplitude become {\it mass-dependent,}
which are of $\mO(M_n^2/E^2)$ and $\mO(M_n^2E^0)$, respectively.
We further found\,\cite{Hang:2021fmp}
that the BCJ-type NLO numerators of the KK gauge boson amplitude can obey the Jacobi identity (after the generalized gauge transformations) and the double-copied KK graviton amplitude at NLO can give the correct structure of the exact NLO KK graviton amplitude, but the numeric coefficients of the double-copied NLO amplitude still differ
somewhat from that of the exact NLO amplitude.\ 
So the NLO double-copy construction needs to be
modified\,\cite{Hang:2021fmp}.\ 
We note that an important reason of this problem is because
the 4-point elastic amplitude of longitudinal KK gauge bosons
contains double-poles with one type of poles
from exchanging the massless zero-mode and another type of mass-poles
from exchanging the level-$(2n)$ KK-mode which contributes to the
mass-dependent NLO amplitude.\  This is beyond the conventional
BCJ double-copy method\,\cite{BCJ}\cite{BCJ-Rev}, so
the deviation from it is expected at the NLO and a modified
BCJ-type double-copy construction was presented for the NLO
KK amplitudes\,\cite{Hang:2021fmp}.
As another reason, we note that
the polarization tensor $\zeta_L^{\mn}$
of the (helicity-zero) longitudinal KK graviton
in \eqrefe{eq:Gpol-L} contains
the sum of three products of two gauge boson polarization vectors
$(\zeta^\mu_{+1}\zeta^\nu_{-1},\,\zeta^\mu_{-1}\zeta^\nu_{+1},\,
\zeta^\mu_{L}\zeta^\nu_{L})$, while the simple double-copy
by using the longitudinal KK gauge boson amplitude alone could only
provide the polarization-vector product of
$\hs\zeta^\mu_{L}\zeta^\nu_{L}$, which does not include
the other two products
$(\zeta^\mu_{+1}\zeta^\nu_{-1},\,\zeta^\mu_{-1}\zeta^\nu_{+1})$
of $\,\zeta_L^{\mn}$
as given by the spin-1 helicity combinations $(+1,-1)$ and $(-1,+1)$.
However, the sum of all three helicity combinations
$(+1\!-\!1,\,-1\!+\!1,\,LL)$
for each external longitudinal KK graviton state
in our present string-based massive double-copy formula
\eqref{eq:KLT-nnnn} is {\it automatically built in}
from the beginning.\
Another key feature of our string-based construction \eqref{eq:KLT-nnnn}
is that it intrinsically includes a set of KK gauge boson sub-amplitudes
with different KK-number combinations as allowed by the condition
\eqref{eq:KK-num-4-point} due to the original string compactification
under $\SS^1$ (without the orbifold $\ZZ$) in Sec.\,\ref{sec:2.1}.
This feature is {\it intrinsically built in} for our string-based double-copy formulation. Using this string-based formulation,
we will derive a precise BCJ-type double-copy formulation
in the future work.

\vspace*{1mm}

We further note that extended BCJ-type double-copy construction
of Ref.\,\cite{Hang:2021fmp} does work elegantly for the LO
amplitudes under the high energy expansion.\
This is highly nontrivial because the
longitudinal KK graviton polarization tensor $\zeta_L^{\mn}$
also contains additional products of transverse polarization vectors
$(\zeta^\mu_{+1}\zeta^\nu_{-1},\,\zeta^\mu_{-1}\zeta^\nu_{+1})$
which can contribute to the LO KK graviton amplitudes
of $\mO(E^2M_n^0)$. Note that our extended massive double-copy
of LO amplitudes is built upon the
KK gauge boson equivalence theorem (KK\,GAET)\,\cite{5DYM2002}\cite{KK-ET-He2004}
which connects the LO longitudinal KK gauge boson amplitude
$\TT_0^{}[4A_L^{an}]$
to its corresponding LO KK Goldstone amplitude
$\tT_0^{}[4A_5^{an}]$:
\begin{equation}
\label{eq:KKET-LO}
\TT_0^{}[4A_L^{an}] \,=\, \tT_0^{}[4A_5^{an}]
= \mO(E^0M_n^0)\hs ,
\end{equation}
where we have introduced the short-hand notations
$\TT_0^{}[4A_L^{an}]\!=\!
 \TT_0^{}[A_L^{an}A_L^{bn}\!\ito A_L^{cn}A_L^{dn}]$ and
$\tT_0^{}[4A_5^{an}]\!=\!
 \tT_0^{}[A_5^{an}A_5^{bn}\!\ito A_5^{cn}A_5^{dn}]\hs$.
We note that the KK Goldstone boson $A_5^{an}$ is just a scalar field
without any polarization vector, and $A_5^{an}$ becomes a massless
physical scalar degree of freedom in the high energy limit. 
Hence, the double-copy construction of the gravitational KK Goldstone
boson $h^{55}_n$-amplitude from the KK Goldstone
$A_n^{a5}$-amplitude is {\it uniquely defined} via the correspondence
$\,A_n^{a5}\otimes A_n^{a5}\to h^{55}_n\,$.
Hence, the LO gravitational KK Goldstone amplitude
$\MT_0^{}[4h_n^{55}]$ as given by the double-copy of
the LO KK Goldstone amplitude $\tT_0^{}[4A_n^{a5}]$ 
is well defined under one-to-one correspondence.
On the other hand, we established the
Gravitational Equivalence Theorem (GRET)\,\cite{Hang:2021fmp}
which connects the LO longitudinal KK graviton amplitude
$\M_0^{}[4h_L^{n}]$
to its corresponding LO gravitational KK Goldstone amplitude
$\MT_0^{}[4h^n_{55}]\hs$:
\begin{equation}
\label{eq:GET-LO}
\M_0^{}[4h_L^{n}]\,=\, \MT_0^{}[4h^n_{55}]
= \mO(E^2M_n^0)\hs ,
\end{equation}
where we have introduced the shorthand notations
$\M_0^{}[4h_L^{n}]\!=\!
 \M_0^{}[h_L^{n}h_L^{n}\!\ito h_L^{n}h_L^{n}]$ 
and
$\MT_0^{}[4h^n_{55}]\!=\!
 \M_0^{}[h^n_{55}h^n_{55}\hsm\ito h^n_{55}h^n_{55}]\hs$.
Thus, the LO equality of the KK\,GAET 
\eqref{eq:KKET-LO} leads to 
the LO equality of the KK\,GRET \eqref{eq:GET-LO} 
by the double-copy construction.
Hence, {\it the success of our extended double-copy construction of
the LO longitudinal KK graviton amplitude $\M_0^{}[4h_L^{n}]$
from the LO longitudinal KK gauge boson
amplitude $\TT_0^{}[4A_L^{an}]$ is fully ensured by the correct
double-copy construction of its corresponding LO gravitational
KK Goldstone amplitude $\MT_0^{}[4h_n^{55}]$ from
the LO gauge KK Goldstone amplitude
$\tT_0^{}[4A_n^{a5}]$} because we can derive the
correspondence \,GAET\,$\Longrightarrow$\,GRET$\hs$
by using the LO double-copy construction. 
This means that we can correctly compute
the LO longitudinal KK graviton amplitude $\M_0^{}[4h_L^{n}]$
by defining an effective LO polarization tensor
for each longitudinal KK graviton:
\\[-6mm]
\begin{equation}
\label{eq:GPol-L0}
\zeta_{L(0)}^{\mn}\,=\, \zeta_L^\mu\hs\zeta_L^\nu \,.
\end{equation}
We have verified this insight by explicitly computing the
four-point elastic scattering amplitude $\M_0^{}[4h_L^{n}]$
of longitudinal KK gravitons with the LO polarization tensor
\eqref{eq:GPol-L0}.
Namely, using the LO longitudinal polarization tensor
\eqref{eq:GPol-L0} can give the same LO longitudinal graviton
amplitude as that of the full longitudinal polarization tensor
\eqref{eq:Gpol-L}.
The difference between the longitudinal KK graviton amplitudes
as computed by using the two types of polarization tensors
\eqref{eq:Gpol-L} and \eqref{eq:GPol-L0} belongs to the
mass-dependent residual term of the GRET\,\cite{Hang:2021fmp},
\begin{equation}
\label{eq:M_delta}
\M_\Delta^{}=\,\dM [4h_L^n]-\dMT [4h^n_{55}]
=\, \mO(E^0M_n^2)\,,
\end{equation}
where $\M \!=\hsm \M_0^{}\hsm +\hsm\dM\hs$ and
$\MT \hsm =\hsm \MT_0^{}\hsm +\hsm\dMT\hs$
under the high energy expansion.
The residual term \eqref{eq:M_delta} is derived from the
GRET identity\,\cite{Hang:2021fmp}, 
$\,\M[4h_L^n]=\MT [4h^n_{55}]+\M_\Delta^{}$,
together with the LO GRET equality \eqref{eq:GET-LO}.

\vspace*{2mm}
\subsection{\hspace*{-3mm}Constructing Inelastic Scattering Amplitudes of Four KK Gravitons}
\label{sec:4.2}

In this subsection, we study the four-point
inelastic scattering channels
$(n,\hs n)\ito (m,\hs m)\hs$ and
$(0,\hs 0)\ito (n,\hs n)\hs$
for the massive KK closed-string amplitudes \eqref{eq:closed-N4}.
We will take the low energy field theory limit
$\,\alpha'\ito\hs 0\,$
and derive the inelastic scattering amplitudes
of four-longitudinal KK gravitons.

\vspace*{1mm}

For inelastic scattering channels
$(n,\hs n)\ito (m,\hs m)\hs$,
we obtain the following extended massive
KLT-like four-point formula, 
which expresses the inelastic scattering amplitude
of longitudinal KK gravitons as the sum of products of the
corresponding color-ordered KK gauge boson scattering amplitudes:
%%%
\begin{align}
\label{eq:KLT-nnmm}
\hspace*{-6mm}
\M[1^n_L \hsx 2^n_L \hsx 3^m_L \hsx 4^m_L]
&\,=\,
\frac{\,\ka^2\,}{\,32\,}\!
\sum_{\{a_{\hsm j}^{},\hsx b_{\hsm j}^{}\}}
\! \vrhohat_{ab}^{} \LB		
%\hspace*{0mm}
s \hsx \TT_{a_{\hsm j}^{}}[1^{+n} \hs 2^{-n} \hs 3^{+m} \hs4^{-m}] \,
\TT_{b_{\hsm j}^{}}[1^{+n} \hs 2^{-n} \hs 4^{-m} \hs3^{+m}]
\right.
\nn\\[-2.5mm]
& \hspace{26.5mm} \left.
+\hsx  s \hsx \TT_{a_{\hsm j}^{}}[1^{+n} \hs 2^{-n} \hs 3^{-m} \hs4^{+m}]  \,
\TT_{b_{\hsm j}^{}}[1^{+n} \hs 2^{-n} \hs 4^{+m} \hs3^{-m}] \RB .
\end{align}
%%%%
Then, using the color-ordered inelastic KK gauge boson 
scattering amplitudes in section\,\ref{sec:3.1}
and our extended massive double-copy formula \eqref{eq:KLT-nnmm},
we construct the four-point inelastic longitudinal KK graviton 
scattering amplitude of $\hs (n,\hs n)\!\ito\hsmx (m,\hs m)\hs$, 
which takes the following compact form:
%%%
\begin{equation}
\label{eq:MLnnmm}
\M[1^n_L \hsx 2^n_L \hsx 3^m_L \hsx 4^m_L] \,=\, 
-\frac{~\ka^2 M_n^2 (X_0^{}\hsm +\! X_1^{}\ct^{}\hsm 
+\! X_2^{}\ctt^{} )
(\hs Y_0^{}\hsm +\hsm Y_2^{}\hs\ctt^{} \hsm 
+\hsm Y_4^{}\hs\ctf^{} \hsm +\hsm Y_6^{}\hs\cts^{})~}
{~1024\hs\bs\[\hsmx (\bs\hsm -\hsm 4\hs\qb\hs\qb'\ct^{})^2
\!+\!16\hs r^2\hsm\]^2\hsmx
\[\hsmx (\bs\hsm +\hsm 4\hs\qb\hs\qb'\ct^{})^2\!
 -\!16\hs r^2\hsm\]~}  \,,	
\end{equation}
%%%
where the coefficients $(X_j^{},\,Y_j)$ in the numerator are 
polynomial functions of the scattering energy and momenta,
%%%
\begin{equation}
\begin{aligned}
X_0^{}  =& \ 3\hs\bs^2 \!-\hsm 4\hs r_+^2\bs \hsm -\! 16\hs r^2 ,~~~
X_1^{} = -16\hs \qb\hs\qb'\bs \hs ,~~~
X_2^{} = 16\hs \qb^2\qqbp ,
\\[1mm]
Y_0^{}  =&\hs  -170\hs\bs^6\!-84\hs r_+^2\hs\bs^5 \!+\hsm
32\hs (39\hs r^4\!+\!135\hs r^2\!+\hsmx 39)\hs\bs
-128\hs (17\hs r^6\!+\!69\hs r^4 \!+\!69\hs r^2\!+\hsmx 17)\hs\bs^3
\\
& +512\hs (39\hs r^4\!+\!51\hs r^2\!+\!39)\hs r^2\hs\bs^2
-51200\hs (r^2\!+\hsmx 1)\hs\bs +57344\hs r^6 \hs ,
\\[1mm]
Y_2^{}  =& \  143\hs\bs^6 \!-\hsm 1804\hs r_+^2\bs^5 \!+\!
16\hs (227\hs r^4\!+\!391\hs r^2 \!+\hsm 227)\hs\bs^4
\!- 64\hs (15\hs r^6 \!-\!121\hs r^4\!-\!121\hs r^2\!+\!15)\hs\bs^3
\\
&-256\hs (61\hs r^4\!+\hsmx 345\hs r^2\!+\hsm 61)\hs r^2\hs\bs^2
\!+\hsm 95232\hs r_+^2\hs r^4\hs\bs -69632\hs r^6 \hs ,
\label{ML-nnmm-XY}
\\[1mm]
Y_4^{}  =& \ 2 \hsm\left[ 13\hs\bs^6 \!-\! 44\hs r_+^2\hs\bs^5
\!-\! 16\hs (23\hs r^4\!+\! 79\hs r^2 \!+\hsm 23)\hs \bs^4
\!+\hsm 64\hs (9\hs r^6\!+\hsm 29\hs r^4 \!+\hsm 29\hs r^2
\!+\hsm 9)\hs\bs^3
\right.
\\
& \left.
+\hs 256\hs (25\hs r^4 \!+\!53 r^2 \!+\!25)\hs r^2\hs\bs^2
\!-\hsmx 23552\hs r_+^2\hs r^4\hs\bs +\hsm 4096\hs r^6\hs \right]\!,
\\[1mm]
Y_6^{} =& \ (\bs^3\!+\!12\hs r^2\hs\bs^2\!-\!48\hs r^4\hs\bs\!-\hsm 64r^6) (\bs^3\!+\!12\hs\bs^2\! -\!48\hs\bs \!-\hsm 64)\hs ,
\end{aligned}
\end{equation}
%%%	
%%%
where $\,\bs\hsm =\hsm s/M_n^2\,$, 
$\,r\!=\!M_m^{}/M_n^{}$, and
$\,r_+^2\!=\!1\hsm + r^2$.
From the above Eq.\eqref{eq:MLnnmm},
we can further derive the LO and NLO scattering amplitudes of
the KK gravitons under high energy expansion:
\beqs
\begin{align} 
\label{eq:MLnnmm-LO}
\M_0[1^n_L \hsx 2^n_L \hsx 3^m_L \hsx 4^m_L] \,&=\, \frac{\,\ka^2\,}{\,64\,}\hs s \hs ( 7 \hsm +\hsm\ctt )
\csc^2 \hsmx \theta
\,=\, \frac{2}{\,3\,}\M_0[1^n_L \hsx 2^n_L \hsx 3^n_L \hsx 4^n_L]\,,
\\[1.5mm]
\label{eq:MLnnmm-NLO}
\dM[1^n_L \hsx 2^n_L \hsx 3^m_L \hsx 4^m_L] \,&=\,
-\frac{\,\kappa^2 M_n^2\,}{128} (1\!+\hsm r^2)
(410\hsm +\hsm 59\hs\ctt\hsmx +38\hs c_{4\theta}^{}\hsmx 
 +5\hs c_{6\theta}^{}) \csc ^4\!\theta \,.
\end{align}
\eeqs
%%%
We note that the \eqrefe{eq:MLnnmm-LO} just
equals $\hs\fr{2}{3}\hs$ times the LO elastic KK graviton amplitude
\eqref{eq:MLnnnn-LO}.
This factor $\hs\fr{2}{3}\hs$ can be understood
in a transparent way from our string-based double-copy formulation.
The reason is that the elastic scattering $\hs (n,n)\ito(n,n)\hs$
contains 6 sub-amplitudes (Fig.\,\ref{fig:1})
and the inelastic scattering $\hs (n,n)\!\to(m,m)$ includes
only 4 sub-amplitudes (Fig.\,\ref{fig:2}).
These sub-amplitude take the same LO form under the high energy limit
which is mass-independent. Hence, we deduce the connection factor
$\hs\fr{4}{6}\hsm =\hsm \fr{2}{3}\hs$ between the two LO amplitudes.

\vspace*{1mm}

Next, we analyze the mixed inelastic channel of
KK gauge boson scattering
$\hs (0,\hs 0)\!\ito\! (n,\hs n)\hs$.
We find that the condition \eqref{KK-num-sum} allows only two
combinations of the external KK states
(as originally defined under the $\SS^1$ compactification of 26d),
$\{0,\hs 0,+n,-n\}$ and $\{0,\hs 0,-n,+n\}$.
Thus, from Eq.\eqref{eq:closed-N4},
we derive the following extended four-point massive KK double-copy
formula which expresses the inelastic amplitude
of longitudinal KK gravitons as the sum of products of the
corresponding color-ordered KK gauge boson amplitudes:
%%%
\begin{align}
\label{eq:KLT-00nn}
\M[1^0_T \hsx 2^0_T \hsx 3^n_L \hsx 4^n_L]
\,=\,
\frac{\,\ka^2\,}{\,32\,}
\sum_{\{a_{\hsm j}^{},\hsx b_{\hsm j}^{}\}}\!
\vrhohat_{ab}^{} \hsx
s \hsx
\TT_{a_{\hsm j}^{}}^{}[1^{0} \hs 2^{0} \hs 3^{+n} \hs4^{-n}] \,
\TT_{b_{\hsm j}^{}}^{}[1^{0} \hs 2^{0} \hs 4^{-n} \hs3^{+n}]  \,,
\end{align}
%%%
where in the amplitude $\M[1^0_T \hs 2^0_T \hs 3^n_L \hs 4^n_L]$
the subscript $\,T\!=\!\pm 2\,$ denotes the helicities of the
massless zero-mode gravitons in the initial state.

\vspace*{1mm}

With the above and using our extended massive double-copy
formula \eqref{eq:KLT-00nn}, we construct the mixed four-point inelastic
graviton amplitudes
of $\hs (0,\hs 0)\hsmx\ito (n,\hs n)\hs$ and derive following form:
%%%%%
\begin{equation}
\label{eq:M00nn}
\M[1^0_{\pm 2} \hsx 2^0_{\mp 2} \hsx 3^n_L \hsx 4^n_L] \,=\,
\frac{~\ka^2\hsm M_n^2\hs (\bs^2\!+\!16\hs\bs\hsm +\! 16)
\hs s^4_\theta~}{16\hs
[(\bs\hsm +\hsm 4)\hsmx -\hsmx (\bs \hsm -\hsm 4)\ctt]}\,.
\end{equation}
%%%
We can further derive the following LO and NLO amplitudes
from Eq.\eqref{eq:M00nn} under high energy expansion:
\\[-9mm]
\beqs
\begin{align}
\M_0[1^0_{\pm 2} \hsx 2^0_{\mp 2} \hsx 3^n_L \hsx 4^n_L] \,&=\, \frac{\,\ka^2 }{~32~}s \hs \st^2  \,,
\\[2mm]
\dM[1^0_{\pm 2} \hsx 2^0_{\mp 2} \hsx 3^n_L \hsx 4^n_L]
&=\,  \frac{~\ka^2\Mn^2\,}{16}  (3\hsm -\hsm 5\hs\ctt^{}) \,.
\end{align}
\eeqs
%%%%%

\vspace*{1mm} 
\subsection{\hspace*{-2mm}Constructing Multi-Point Scattering Amplitudes of KK Gravitons}
\label{sec:4.3}

Our above analyses of the four-point scattering amplitudes can be further
extended to the $N$-point amplitudes with $\hs N\!\geqq\! 5\hsx$.\ 
We first consider a five-point inelastic scattering process 
$(2n,n)\ito (n,n,n)$, where all the external KK states
are set to be $\ZZ\hsmx$ even.
We find that the condition \eqref{KK-num-sum} of KK number conservation
allows the following combinations of the KK numbers of the 
external-state gauge bosons for the sub-amplitudes: 
%%%%
\begin{align}
\label{eq:KK-num-5-point}
&\{+2n,+n,-n,-n,-n\} \hs , \hspace{2mm}
\{+2n,-n,+n,-n,-n\} \hs , \hspace{2mm}
\{+2n,-n,-n,+n,-n\} \hs ,  
\nn\\
& \{+2n,-n,-n,-n,+n\} \hs , \hspace{2mm}
\{\rm{all \ permutations \ of\  \( +,-\)}\} \,,
\end{align}
%%%
where we assign the KK number of particle-1 as $+2n$
for convenience.

\vspace*{1mm}

Thus, using the general formula \eqref{eq:Amp-N-Gr} of the 
$N$-point KK graviton amplitudes, 
we derive the five-point longitudinal KK graviton 
scattering amplitude as follows:
%%%
%\begin{equation}
\begin{align}
\hspace*{-5mm}
\M[1^{2n}_L\hs2^{n}_L\hs3^{n}_L\hs4^{n}_L\hs5^n_L]
=  
\frac{\ka^3}{\,128\sqrt{2\,}\,}\!
\sum_{\{a_{\hsm j}, b_{\hsm j}\}}  
& \vrhohat_{ab}\,\Big\{
\hspace*{2mm} 
\nn\\
\hspace*{-10mm}
\big\{ (s_{12}\!-\!9\Mn^2)(s_{13}\!-\!\Mn^2)
& \hs\TT_{a_{\hsm j}}[1^{+2n} 2^{+n} 3^{-n}4^{-n}5^{-n}] \hsx 
\TT_{b_{\hsm j}}[1^{+2n} 3^{-n}2^{+n} 5^{-n}4^{-n}]
\nn\\
\hspace*{-10mm}
+(s_{12}\!+\!s_{23}\!-\!9\Mn^2)(s_{13}\!-\!\Mn^2)
& \hs\TT_{a_{\hsm j}}[1^{+2n} 2^{+n} 3^{-n}4^{-n}5^{-n}] \hsx 
\TT_{b_{\hsm j}}[1^{+2n} 2^{+n} 3^{-n}5^{-n}4^{-n}] \big\}
\nn\\
\hspace*{-10mm}
+\big\{ (s_{12}\!-\!\Mn^2)(s_{13}\!-\!9\Mn^2)
& \hs\TT_{a_{\hsm j}}[1^{+2n} 2^{-n}3^{+n} 4^{-n}5^{-n}] \hsx 
\TT_{b_{\hsm j}}[1^{+2n} 3^{+n} 2^{-n}5^{-n}4^{-n}]
\nn\\
\hspace*{-10mm}
+(s_{12}\!+\!s_{23}\!-\!\Mn^2)(s_{13}\!-\!9\Mn^2)
& \hs\TT_{a_{\hsm j}}[1^{+2n} 2^{-n}3^{+n} 4^{-n}5^{-n}] \hsx 
\TT_{b_{\hsm j}}[1^{+2n} 2^{-n}3^{+n} 5^{-n}4^{-n}] \big\}
\label{eq:Amp-5pt}
\\
\hspace*{-10mm}
+\big\{ (s_{12}\!-\!\Mn^2)(s_{13}\!-\!\Mn^2)
& \hs\TT_{a_{\hsm j}}[1^{+2n} 2^{-n}3^{-n}4^{+n} 5^{-n}] \hsx 
\TT_{b_{\hsm j}}[1^{+2n} 3^{-n}2^{-n}5^{-n}4^{+n}]
\nn\\
\hspace*{-10mm}
+(s_{12}\!+\!s_{23}\!-\!5\Mn^2)(s_{13}\!-\!\Mn^2)
& \hs\TT_{a_{\hsm j}}[1^{+2n} 2^{-n}3^{-n}4^{+n} 5^{-n}] \hsx 
\TT_{b_{\hsm j}}[1^{+2n} 2^{-n}3^{-n}5^{-n}4^{+n}] \big\}
\nn\\
\hspace*{-10mm}
+\big\{(s_{12}\!-\!\Mn^2)(s_{13}\!-\!\Mn^2)
& \hs\TT_{a_{\hsm j}}[1^{+2n} 2^{-n}3^{-n}4^{-n}5^{+n}] \hsx 
\TT_{b_{\hsm j}}[1^{+2n} 3^{-n}2^{-n}5^{+n} 4^{-n}]
\nn\\
\hspace*{-10mm}
+(s_{12}\!+\!s_{23}\!-\!5\Mn^2)(s_{13}\!-\!\Mn^2)
& \hs\TT_{a_{\hsm j}}[1^{+2n} 2^{-n}3^{-n}4^{-n}5^{+n}] \hsx 
\TT_{b_{\hsm j}}[1^{+2n} 2^{-n}3^{-n}5^{+n} 4^{-n}] 
\big\}
\nn\\
+ \, \rm{all~permutations~of}~ (+n,-n) \Big\} \,.  
\hspace*{-10.5mm} &\nn
\end{align}
%\end{equation}
%%%

Then, we consider the six-point scattering process
with all the external states being $\ZZ$ even.
The KK numbers of the six external states are chosen as   
$(\hat{n}_1^{},\hat{n}_2^{},\hat{n}_3^{},\hat{n}_4^{}, 
\hat{n}_5^{}, \hat{n}_6^{}) \hsm\!=\!(n,n,n,n,n,n)\hs$.
From the condition \eqref{KK-num-sum} of the KK number conservation,
we deduce the following allowed combinations of the KK numbers of the 
external-state gauge bosons of the sub-amplitudes: 
%%%
\begin{equation}
\begin{aligned}
&\{+n,\hs+n,\hs+n,\hs-n,\hs-n,\hs-n\} \hs , \hspace{2mm}
\{+n,\hs+n,\hs-n,\hs+n,\hs-n,\hs-n\} \hs , \hspace{2mm}
\\
&\{+n,\hs+n,\hs-n,\hs-n,\hs+n,\hs-n\} \hs , \hspace{2mm}
\{+n,\hs+n,\hs-n,\hs-n,\hs-n,\hs+n\} \hs , \hspace{2mm}
\\
&\{+n,\hs-n,\hs+n,\hs+n,\hs-n,\hs-n\} \hs , \hspace{2mm}
\{+n,\hs-n,\hs+n,\hs-n,\hs+n,\hs-n\} \hs , \hspace{2mm}
\\
&\{+n,\hs-n,\hs+n,\hs-n,\hs-n,\hs+n\} \hs , \hspace{2mm}
\{+n,\hs-n,\hs-n,\hs+n,\hs+n,\hs-n\} \hs , \hspace{2mm}
\\
&\{+n,\hs-n,\hs-n,\hs+n,\hs-n,\hs+n\} \hs , \hspace{2mm}
\{+n,\hs-n,\hs-n,\hs-n,\hs+n,\hs+n\} \hs , \hspace{2mm}
\\
& \{\rm{all \ permutations \ of\  \( +,-\)}\} \hs.
\end{aligned}
\end{equation}
%%%
Using the above decomposition and Eq.\eqref{eq:Amp-N-Gr} of the 
general $N$-point KK graviton amplitudes, 
we obtain the six-point longitudinal KK graviton 
scattering amplitude as follows:
%%%
\begin{align}
\M[1^{n}_L\hs2^{n}_L\hs3^{n}_L\hs4^{n}_L\hs5^n_L\hs6^n_L]
= \frac{\ka^4}{\,1024\,}\!
\sum_{\{a_{\hsm j}, b_{\hsm j}\}}  
& \vrhohat_{ab}\,\Big\{
\hspace*{2.5mm} 
\nn\\
\hspace*{-5mm}
s_{14}^{}\big\{(s_{12}^{}\!-\!4\Mn^2)(s_{13}^{}\!-\!4\Mn^2)
&\hs\TT_{a_{\hsm j}}[1^{+n}2^{+n}3^{+n}4^{-n}5^{-n}6^{-n}]\hs 
\TT_{b_{\hsm j}}[1^{+n}4^{-n}3^{+n}2^{+n}6^{-n}5^{-n}]
\nn\\
\hspace*{-5mm}
+(s_{12}^{}\!-\!4\Mn^2)(s_{13}^{}\!+\!s_{34}^{}\!-\!4\Mn^2)
&\hs\TT_{a_{\hsm j}}[1^{+n}2^{+n}3^{+n}4^{-n}5^{-n}6^{-n}]\hs 
\TT_{b_{\hsm j}}[1^{+n}3^{+n}4^{-n}2^{+n}6^{-n}5^{-n}]
\nn\\
\hspace*{-5mm}
+(s_{13}^{}\!-\!4\Mn^2)(s_{12}^{}\!+\!s_{23}^{}\!-\!8\Mn^2)
&\hs\TT_{a_{\hsm j}}[1^{+n}2^{+n}3^{+n}4^{-n}5^{-n}6^{-n}]\hs 
\TT_{b_{\hsm j}}[1^{+n}4^{-n}2^{+n}3^{+n}6^{-n}5^{-n}]
\nn\\
\hspace*{-3mm}
{-(s_{25}^{}\!+\!s_{26}^{})(s_{13}^{}\!-\!4\Mn^2)}
&\hs\TT_{a_{\hsm j}}[1^{+n}2^{+n}3^{+n}4^{-n}5^{-n}6^{-n}]\hs 
\TT_{b_{\hsm j}}[1^{+n}2^{+n}4^{-n}3^{+n}6^{-n}5^{-n}]
\label{eq:Amp-6pt}
\\
\hspace*{-3mm}
+(s_{12}^{}\!+\!s_{24}^{}\!-\!4\Mn^2)
(s_{13}^{}\!+\!s_{34}^{}\!-\!4\Mn^2)
&\hs\TT_{a_{\hsm j}}[1^{+n}2^{+n}3^{+n}4^{-n}5^{-n}6^{-n}]\hs 
\TT_{b_{\hsm j}}[1^{+n}3^{+n}2^{+n}4^{-n}6^{-n}5^{-n}]
\nn\\
\hspace*{-3mm}
{-(s_{25}^{}\!+\!s_{26}^{})(s_{13}^{}\!+\!s_{34}^{}\!-\!4\Mn^2)}
&\hs\TT_{a_{\hsm j}}[1^{+n}2^{+n}3^{+n}4^{-n}5^{-n}6^{-n}]\hs 
\TT_{b_{\hsm j}}[1^{+n}2^{+n}3^{+n}4^{-n}6^{-n}5^{-n}] \big\}
\nn\\[-0.5mm]
\hspace*{-3mm}
+ \, \rm{all \ permutations \ of} \ (+n,-n)   \Big\} \,.
\hspace*{-0mm} &
\nn
\end{align}
The above five-point and six-point KK gauge boson
amplitudes and KK graviton amplitudes are worth of further systematic 
studies and we will pursue these in the future works.

\vspace*{2mm} 
\section{\hspace*{-3mm}Conclusions}
\label{sec:5}

The Kaluza-Klein (KK) compactification\,\cite{KK} 
of higher dimensional spacetime
is a fundamental ingredient of the major directions for 
new physics beyond the standard model (SM), including the string/M
theories\,\cite{string} and extra dimensional field theories 
with large or small extra dimensions\,\cite{Exd}.\  
Studying the double-copy construction
of graviton scattering amplitudes from gauge boson scattering amplitudes has pointed to profound deep connections 
between the gauge forces and gravitational force in nature.

\vspace*{1mm}

So far substantial efforts have been made to formulate and test
the double-copy constructions between the massless gauge theories and
massless general relativity (GR)\,\cite{BCJ-Rev}.
But the extensions of conventional double-copy method to 
massive gauge/gravity theories are generally difficult,
because most of such theories (including the massive Yang-Mills 
theory and massive Fierz-Pauli gravity) violate explicitly the 
gauge symmetry and diffeomorphism invariance (which are the key for successful double-copy construction).\  
The two important candidates with promise include the compactified KK 
gauge/gravity theories and the topologically massive Chern-Simons 
gauge/gravity theories. 
The extended BCJ-type double-copy construction for realistic 
massive KK gauge/gravity theories was found\,\cite{Hang:2021fmp} 
to be highly nontrivial even for the four-point KK scattering 
amplitudes at tree level, and proper modifications of the 
conventional BCJ method are generally needed 
for the KK scattering amplitudes
at the next-to-leading order (NLO) of the high energy expansion\,\cite{Hang:2021fmp}.

\vspace*{1mm}  

In this work, we studied the scattering amplitudes of 
massive KK states of open and closed bosonic strings 
under toroidal compactification.\ 
The essential advantage of the compactified KK string theory 
is that {\it the connection between the KK closed-string amplitudes 
and the proper products of KK open-string amplitudes 
can be intrinsically built in from the start.}\
For the present study, we take the bosonic string theory as a
computational tool for establishing the massive KLT-like relations 
of KK string states and for deriving the low energy KK graviton 
scattering amplitudes. 

\vspace*{1mm} 

In section\,\ref{sec:2.1}, we set up the toroidal compactification
for the 26d bosonic string theory where the 21 of the extra spatial 
dimensions have very small compactification radii of $\mO(\MPl^{-1})$
and get decoupled in our effective string theory below the Planck scale.
Thus, we only deal with the KK strings in a single compactified 25th 
spatial dimension under $\hs\SS^1$ with relatively larger radius $R\hsx$. 
With these, in section\,\ref{sec:2.2} 
we studied vertex operators of the KK open and closed strings.
In particular, these include
a class of vertex operators having $\ZZ$-even parity 
whose scattering amplitudes will give, in the field-theory limit, the corresponding scattering amplitudes of the KK gauge bosons 
and of the KK gravitons in the KK gauge/gravty theories under the 
5d compactification of $\hs\SS^1\!/\ZZ\hs$.\ 
Then, we formulated the $N$-point scattering amplitudes 
\eqref{eq:close-amp} of the KK open and closed strings 
in section\,\ref{sec:2.3},
and further derived the corresponding scattering amplitudes 
\eqref{eq:Amp-N-Gr} 
of the KK gauge bosons and gravitons in the low energy field-theory 
limit in section\,\ref{sec:2.4}.\ 
We observed that any KK amplitude with external states being $\ZZ$
even (odd) should be decomposed into a sum of relevant sub-amplitudes
whose external states have KK numbers obey the conservation condition
\eqref{KK-num-sum}.  

\vspace*{1mm}  

In section\,\ref{sec:3}, using the formulas of section\,\ref{sec:2}
we computed explicitly the four-point color-ordered scattering 
amplitudes of KK open strings and derived the corresponding gauge boson
scattering amplitudes in the field theory limit.\
The four-point elastic KK gauge boson scattering amplitudes 
are given in Eqs.\eqref{eq:T1234-KK}-\eqref{eq:Tnnnn-Pn}
of section\,\ref{sec:3.1}, 
while the four-point inelastic KK gauge boson scattering amplitudes 
are given in Eqs.\eqref{eq:Tnnmm-L1234}-\eqref{eq:Tnnmm-L1243}
and Eq.\eqref{eq:T00nn-LL} of section\,\ref{sec:3.2}. 
Then, in section\,\ref{sec:3.3}, 
we analyzed the structure of the color-ordered 
scattering amplitudes of massive KK gauge bosons.\
We demonstrated that the tree-level massive KK gauge boson amplitudes 
can be obtained from the corresponding 
color-ordered amplitudes of the massless zero-mode gauge bosons
by making proper shifts of the Mandelstam variables.\ 
This serves as an elegant and efficient method to compute 
any color-ordered massive KK gauge boson amplitudes.

In section\,\ref{sec:4}, we applied the massive KLT-like
relation \eqref{eq:closed-N4} of four-point KK string amplitudes
by taking the field theory limit,
and derived the double-copy formulas
\eqref{eq:KLT-nnnn}, \eqref{eq:KLT-nnmm} and \eqref{eq:KLT-00nn} 
for constructing the four-point 
KK graviton scattering amplitudes.
These give an explicit prescription 
on how to construct the exact four-point KK graviton 
amplitudes from the sum of relevant products of the corresponding color-ordered KK gauge boson amplitudes.\
With these, we computed the exact tree-level four-point 
elastic KK graviton scattering amplitudes \eqref{eq:MLnnnn}-\eqref{eq:4hL-Pn}
in section\,\ref{sec:4.1},
and the exact tree-level four-point 
inelastic KK graviton scattering amplitudes
\eqref{eq:MLnnmm}-\eqref{ML-nnmm-XY} and
\eqref{eq:M00nn} in section\,\ref{sec:4.2}.
Finally, in section\,\ref{sec:4.3},
we used our general string-based double-copy construction 
formula \eqref{eq:Amp-N-Gr} to obtain 
the five-point and six-point scattering amplitudes of massive 
KK gravitons as given by Eq.\eqref{eq:Amp-5pt} and 
Eq.\eqref{eq:Amp-6pt}.

\newpage
%\vspace*{10mm}
\noindent
{\bf\large Acknowledgments:}\\
We thank Henry Tye for valuable discussions.\ 
We also thank Ziqi Yan for discussing Ref.\,\cite{wKLT}. 
The works of HJH,\ YFH and YL were supported in part 
by National Natural Science Foundation 
of China (under grants Nos.\,11835005, 12175136), 
by National Key R\,\&\,D Program of China 
(under grant No.\,2017YFA0402204),
and by the CAS Center for Excellence in Particle Physics (CCEPP).
The research of SH was supported in part by National Natural Science Foundation of China (under Grant Nos.\,11935013, 11947301, 12047502, 12047503).

\vspace*{10mm}

%\newpage
\appendix

\noindent
{\large\bf Appendix:}
%\vspace*{7mm}

\section{\hspace*{-2mm}Kinematics of Four-Point Scattering Amplitudes
of KK States}
\label{app:A}

In this Appendix we present the kinematics of four-particle scattering
processes of KK states in the (3+1)d spacetime. 
We choose the Minkowski metric tensor 
$\,\eta^{\mn} \!=\! \eta_{\mn}^{} \!=\! \diag(-1,1,1,1)$.

\vspace*{1mm}

For the four-particle elastic scattering of KK states 
$(n,\hs n)\ito (n,\hs n)$, 
we define the 4-momenta of the external KK states
in the center-of-mass frame as follows:
%%%
\begin{equation}
\begin{alignedat}{3}
\label{eq:Momenta}
p_1^\mu & =  -E\hs (1, 0, 0, \be)  \hs ,  \hspace*{3mm}
& \hspace*{5mm}
& p_2^\mu = -E\hs (1, 0, 0, -\be)  \hs ,
\\[1mm]
p_3^\mu &= E\hs ( 1, \be\st, 0, \be\ct) \hs ,
& \hspace*{8mm}
& p_4^\mu = E\hs ( 1, -\be\st, 0, -\be\ct) \hs ,
\end{alignedat}
\end{equation}
%%%
where $(\st,\ct)=(\sin\hsm\theta ,\hs \cos\hsm\theta)$ and
$\hs\be\!=\!(1\!-\!\Mn^2/\hsm E^2)^{1/2}\hs$.
With the above, we define the three Mandelstam variables:
%%%%
\begin{equation}
\begin{aligned}
\label{eq:s-t-u}
s &=-\( p_{1} \!+\hsm p_{2} \)^{2} \!= 4E^2 \,, \quad
\\[1mm]
t &=-\( p_{1} \!+\hsm p_4 \)^{2} \!= -\frac{1}{2}s\hs\be^2
(1\!+\!\ct) \,, \quad
\\[1mm]
u &=-\( p_{1}^{} \!+\hsm p_3^{} \)^{2} \!=
-\frac{1}{2}s\hs\be^2 (1\!-\!\ct) \,,
\end{aligned}
\end{equation}
where  $\hs\be\!=\!(1\!-\!\Mn^2/\hsm E^2)^{1/2}\hs$.\
With the on-shell condition
$\,E^2 \!=\! E^2\be^2 \!+\! \Mn^2\,$, we may
define the following set of Mandelstam variables:
\\[-6mm]
%%%%%%
\begin{align}
\label{eq:s0-t0-u0}
\sz \,=\, 4E^2\be^2   \,,  \quad~~
\tz \,=\, -\frac{1}{2}\sz(1\!+\!\ct)  \,, \quad~~
\uz \,=\, -\frac{1}{2}\sz(1\!-\!\ct)  \,,
\end{align}
where we have introduced the notation 
$\hs\sz\hsm =\hsm s\hsm -\hsm 4\Mn^2$\,,
and thus $(\sz,\,\tz,\,\uz)\hsm =\hsm (s\be^2\hsm ,\,t,\,u)$.
Summing up the Mandelstam variables \eqref{eq:s-t-u}
and \eqref{eq:s0-t0-u0} gives the following relations: 
\begin{equation}
s+t+u =4 \Mnn\,, \hspace*{8mm}
\sz +\tz +\uz =0 \,.
\end{equation}
Then, we define the following polarization vectors 
for the external KK gauge bosons in 
the center-of-mass frame:
%%%
\begin{equation}
\begin{alignedat}{3}
\label{eq:Pol}
\zeta^\mu_{1,\hs\pm1} &=\frac{1}{\sqrt{2}}\(0,1, \pm \ii,0\)\!,
\hspace*{8mm}
&& \zeta^\mu_{1,L} =  \frac{E}{\Mn}\(\be,0,0,1\)\!,
\\
\zeta^\mu_{2,\hs\pm1} &=\frac{1}{\sqrt{2}}\(0,1, \mp \ii,0\)\!,
\hspace*{8mm}
&&\zeta^\mu_{2, L} = \frac{E}{\Mn}\(\be,0,0,-1\) \!,
\\
\zeta^\mu_{3,\hs\pm1} &=\frac{1}{\sqrt{2}}\(0, \mp \ii\hs\ct,1, \pm \ii\hs\st\)\!,
\hspace*{8mm}
&&\zeta^\mu_{3,L} = \frac{E}{\Mn} \(\be, \st, 0, \ct\) ,
\\
\zeta^\mu_{4,\hs\pm1} &=\frac{1}{\sqrt{2}}\(0, \pm \ii\hs\ct,1, \mp \ii\hs\st\)\!,
\hspace*{8mm}
&&\zeta^\mu_{4,L} = \frac{E}{\Mn} \(\be, -\st, 0, -\ct \) \!.
\end{alignedat}
\end{equation}
%%%

\hspace*{1mm}

Next, we consider the inelastic KK scattering process of
$\hs (n,\hs n) \!\ito\! (m,\hs m)\hs$.\ 
Thus, the 4-momenta of the external states 
in the center-of-mass frame can be defined as follows:
%%%
\begin{equation}
\begin{alignedat}{3}
\label{eq:Momenta-2}
p_1^\mu &=  -E(1, 0, 0, \be)  \hs ,  \hspace*{3mm}
& \hspace*{7mm}
&p_2^\mu = -E(1, 0, 0, -\be)  \hs ,
\\[1mm]
p_3^\mu &= E ( 1, \be'\hsm\st, 0, \be'\hsm\ct) \hs ,
& \hspace*{7mm}
&p_4^\mu = E ( 1, -\be'\hsm\st, 0, -\be'\hsm\ct) \hs ,
\end{alignedat}
\end{equation}
%%%
where $\hs\be\!=\!(1\!-\!\Mn^2/\hsm E^2)^{1/2}\hs$ and 
$\hs\be'\!=\!(1\!-\!M_m^2/\hsm E^2)^{1/2}$. 
With these, we can define the Mandelstam variables:
%%%
\begin{equation}
\begin{aligned}
s &=-\( p_{1} \!+\hsm p_{2} \)^{2} \!= 4E^2 , \qquad
\\[1mm]
t &=-\( p_{1} \!+\hsm p_4 \)^{2} \!= 
-\frac{1}{4}s \hs (\be^2 \!+\! \be'^2  \!+\!  2\be\be' \ct) \hs , 
\\[1mm]
u &=-\( p_{1}^{} \!+\hsm p_3^{} \)^{2} \!=
-\frac{1}{4}s \hs (\be^2 \!+\! \be'^2  \!-\!  2\be\be' \ct) \hs ,
\end{aligned}
\end{equation}
%%%
from which we deduce
$\,s + t + u\hsm =2(M_n^2\hsm +\! M_m^2)\hs$.\ 
The corresponding longitudinal polarization vectors of the
KK gauge bosons are given by
%%%
\begin{equation}
\begin{alignedat}{3}
\label{eq:Polnnmm}
\zeta^\mu_{1,L} &=  \frac{E}{\Mn}({\be},0,0,1)\hs ,
 && \zeta^\mu_{2, L} = \frac{E}{\Mn}({\be},0,0,-1)\hs , \quad
\\[1mm]
\zeta^\mu_{3,L} &= \frac{E}{M_m}{({\be'}, \st, 0, \ct)}\hs , 
\quad~~
&& \zeta^\mu_{4,L} = \frac{E}{M_m}({\be'}, -\st, 0, -\ct )\hs ,
\end{alignedat}
\end{equation}
%%%
while their transverse polarization vectors are mass-independent
and remain the same as in Eq.\eqref{eq:Pol}. 
For another inelastic scattering channel
$\hs (0,\hs 0) \ito (m,\hs m)\hs$, we just set the
initial state masses to be zero ($M_n^{}\!\ito 0$) in the above
setup and only allow transverse polarizations for the
massless initial state gauge bosons.

\vspace*{1mm}
\section{\hspace*{-2mm}Full Scattering Amplitudes of KK Gauge 
and Goldstone Bosons}
\label{app:B}

In this Appendix, for the sake of comparison we present systematically 
the four-point elastic and inelastic scattering amplitudes of 
KK gauge bosons in the 5d KK Yang-Mills gauge theories under the orbifold compactification of $\,\SS^1\!/\ZZ\hs$.

\vspace*{1.5mm} 
\subsection{\hspace*{-2mm}Elastic KK Gauge and Goldstone Boson Scattering Amplitudes}
\vspace*{1mm} 
\label{app:B1}

According to Ref.\,\cite{Hang:2021fmp}, 
we summarize the four-point elastic
scattering amplitudes of longitudinal KK gauge bosons and 
of KK Goldstone bosons as follows:
\beqs
\label{eq:Amp-4AL-4A5}
\begin{align}
\TT[A_L^{a\hs n}A_L^{b\hs n}\hsm\ito A_L^{c\hs n}A_L^{d\hs n}]
\,=~\,& g^2 (\CC_s \KK_s^{\rm{el}} + \CC_t \KK_t^{\rm{el}}  + \CC_u \KK_u^{\rm{el}} )\,,
\\[1.mm]
\tT[A_5^{a\hs n}A_5^{b\hs n}\hsm\ito A_5^{c\hs n}A_5^{d\hs n}]
\,=~\,& g^2 (\CC_s \KKt_s^{\rm{el}} + \CC_t \KKt_t^{\rm{el}} + \CC_u \KKt_u^{\rm{el}})\,,
\end{align}
\eeqs
where $\{\KK_j^{\rm{el}}\}$ denote the kinematic factors for KK gauge bosons,
\beqs
\label{eq:K-KT-exact}
\begin{align}
\label{eq:Ks-exact}
\KK_s^{\rm{el}} &\dis =
-\frac{\,(4\bs^2 \!-\! 5\bs \!-\! 8)\ct\,}{2\bs} \,,
\\[1.5mm]
\label{eq:Kt-exact}
\KK_t^{\rm{el}} &\dis =
-\frac{\,Q_0^{}\!+ Q_1^{}\ct\! + Q_2^{}\ctt\!+Q_3^{}c_{3\theta}^{}\,}
{\,2(\bs \!-\! 4)[(3\bs\!+\!4)\!+\!4\bs\hs\ct\!+\!(\bs\!-\!4)\ctt] \,} \,,
\\[1.5mm]
\label{eq:Ku-exact}
\KK_u^{\rm{el}} &\dis =
\frac{\,Q_0^{}\!- Q_1^{}\ct\! + Q_2^{}\ctt\!-Q_3^{}c_{3\theta}^{}\,}
{\,2(\bs \!-\! 4)[(3\bs\!+\!4)\!-\!4\bs\hs\ct\!+\!(\bs\!-\!4)\ctt] \,} \,,
\end{align}
\eeqs
and $\{\KKt_j^{\rm{el}}\}$ denote the kinematic factors fro KK Goldstone bosons,
\beqs
%\label{eq:K-KT-exact}
\begin{align}
\label{eq:Ks-exact}
\KKt_s^{\rm{el}} &\dis = 
-\frac{\,(3\bs \!-\! 8)\ct\,}{2\bs} \,,
\\[1mm]
\label{eq:Kt-exact}
\KKt_t^{\rm{el}} &\dis =
\frac{\,\Qt_0^{}\!+ \Qt_1^{}\ct\! + \Qt_2^{}\ctt\,}
{\,2(\bs \!-\! 4)[(3\bs\!+\!4)\!+\!4\bs\hs\ct\!+\!(\bs\!-\!4)\ctt] \,} \,,
\\[1.5mm]
\label{eq:Ku-exact}
\KKt_u^{\rm{el}} &\dis =
-\frac{\,\Qt_0^{}\!- \Qt_1^{}\ct\! + \Qt_2^{}\ctt\,}
{\,2(\bs \!-\! 4)[(3\bs\!+\!4)\!-\!4\bs\hs\ct\!+\!(\bs\!-\!4)\ctt]\,} \,,
\end{align}
\eeqs
with the functions $\{Q_j^{},\Qt_j^{}\}$ expressed as 
\begin{equation}
\begin{alignedat}{3}
& Q_0^{} =  8\bs^3 \!- \! 63\bs^2  \!+ \! 72\bs  \!+ \! 80 \,, \hspace*{8mm}
&&\Qt_0^{}= 15\bs^2 \!+\! 24\bs \!-\! 80\,,
\\
&Q_1^{} = 2 (7\bs^3  \!- \! 44 \bs^2  \!+ \! 80\bs  \!- \! 64) \,,
\hspace*{8mm}
&&\Qt_1^{} = 4(3\bs^2 \!-\! 20\bs \!+\! 32)\,,
\\
& Q_2^{} = 8 \bs^3  \!- \! 45\bs^2  \!+ \! 8\bs  \!+ \! 48 \,,
\hspace*{8mm}
&&\Qt_2^{} = -3(\bs\!-\!4)^2  \,,
\\
&Q_3^{} = 2 \bs(\bs^2  \!- \! 10\bs  \!+ \! 24)  \,.
\end{alignedat}
\end{equation}
Making the high energy expansion of $\,1/s\hs$, 
we derive the following LO scattering amplitudes:
%%%
\beqs
\begin{alignat}{3}
\KK_s^{{\rm{el}}\hs0} &= \frac{\,5\ct\,}{2} \,,
\hspace*{12mm}
& \KKt_s^{{\rm{el}}\hs0}  &= -\frac{\,3\ct\,}{2} \,,
\\[1mm]
\KK_t^{{\rm{el}}\hs0}  &=\frac{\,13 \!+\! 5 \ct \!+\! 4 \ctt\,}{2(1 \!+\! \ct)\,}  \,, \hspace*{12mm}
&\KKt_t^{{\rm{el}}\hs0} &=\frac{\,3(3 \!-\! \ct)\,}{2(1 \!+\! \ct)\,}  \,,
\\[1mm]
\KK_u^{{\rm{el}}\hs0}  &=-\frac{\,13 \!-\! 5 \ct \!+\! 4 \ctt\,}{2(1 \!-\! \ct)\,} \,, \hspace*{12mm}
&\KKt_u^{{\rm{el}}\hs0}  &=-\frac{\,3(3\!+\!\ct)\,}{2(1 \!-\!\ct)\,} \,,
\end{alignat}
\eeqs
%%%
and the NLO scattering amplitudes:
%%%
\beqs
\begin{alignat}{3}
\dKK_s^{\rm{el}} &= \frac{4\ct}{\bs} \,, \hspace*{12mm}
&\dKKt_s^{\rm{el}} & =\frac{4\ct}{\bs} \,,
\\[1mm]
\dKK_t^{\rm{el}} &= -\frac{\,8 (2\!-\!3\ct\!-\!2\ctt\!-\!\cttt)\,}{(3\!+\!4\ct \!+\! \ctt) \bs\,} \,, \hspace*{12mm}
&\dKKt_t^{\rm{el}}&= \frac{32\ct}{(3\!+\!4\ct \!+\! \ctt) \bs} \,,
\\[1mm]
\dKK_u^{\rm{el}} &= \frac{\,8 (2\!+\!3\ct\!-\!2\ctt\!+\!\cttt)\,}{(3\!-\!4\ct \!+\! \ctt) \bs\,}
\,, \hspace*{12mm}
&\dKKt_u^{\rm{el}}&= \frac{32\ct}{(3\!-\!4\ct \!+\! \ctt) \bs} \,.
\end{alignat}
\eeqs
Note that the above expansion of $\hs 1/s\hs$ differs from the
expansion of $\hs 1/\sz\hs$ [cf.\ Eq.\eqref{eq:s0-t0-u0}]
as adopted in Ref.\,\cite{Hang:2021fmp}.
We also note that in each channel of $(s,t,u)$
the LO longitudinal KK gauge boson amplitude differs from 
the LO KK Goldstone boson amplitude by the same amount:\ 
$\hs\KK_j^{{\rm{el}}\hs0} -\hs\KKt_j^{{\rm{el}}\hs0}\! = 4\hs\ct\,$.
Hence, due to the Jacobi identity
the elastic KK longitudinal gauge boson amplitude and KK Goldstone boson amplitude
are equal at the LO, 
$\,\TT_{0L}^{}[4A_L^{a\hs n}]\!=\!
 \tT_{05}^{}[4A_5^{a\hs n}]\hs$, 
which verifies the KK gauge boson equivalence theorem 
(KK\,GAET)\,\cite{5DYM2002}\cite{KK-ET-He2004}.\footnote{%
The four-point KK gauge boson scattering amplitudes were also
computed\,\cite{5dSM} for the 5d SM 
under the orbifold compactification of $\hs\SS^1\!/\ZZ\hs$.}

\vspace*{1mm}

Then, we further define the BCJ-type numerators:
\beqs
\begin{alignat}{3}
& \NN_j^{\rm{el}} =
s_j \hs \KK_j^{\rm{el}}  \hs,
\hspace*{10mm}
&& \NN_j^{\rm{el}} = \NN_j^{\rm{el}\hs0}\!+\da\NN_j^{\rm{el}}
= s_j (\KK_j^{\rm{el}\hs 0}\!+\da\KK_j^{\rm{el}})\hs,
\\[1mm]
& \NNt_j^{\rm{el}} =
s_j \hs \KKt_j^{\rm{el}}  \hs,
\hspace*{10mm}
&& \NNt_j^{\rm{el}} = \NNt_j^{\rm{el}\hs 0}\!+\da\NNt_j^{\rm{el}}
= s_j (\KKt_j^{\rm{el}\hs0}\!+\da\KKt_j^{\rm{el}}) \hs,
\end{alignat}
\eeqs
%%%
where $\,j\!\in\!(s,\hs t,\hs u)\,$, and 
we have decomposed the numerators
$\{\NN_j^{\rm{el}}\hsm ,\,\NNt_j^{\rm{el}}\}$ 
into the LO and NLO parts under high energy expansion.\ 
With these, we can reformulate the scattering amplitudes
\eqref{eq:Amp-4AL-4A5} as follows:
\beqs
\label{eq:AmpNj-4AL-4A5}
\begin{align}
\TT[A_L^{a\hs n}A_L^{b\hs n}\hsm\ito A_L^{c\hs n}A_L^{d\hs n}]
\,=~\,& g^2 \!\(\!\frac{\,\CC_s\hs\NN_s^{\rm{el}}\,}{s} + 
\frac{\,\CC_t\hs \NN_t^{\rm{el}}\,}{t}  
+ \frac{\,\CC_u\,\NN_u^{\rm{el}}\,}{u} \!\)\!,
\\[1.mm]
\tT[A_5^{a\hs n}A_5^{b\hs n}\hsm\ito A_5^{c\hs n}A_5^{d\hs n}]
\,=~\,& g^2 \!\(\!\hsmx 
\frac{\,\CC_s\hs\NNt_s^{\rm{el}}\,}{s} + 
\frac{\,\CC_t\hs\NNt_t^{\rm{el}}\,}{t} + 
\frac{\,\CC_u\hs\NNt_u^{\rm{el}}\,}{u}
\hsm\!\) \!.
\end{align}
\eeqs
Then, we find that the LO numerators 
$\hs\{\NN_j^{\rm{el} \hs 0},\, \NNt_j^{\rm{el} \hs 0}\}\hs$ 
and the NLO numerators 
$\hs\{\dNN_j^{\rm{el}}, \, \dNNt_j^{\rm{el}}\}$ 
are both mass-dependent and
their sums violate the kinematic Jacobi identity
by terms of $\mO(E^0\Mnn)$ and smaller: 
\beqs
\label{eq:Nj-sum-nnnn}
\begin{align}
\label{eq:sum-Nj-LO}
&\sum_j\hsm\NN_j^{\rm{el} \hs 0} = 10\hs\ct \Mnn  \,,
\hspace*{10mm}
\sum_j\hsm\NNt_j^{\rm{el} \hs 0} = -6\hs\ct \Mnn  \,,
\\ 
\label{eq:Nj-sumLO-nnnn}
& \sum_j\! \da\NN_j^{\rm{el}} =
-2\Mnn\!\left[\!(7\hsmx +\hsmx\hs\ctt)
\hsm - 
\frac{\,4\hs (31\!+\hsm\ctf)\,}{\bs\hs s_\theta^2}
\hsm\right]\!\!\frac{~\ct~}{~s_\theta^2~} \hs ,
\\
& \sum_j\! \da\NNt_j^{\rm{el}} =
-2\Mnn\!\(\!1\hsm -\hsm
\frac{\,16\,}{~\bs\hs s_\theta^2~}\!\)\!\hsm 
\frac{(7\!+\hsm\ctt)\ct~}{s_\theta^2}\hs .
\end{align}
\eeqs
We note that all the $\mO(E^2)$ terms in the LO amplitudes are
mass-independent and obey the kinematic Jacobi identity as shown
in Eq.\eqref{eq:sum-Nj-LO}, while all the Jacobi-violating terms
in the LO/NLO amplitudes are mass-dependent 
and have $\mO(E^0\Mnn)$ or smaller.\ 
Because of these Jacobi-violating terms, 
the conventional BCJ double-copy method of the massless gauge theories
cannot be naively applied to the case of the elastic scattering 
amplitudes of KK gauge (Goldstone) bosons. However, we note that
the amplitudes \eqref{eq:AmpNj-4AL-4A5} are invariant 
under the generalized gauge transformations 
of the kinematic numerators:
\begin{equation}
\label{eq:GGtransf}
\NN_{\hsm j}^{\rm{el}\hs\pp} = \NN_{\hsm j}^{\rm{el}} + s_{\hsm j}^{}\hs\Delta^{\rm{el}} \hs ,
~~~~~~~
\NNt_{\hsm j}^{\rm{el}\hs\pp} = \NNt_{\hsm j}^{\rm{el}} + s_{\hsm j}^{}\hs
\widetilde{\Delta}^{\rm{el}} \hs .
\end{equation}
%%%
In the above, the gauge parameters 
$(\Delta^{\rm{el}},\,\widetilde{\Delta}^{\rm{el}})$ can be solved 
by requiring the gauge-transformed
numerators to satisfy the Jacobi identities:
$\sum_j\hsm\NN_{\hsm j}^{\rm{el}\hs\pp}\hsm\!=\hsm 0\hs$ and
$\sum_j\hsm\NNt_{\hsm j}^{\rm{el}\hs\pp}\hsm\!=\hsm 0\,$.\
Thus, we derive the following general solutions:
%%%
\begin{equation}
\label{eq:sol-Delta-tDelta}
\Delta^{\rm{el}} =
-\frac{1}{\,4\Mnn\,}\!\sum_j \NN_j^{\rm{el}}\hs ,
~~~~~~~
\widetilde{\Delta}^{\rm{el}}=
-\frac{1}{\,4\Mnn\,}\!\sum_j \NNt_j^{\rm{el}}\hs .
\end{equation}
%%%
Expanding both sides of \eqrefe{eq:sol-Delta-tDelta}, we derive the gauge parameters 
$(\Delta^{\rm{el}},\,\widetilde{\Delta}^{\rm{el}})=
(\Delta_0^{\rm{el}}\hsm +\hsm\Delta_1^{\rm{el}},\,
 \widetilde{\Delta}_0^{\rm{el}}\hsm +\hsm\widetilde{\Delta}_1^{\rm{el}})\, 
$ at the LO and NLO$\hs$: 
\begin{equation}
\begin{alignedat}{3}
\Delta_0^{\rm{el}} &=
\frac{1}{\,4\,}(9\hsm +\hsm 7\ctt)\hs\ct\csc^2\!\theta \hs ,
\qquad
&& \widetilde{\Delta}_0^{\rm{el}} =
\frac{1}{\,4\,}(17\!-\hsm\ctt)\ct\csc^2\!\theta \hs ,
\label{eq:sol-Delta01}
\\[2mm]
\Delta_1^{\rm{el}}& =
- \frac{~2\hs (31\! +\hsm\ctf )
\hs\ct\csc^4\!\theta~}{\bs}\,, \qquad
&& \widetilde{\Delta}_1^{\rm{el}}\! =
-\frac{~8\hs (7\! +\hsm \ctt)\hs\ct\csc^4\!\theta~}{\bs}\, .
\end{alignedat}
\end{equation}
Then, we can extend the conventional BCJ method and 
apply the color-kinematics duality to the following
gauge-transformed scattering amplitudes:
\beqs
\label{eq:AmpNj-4AL-4A5x}
\begin{align}
\TT[A_L^{a\hs n}A_L^{b\hs n}\hsm\ito A_L^{c\hs n}A_L^{d\hs n}]
\,=~\,& g^2 \!\(\!\frac{\,\CC_s\hs\NN_s^{\rm{el}\,\pp}}{s} + 
\frac{\,\CC_t\hs \NN_t^{\rm{el}\,\pp}\,}{t}  
+ \frac{\,\CC_u\,\NN_u^{\rm{el}\,\pp}\,}{u} \!\)\!,
\\
\tT[A_5^{a\hs n}A_5^{b\hs n}\hsm\ito A_5^{c\hs n}A_5^{d\hs n}]
\,=~\,& g^2 \!\(\!\hsmx 
\frac{\,\CC_s\hs\NNt_s^{\rm{el}\,\pp}\,}{s} + 
\frac{\,\CC_t\hs\NNt_t^{\rm{el}\,\pp}\,}{t} + 
\frac{\,\CC_u\hs\NNt_u^{\rm{el}\,\pp}\,}{u}
\hsm\!\) \!.
\end{align}
\eeqs
We find that this extended BCJ-type double-copy construction 
gives the correct LO KK graviton (Goldstone) amplitudes
at $\mO(E^2M_n^0)$, 
and also gives the correct structure of the NLO KK    
graviton (Goldstone) amplitudes at $\mO(E^0M_n^2)$
although the coefficients do not exactly match 
that of the original KK graviton (Goldstone) amplitudes at the NLO.
So, we need proper modifications on the extended double-copy 
construction of the massive NLO KK gauge/gravity amplitudes, 
as shown in Ref.\,\cite{Hang:2021fmp}.  
In the current study, we have demonstrated
in sections\,\ref{sec:3}-\ref{sec:4}  
that the double-copy construction
for the massive KK gauge/gravity amplitudes can be successfully 
realized by using the KK string-based formulation of the 
extended massive KLT-like relations, 
which hold for the exact $N$-point tree-level amplitudes
without making the high energy expansion.

\vspace*{1.5mm} 
\subsection{\hspace*{-2mm}Inelastic Scattering Amplitudes of 
KK Gauge and Goldstone Bosons}
\vspace*{1mm} 
\label{app:B2}

In this Appendix, we consider the KK YM gauge theory under
the 5d compactification of $\SS^1/\ZZ\hs$.
We present the full four-point scattering amplitudes 
for the inelastic channels 
$(n,n) \ito (m,m)$ and $(0,0) \ito (n,n)\hs$
at tree level, which were not given previously in 
Ref.\,\cite{Hang:2021fmp}.

\vspace*{1mm} 
\subsubsection{\hspace*{-2.5mm}Inelastic Scattering Amplitudes of 
\boldmath{$(n,n) \ito (m,m)$}}
\label{app:B2.1}

For the inelastic scattering process $(n,n) \ito (m,m)$, we compute 
the four-point scattering amplitudes of the 
longitudinal KK gauge and Goldstone bosons:
%%%%
\beqs
\label{eq:Tnnmm}
\begin{align}
\label{eq:Tnnmm-AL}
\TT[A_L^{a\hs n}A_L^{b\hs n}\hsm\ito A_L^{c\hs m}A_L^{d\hs m}]
\,&=\, g^2(\CC_s \KK_s^{\rm{in}} \!+ \CC_t \KK_t^{\rm{in}}
\!+ \CC_u \KK_u^{\rm{in}})  \hs,
\\[1mm]
\label{eq:Tnnmm-A5}
\tT[A_5^{a\hs n}A_5^{b\hs n}\hsm\ito A_5^{c\hs m}A_5^{d\hs m}]
\,&=\, g^2 (\CC_s \KKt_s^{\rm{in}} + \CC_t \KKt_t^{\rm{in}} 
+ \CC_u \KKt_u^{\rm{in}})  \hs,
\end{align}
\eeqs
where $\{\KK_j^{\rm{in}}\}$ denote the kinematic factors for KK gauge bosons,
\beqs
\label{eq:K-KT-nnmm}
\begin{align}
\KK_s^{\rm{in}}  & =
-\frac{\,2\hs\qb\hs\qb'(\bs\hs\rrp \!+\! 2\hs r^2)\hs\ct^{}\,}
{\bs\hs r^2} \hs,
\\[1.5mm]
\KK_t^{\rm{in}} & =
\frac{\,-(Q_0^{}\!+ Q_1^{}\ct^{}\! + Q_2^{}\ctt^{}
\!+\! Q_3^{}\cttt^{} )\,}
{\,r^2(3\bs^2 \!-\! 4\bs\hs\rrp \!-\!16\hs r^2
\!+\! 16\bs\hs\qb\hs\qb'\ct^{} \!+\! 4\hs\qb\hs\qb'\ctt^{})\,} \hs,
\\[1.5mm]
\KK_u^{\rm{in}} & =
\frac{\,Q_0^{}\!- Q_1^{}\ct^{}\! + Q_2^{}\ctt^{}
\!-\! Q_3^{}\cttt^{}\,}
{\,r^2(3\hs\bs^2 \!-\! 4 \bs\hs\rrp \!-\!16\hs r^2
\!-\! 16\hs\bs\hs\qb\hs\qb'\ct^{} \!+\! 4\qb\hs\qb'\ctt^{})\,}  \hs,
\end{align}
\eeqs
and $\{\KKt_j^{\rm{in}}\}$ denote the kinematic factors for KK Goldstone bosons,
\beqs
%\label{eq:K-KT-nnmm}
\begin{align}
\KKt_s^{\rm{in}} &=
-\frac{\,4\hs\qb\hs\qb'\ct^{}\,}{\bs} \hs,
\\[1.5mm]
\KKt_t^{\rm{in}} &=
\frac{\,\Qt_0^{}\!+\Qt_1^{}\ct^{}\! + \Qt_2^{}\ctt^{} \,}
{\,2\hs (\bs^2 \!-\! 16\hs r^2 \!+\! 8\bs\hs\qb\hs\qb'\ct^{} 
\!+\! 16\hs\qb^2\hs\qb'^2\cct)\,}  \hs,
\\[1.5mm]
\KKt_u^{\rm{in}} &=
\frac{\,-(\Qt_0^{}\!-\Qt_1^{}\ct^{}\! + \Qt_2^{}\ctt^{} )\,}
{\,2\hs (\bs^2 \!-\! 16\hs r^2 \!-\! 8\bs\hs\qb\hs\qb'\ct^{} 
\!+\! 16\hs\qb^2\qb'^2\cct)\,}  \hs.
\end{align}
\eeqs
In the above, 
the polynominal functions $\{Q_j^{},\hs\Qt_j^{}\}$ are expressed as
%%%
\begin{equation}
\begin{alignedat}{3}
Q_0&= \bs^3\rrp- \bs^2\hs (4\hs r^4\!+\!13\hs r^2\!+\!4) 
+20\hs\bs\hs r^2 \rrp - 16\hs r^4 \hs ,
\hspace*{9mm}
&&\Qt_0 = 5\hs\bs^2 \!-\! 16\hs r^2  \hs ,
\\[1mm]
Q_1 & = \bs\hs\qb\hs\qb'\hs [7\hs\bs\hs\rrp - 4\hs (\rrp\!+\!2)
(3\hs r^2\!+\!1)] \hs ,
\hspace*{9mm}
&& \Qt_1 = 16\hs\qb\hs\qb'(\bs\hsm -\hsm r_+^2) \hs ,
\\[1mm]
Q_2&= \bs^3\rrp - 
\bs^2\hs (4\hs r^4\!+\hsm 7\hs r^2\!+\hsm 4) 
+4\hs \bs\hs r^2\rrp + 16\hs r^4 
\hs ,
\hspace*{9mm}
&&\Qt_2= -16 \hs\qb^2 \qb'^2 \hs ,
\\[1mm]
Q_3 &= \bs\hs\qb\hs\qb'\rrp\hs (\bs - 4\rrp) \hs ,
\end{alignedat}
\end{equation}
%%%
where we have introduced the notations
$\,r\!=\hsm M_m^{}/M_n^{} \hs$ and
$\,r_+^2 \!=\! 1\!+\hsm r^2$,\, 
and other kinematic quantities are also defined in Eq.\eqref{eq:qsr}.

\vspace*{1mm}

Then, we make the high energy expansions for the above amplitudes
at the LO and NLO:
\beqs
\begin{alignat}{3}
&\TT[A_L^{a\hs n}A_L^{b\hs n}\hsm\ito A_L^{c\hs m}A_L^{d\hs m}] = \TT_{0L}\!+\da\TT_L^{} \hs ,
& \hspace*{6mm}
& \tT[A_5^{a\hs n}A_5^{b\hs n}\hsm\ito A_5^{c\hs m}A_5^{d\hs m}] = \tT_{05}\!+\da\tT_5^{} \hs ,
\\[1mm]
%%%%%%%%%
& \TT_{0L} = g^2({\CC_s\KK_s^{\rm{in}\hs 0}} \!+ 
{\CC_t\KK_t^{\rm{in}\hs 0}} \!+ {\CC_u\KK_u^{\rm{in}\hs 0}}) 
\hs, 
%%%%%%%%%
\hspace*{6mm}
&& \tT_{05}= g^2 ({\CC_s\KKt_s^{\rm{in}\hs 0}} \!+ 
{\CC_t\KKt_t^{\rm{in}\hs 0}} \!+ {\CC_u\KKt_u^{\rm{in}\hs 0}}) 
\hs, 
\\[1mm]
%%%%%%%%
&\dT_L^{} = g^2({\CC_s\da\KK_s^{\rm{in}}} \!+ 
{\CC_t\da\KK_t^{\rm{in}}} \!+ {\CC_u\da\KK_u^{\rm{in}}}) 
\hs, \quad
%%%%%%%%%%
&&\da\tT_5^{} =\, g^2({\CC_s\da\KKt_s^{\rm{in}}} 
\!+ {\CC_t\da\KKt_t^{\rm{in}}} 
\!+ {\CC_u\da\KKt_u^{\rm{in}}}) \hs.
\end{alignat}
\eeqs
We derive the LO inelastic scattering amplitudes as follows:
\beqs
\begin{alignat}{3}
\KK_s^{\rm{in}\hs 0} &=\, \ct \,,
\hspace*{14mm}
&& \KKt_s^{\rm{in}\hs 0} =\, -\ct^{}  \,,
\\[2.0mm]
\KK_t^{\rm{in}\hs 0} &=\, 
\frac{~4\!+\!\ct^{}\!+\!\ctt\,}{\,1\!+\hsm\ct^{}\,} \,,
\hspace*{14mm}
&& \KKt_t^{\rm{in}\hs 0} =\, 
\frac{~3\hsm -\hsm\ct^{}\,}{\,1\!+\hsm\ct^{}\,} \,,
\\[.5mm]
\KK_u^{\rm{in}\hs 0} &=\, 
-\frac{~4\!-\!\ct^{}\!+\!\ctt\,}{\,1\!-\hsm\ct^{}\,}\,, 
\hspace*{14mm}
&& \KKt_u^{\rm{in}\hs 0} =\,
-\frac{~3\hsm +\hsm\ct^{}\,}{\,1\!-\hsm\ct^{}\,}\,,
\end{alignat}
\eeqs
where we have dropped a common mass-dependent term 
$\,(r^2 + r^{-2})\hs\ct^{}\,$ in each   
$\,\KK_j^{\rm{in}\hs 0}\,$ by using the Jacobi identity
$\,\CC_s+\CC_t +\CC_u\!=\hsm 0\hs$.\ 
Thus, the remaining full LO amplitude is still mass-independent,
as we would expect.
Furthermore, we find that in each channel of $(s,t,u)$
the longitudinal KK gauge boson amplitude
and KK Goldstone boson amplitude differ
by the same amount at the LO (which is $r$-independent):
\begin{equation}
\KK_s^{\rm{in}\hs 0}\!-\KKt_s^{\rm{in}\hs 0} 
=\, \KK_t^{\rm{in}\hs 0}\!-\KKt_t^{\rm{in}\hs 0} 
=\, \KK_u^{\rm{in}\hs 0}\!-\KKt_u^{\rm{in}\hs 0}
=\, 2 \hs \ct \,.
\end{equation}
Hence, due to the Jacobi identity
the inelastic longitudinal KK gauge and Goldstone boson amplitudes
are equal, 
$\,\TT_{0L}^{}[A_L^{a\hs n/m}]\hsm =\hsm 
 \tT_{05}^{}[A_5^{a\hs n/m}]\hs$, 
in accord with the KK gauge boson equivalence theorem 
(KK\,GAET)\,\cite{5DYM2002}\cite{KK-ET-He2004}.

\vspace*{1mm}

We further compute the inelastic scattering amplitudes at the NLO
and derive these scattering amplitudes as follows:
\beqs
\begin{align}
\da\KK_s^{\rm{in}}	&\dis \,=\, \frac{~r_+^2\hs 
(1\!+\hsm r^4)\hs\ct^{}~}{r^2\hs\bs} \,,
\hspace*{-3mm}
%%%%%%
& \da\KKt_s^{\rm{in}} &\dis  \,=\, 
\frac{~2\hs r_+^2\hs\ct^{}~}{\bs} \,,
\\[1.5mm]
%%%%
\da\KK_t^{\rm{in}}	&\dis \,=\, 
\frac{\,r_+^2\hs (R_0 \!+\! R_1 \ct^{} \!+\! 
R_2 \ctt^{} \!+\! R_3\hs\cttt^{})\,}
{2\hs r^2\hs (3\!+\!4\ct^{}\!+\!\ctt)\hs\bs} \,,
%%%%%%%
& \da\KKt_t^{\rm{in}} &\dis \,=\,-
\frac{~4\hs r_+^2\hs (1\!-\!3\hs\ct^{})\,}
{~(3\!+\!4\ct^{}\!+\!\ctt)\hs\bs~} \,,
\\[1mm]
%%%
\da\KK_u^{\rm{in}}	&\dis \,=\,
-\frac{~r_+^2\hs 
(R_0 \!-\! R_1\ct^{} \!+\! R_2\ctt^{}\!-\! R_3\cttt^{})~}
{2\hs r^2\hs (3\!-\!4\ct^{}\!+\!\ctt)\hs\bs}\,,
%%%%%%
& \da\KKt_u^{\rm{in}} &\dis \,=\,
\frac{\,4\hs r_+^2\hs 
(1\!+\hsm 3\hs\ct^{})\,}
{~(3\!-\!4\ct^{}\!+\!\ctt)\hs\bs~}  \,,
\end{align}
\eeqs
where we have adopted the following notations:
%%%
\begin{alignat}{3}
R_0^{} &= 4(1\hsmx -\hsm 6\hs r^2 \!+\hsm r^4) \hs, 
\hspace*{7mm}
&& R_1^{}  = 7\hsm +\hsm 10\hs r^2 \!+\hsm 7\hs r^4  \hs,
\nn\\[1mm]
R_2^{}  &=  4\hs (1\hsmx +\hsm r^2)^2  \hs, 
\hspace*{7mm}
&& R_3  =  1\hsmx +\hsm 6\hs r^2 \!+\hsm r^4 \hs,
\\[1mm]
r &= M_m^{}/M_n^{} \hs,
\hspace*{7mm}
&& \hs r_+^2 = 1\hsmx +r^2 \hs.
\nn 
\end{alignat}
%%%
We further define the following LO and NLO inelastic numerators:
\beqs 
\label{eq:N-inelastic}
\begin{alignat}{3}
\label{eq:N-ineLO}
\NN_j^{\rm{in}\hs 0} &= s_j^{}\hs \KK_j^{\rm{in}\hs 0},
\hspace*{10mm}
&&
\dNN_j^{\rm{in}\hs 0} &= s_{j}^{}\dKK_j^{\rm{in}\hs 0},
\\[1mm]
\label{eq:N-ineNLO}
\NNt_j^{\rm{in}\hs 0} &= s_j^{} \hs \KKt_j^{\rm{in}\hs 0},
&&
\dNNt_j^{\rm{in}\hs 0} &= s_{j}^{}\dKKt_j^{\rm{in}\hs 0}.
\end{alignat} 
\eeqs 
Then, we compute their sums at the LO and NLO. 
We find that the sums of these numerators 
violate the kinematic Jacobi identities,
$\sum_j\NN_j^{\rm{in}}\!\!\neq\!\hsm 0\,$ and
$\sum_j\NNt_j^{\rm{in}}\!\!\neq\! 0\,$.\
Thus, to recover the kinematic Jacobi identity, 
we make the following generalized gauge transformations 
for the inelastic numerators: 
\begin{equation}
\label{eq:GGtransfIE}
\NN_{\hsm j}^{\rm{in}\hs\pp} = \NN_{\hsm j}^{\rm{in}} + s_{\hsm j}^{}\hs\Delta^{\rm{in}} \,,
\hspace*{10mm}
\NNt_{\hsm j}^{\rm{in}\hs\pp}= \NNt_{\hsm j}^{\rm{in}} + s_{\hsm j}^{}\hs
\widetilde{\Delta}^{\rm{in}} \,,
\end{equation}
%%%
under which the scattering amplitudes
\eqref{eq:Tnnmm-AL}-\eqref{eq:Tnnmm-A5} are invariant.
%%%
Then, imposing the kinematic Jacobi identities on the
gauge-transformed amplitudes 
$\,\sum_j\hsm\NN_{\hsm j}^{\rm{in}\hs\pp}\!\!=\!0\,$
and
$\,\sum_j\hsm\NNt_{\hsm j}^{\rm{in}\hs\pp}\!\!=\!0\,$,
we derive the general solutions of the gauge parameters 
$(\Delta^{\rm{in}},\,\widetilde{\Delta}^{\rm{in}})$ 
as follows: 
%%%
\begin{equation}
\label{eq:sol-Delta-tDeltaIE}
\Delta^{\rm{in}} =
-\frac{1}{\,2(\Mnn\hsm +\hsmx M_m^2)\,}\!\sum_j \NN_j^{\rm{in}} 
\,,~~~~
\hspace*{4mm} 
\widetilde{\Delta}^{\rm{in}} =
-\frac{1}{\,2(\Mnn\hsm +\hsmx M_m^2)\,}\!\sum_j \NNt_j^{\rm{in}} 
\,.~~~
\end{equation}
%%%

For the simplicity of illustration, we set $\hsx r\!=2\hsx$
and compute explicitly the sums of the inelastic numerators 
to the $\mO(E^{-2})\hs$:
%%%
\beqs 
\label{eq:Sum-Nj-Nj'-nnmm}
\begin{align}
\sum_j \NN_j^{\rm{in}}  &=
-\frac{1}{\,4\,} M_n^2\hs r_+^2\!
\left[\hsm (7\!+\!25\ctt )\hsm -\hsm  
\frac{\,(20867\hs\ct \!-\hsm 900\hs\ctt \!+\hsm 1025\hs\ctf)\,}
{20\hs\sst\hsx\bs}\hsm\right] 
\!\ct\csc^2\!\theta \,,
\\
\sum_j \NNt_j^{\rm{in}} &= 
-8 M_n^2\hs r_+^2\! 
\left[\hsm 1- \frac{~(565\hsm +\hsm 91\hs\ctt)~}
{20\hs\sst\hsx\bs}\hsm\right]
\!\ct\csc^2\!\theta \,,
\end{align}
\eeqs 
%%%
where $\,r\hsm =\hsm M_m^{}/M_n^{}\,$ 
and $\,r_+^2\! = 1\hsmx +r^2$.
With the above, we make high energy expansion of 
the general solutions \eqref{eq:sol-Delta-tDeltaIE} 
and derive explicitly the LO and NLO of gauge parameters:
%%%
\beqs
\begin{align}
\label{eq:Delta-nnmm}
\Delta_0^{\rm{in}} &= \frac{1}{8}(7\!+\hsm 25\hs\ctt)
\ct \csc^2\hsmx\theta  \,, 
\\[1.5mm]
\widetilde\Delta_0^{\rm{in}} &= 4\ct \csc^2 \hsmx \theta  \,,
\\[1.5mm]
\Delta_1^{\rm{in}} &=
-\frac{~(20867\hs\ct \!-\hsm 900\hs\ctt \!+\hsm 1025\hs\ctf)
\hs\ct\hsm\csc^4\!\theta~}{160\hs\bs} \,, 
\\[1mm]
\widetilde{\Delta}_1^{\rm{in}} &=
-\frac{~(565\hsm + 91\hs\ctt)
\hs\ct\csc^4\hsmx\theta~}{5\hs\bs} \,,
\end{align}
\eeqs
where we have set $\hsx r\!=2\hsx$ for illustration.

\vspace*{1.5mm} 
\subsubsection{\hspace*{-2.5mm}Inelastic Scattering Amplitudes of 
\boldmath{$(0,0) \ito (n,n)$}}
\label{app:B2.2}

Next, we study another inelastic channel $(0,0)\ito (n,n)$.\ 
We compute the following full tree-level scattering amplitudes of
KK gauge bosons and of KK Goldstone bosons:
%%%
\beqs
\begin{align}
\TT[A^0_{\pm 1} A^0_{\mp 1} \ito A^n_L A^n_L] 
&\,=\, g^2 ({\CC_s \KK^{\rm{in}}_s} \!+ 
{\CC_t \KK^{\rm{in}}_t} \!+ 
{\CC_u \KK^{\rm{in}}_u})  \,,
\\[1mm]
\tT[A^0_{\pm 1} A^0_{\mp 1} \ito A^n_5 A^n_5] 
&\,=\, g^2 ({\CC_s \KKt^{\rm{in}}_s} \!+ 
{\CC_t \KKt^{\rm{in}}_t} \!+ 
{\CC_u \KKt^{\rm{in}}_u})  \,,
\end{align}
\eeqs
where the sub-amplitudes $\{\KK_j^{\rm{in}}\}$ and
$\{\KKt_j^{\rm{in}}\}$ are given by
%%%
\beqs
\label{eq:K-KT-00nn}
\begin{alignat}{3}
\KK_s^{\rm{in}}  &=\,  0 \,,  \hspace*{14mm}
&& \KKt^{\rm{in}}_s =\, 0  \,,  
\\[1.5mm]
\KK_t^{\rm{in}} &=\, 
\frac{-(\bs \hsm +\hsm 4)\hs\sst}
{~\bs\hsm +\hsm [\bs\hs (\bs\hsm -\hsm 4)]^{1/2}\ct^{}~}\,,  
\hspace*{14mm}
&& \KKt_t^{\rm{in}} 
\,=\, \frac{-(\bs \hsm -\hsm 4)\hs\sst}
{\bs \hsm +\hsm [\bs\hs (\bs\hsm -\hsm 4)]^{1/2}\ct^{}~}  \,,
\\[0mm]
\KK_u^{\rm{in}} 
&=\, \frac{(\bs \hsm +\hsm 4)\hs \sst}
{~\bs \hsm -\hsm [\bs\hs (\bs\hsm -\hsm 4)]^{1/2}\hs\ct^{}~}
\,,  \hspace*{14mm}
&& \KKt_u^{\rm{in}} \,=\, 
\frac{(\bs\hsm -\hsm 4)\hs \sst}
{~\bs\hsm -\hsm [\bs\hs (\bs\hsm -\hsm 4)]^{1/2}\hs\ct^{}~}  \,.
\end{alignat}
\eeqs
%%%
Then, we make the high energy expansions of the above amplitudes
at the LO and NLO:
\beqs
\begin{alignat}{3}
&\TT[A^0_{\pm 1} A^0_{\mp 1} \ito A^n_L A^n_L] = \TT_{0L}\!+\da\TT_L \,,
& \hspace*{8mm}
&\tT[A^0_{\pm 1} A^0_{\mp 1} \ito A^n_5 A^n_5] = \tT_{05}\!+\da\tT_5 \,,
\\[1mm]
%%%%%%%%%
&\TT_{0L} = g^2( \CC_t\hs\KK_t^{\rm{in}\hs 0} + 
\CC_u\hs\KK_u^{\rm{in}\hs 0}) \,, 
%%%%%%%%%
\hspace*{8mm}
&&\tT_{05}= g^2 ( \CC_t\KKt_t^{\rm{in}\hs 0} + 
\CC_u\KKt_u^{\rm{in}\hs 0}) \,, 
\\[1mm]
%%%%%%%%
&\da\TT_L= g^2 (\CC_t\hs\da\KK_t^{\rm{in}} + 
\CC_u\hs\da\KK_u^{\rm{in}}) \,, 
%%%%%%%%%%
& \hspace*{8mm}
&\da\tT_5 = g^2(\CC_t\hs\da\KKt_t^{\rm{in}}+ \CC_u\hs\da\KKt_u^{\rm{in}}) \,.
\end{alignat}
\eeqs
Since the $s$-channel sub-amplitudes vanish, 
$\hs\KK_s^{\rm{in}}\!=\KKt_s^{\rm{in}}\!=0\hs$, 
we derive the following LO inelastic sub-amplitudes
of $(t,\,u)$ channels,  
%
%\beqs
\begin{align}
\KK_t^{\rm{in}\hs 0}  =\, 
\KKt_t^{\rm{in}\hs 0} = -(1 \!-\! \ct ) \,, \qquad
\KK_u^{\rm{in}\hs 0}  =\, 
\KKt_u^{\rm{in}\hs 0} = 1 \!+\hsm \ct  \,,
\end{align}
%\eeqs
%
and the following NLO inelastic sub-amplitudes, 
\beqs
\begin{alignat}{3}
\da\KK_t^{\rm{in}}	 \,&=\, 
-\frac{~1 \!+\! 2\hs\ct \!-\! 3\hs\ctt~}{(1\!+\!\ct)\bs}
\,, \hspace*{10mm}
&& \da\KKt_t^{\rm{in}} = 
\frac{~3 \!-\! 2\hs\ct \!-\! \ctt~}{(1\!+\!\ct)\bs}  \,,
\\[1mm]
\da\KK_u^{\rm{in}}	\,&=\, 
\frac{~1 \!-\! 2\hs\ct \!-\! 3\hs\ctt~}{(1\!-\!\ct)\bs} \,, 
\hspace*{10mm}
&& \da\KKt_u^{\rm{in}} 
\,=\, - \frac{~3 \!+\! 2\hs\ct \!-\! \ctt~}{(1\!-\!\ct)\bs}  \,.
\end{alignat}
\eeqs
%%%
Then, we define the LO and NLO numerators of the inelastic amplitude 
as in Eq.\eqref{eq:N-inelastic}. 
With these, we compute the sums of the inelastic numerators 
to the $\mO(E^{-2})$:
%%%
\beqs
\begin{align}
\sum_j \NN_j^{\rm{in}}  &=\, 
2\ct\Mnn\!\left[ 1- \frac{~(1\!+\hsm 3 \ctt)
\hsmx\csc^2\!\theta~}{\,\bs\,}\right]\!,
\\
\sum_j \NNt_j^{\rm{in}} &=\,  
2\ct\Mnn\!\left[ 1- \frac{~(5\hsm -\hsm \ctt)\hsmx
\csc^2\!\theta~}{\bs}\right]\!.
\end{align}
\eeqs
%%%
To recover the kinematic Jacobi identities, 
we make the generalized gauge-transformations 
\eqref{eq:GGtransfIE} on the numerators such that 
$\,\sum_j\hsm\NN_{\hsm j}^{\rm{in}\hs\pp}\!\!=\!0\,$
and
$\,\sum_j\hsm\NNt_{\hsm j}^{\rm{in}\hs\pp}\!\!=\!0\,$.
Thus, we derive the following general solutions 
of the gauge parameters 
$(\Delta^{\rm{in}},\,\widetilde{\Delta}^{\rm{in}})$: 
%%%
\begin{equation}
\label{eq:sol-Delta-00nn}
\Delta^{\rm{in}} =
-\frac{1}{\,2\Mnn\,}\!\sum_j \NN_j^{\rm{in}} 
\,,~~~~
\hspace*{4mm} 
\widetilde{\Delta}^{\rm{in}} =
-\frac{1}{\,2\Mnn\,}\!\sum_j \NNt_j^{\rm{in}} 
\,.~~~
\end{equation}
%%%
Finally, under high energy expansion,  
we derive the LO and NLO gauge parameters as follows:
%%%
\begin{align}
\Delta_0^{\rm{in}} =  \widetilde\Delta_0^{\rm{in}} = - \ct\,, 
\qquad
\Delta_1^{\rm{in}} = 
\frac{~\ct (1\!+\hsm 3\hs\ctt)~}
{\,\sst\,\bs\,}\,,
\qquad
\widetilde{\Delta}_1^{\rm{in}} = 
\frac{~\ct (5\!-\hsm\ctt)~}
{\,\sst\,\bs\,}\,.
\end{align}

\vspace*{4mm}

%\newpage 

\end{document}